\documentclass[twocolumn,twocolappendix]{aastex63}

\usepackage{multirow}

\newcommand{\fdeg}{\mbox{$. \! ^{\circ}$}}

\graphicspath{{./}}

\received{2019 December 17}
\revised{2020 January 21}
\accepted{2020 January 22}
\submitjournal{ApJ}

\shorttitle{Ages of single and multiple stellar populations in seven bulge
globular clusters}
\shortauthors{Oliveira et al.}
\defcitealias{2015AJ....149...91P}{Paper I}
\defcitealias{2015ApJ...808...51M}{Paper III}
\defcitealias{2015MNRAS.451..312N}{Paper IV}
\defcitealias{2015MNRAS.454.4197R}{Paper V}
\defcitealias{2016ApJ...822...44B}{Paper VII}
\defcitealias{2017MNRAS.464.3636M}{Paper IX}
\defcitealias{2018MNRAS.475.4088L}{Paper XII}
\defcitealias{2018MNRAS.481.5098M}{Paper XVI}
\defcitealias{2018MNRAS.481.3382N}{Paper XVII}
\defcitealias{2019ApJ...873..109L}{Paper XVIII}
\defcitealias{1996AJ....112.1487H}{H10}
\defcitealias{2009A&A...508..695C}{C09}
\begin{document}

\title{The {\it Hubble Space Telescope} UV Legacy Survey of Galactic Globular Clusters. XX. Ages of single and \\ multiple stellar populations in seven bulge globular clusters}

\correspondingauthor{Raphael A.~P. Oliveira, Stefano O. Souza}
\email{rap.oliveira@usp.br, stefano.souza@usp.br}

\author[0000-0002-4778-9243]{R.~A.~P. Oliveira}
\affiliation{Universidade de S\~ao Paulo, IAG, Rua do Mat\~ao 1226, Cidade
Universit\'aria, S\~ao Paulo 05508-900, Brazil}

\author[0000-0001-8052-969X]{S.~O. Souza}
\affil{Universidade de S\~ao Paulo, IAG, Rua do Mat\~ao 1226, Cidade
Universit\'aria, S\~ao Paulo 05508-900, Brazil}

\author[0000-0002-7435-8748]{L.~O. Kerber}
\affil{Universidade de S\~ao Paulo, IAG, Rua do Mat\~ao 1226, Cidade
Universit\'aria, S\~ao Paulo 05508-900, Brazil}
\affil{Universidade Estadual de Santa Cruz, Rodovia Jorge Amado km 16,
Ilh\'eus 45662-000, Brazil}

\author[0000-0001-9264-4417]{B. Barbuy} 
\affil{Universidade de S\~ao Paulo, IAG, Rua do Mat\~ao 1226, Cidade
Universit\'aria, S\~ao Paulo 05508-900, Brazil}

\author[0000-0001-7939-5348]{S. Ortolani}
\affil{Dipartimento di Fisica e Astronomia `Galileo Galilei', Universit\`a
di Padova, Vicolo dell'Osservatorio 3, Padova I-35122, Italy}
\affil{Istituto Nazionale di Astrofisica\,--\,Osservatorio Astronomico di
Padova, Vicolo dell'Osservatorio 5, Padova I-35122, Italy}

\author[0000-0002-9937-6387]{G. Piotto}
\affil{Dipartimento di Fisica e Astronomia `Galileo Galilei', Universit\`a
di Padova, Vicolo dell'Osservatorio 3, Padova I-35122, Italy}
\affil{Istituto Nazionale di Astrofisica\,--\,Osservatorio Astronomico di
Padova, Vicolo dell'Osservatorio 5, Padova I-35122, Italy}

\author[0000-0003-1149-3659]{D. Nardiello}
\affil{Dipartimento di Fisica e Astronomia `Galileo Galilei', Universit\`a
di Padova, Vicolo dell'Osservatorio 3, Padova I-35122, Italy}
\affil{Istituto Nazionale di Astrofisica\,--\,Osservatorio Astronomico di
Padova, Vicolo dell'Osservatorio 5, Padova I-35122, Italy}
\affil{Aix Marseille Univ, CNRS, CNES, LAM, Marseille, France}

\author[0000-0002-5974-3998]{A. P\'erez-Villegas} 
\affil{Universidade de S\~ao Paulo, IAG, Rua do Mat\~ao 1226, Cidade
Universit\'aria, S\~ao Paulo 05508-900, Brazil}

\author[0000-0002-2569-4032]{F.~F.~S. Maia} 
\affil{Universidade Federal do Rio de Janeiro,  Av. Athos da Silveira Ramos,
149 - Cidade Universit\'aria, Rio de Janeiro 21941-909, Brazil}

\author[0000-0003-3336-0910]{E. Bica}
\affil{Universidade Federal do Rio Grande do Sul, Departamento de Astronomia,
CP 15051, Porto Alegre 91501-970, Brazil}

\author[0000-0001-5870-3735]{S. Cassisi}
\affil{INAF - Astronomical Observatory of Abruzzo, Via M. Maggini, sn, 64100
Teramo, Italy}
\affil{INFN - Sezione di Pisa, Largo Pontecorvo 3, 56127 Pisa, Italy}

\author[0000-0003-4697-0945]{F. D'Antona}
\affil{INAF - Osservatorio Astronomico di Roma, Via Frascati 33, I-00040,
Monte Porzio Catone, Roma, Italy}

\author[0000-0003-1713-0082]{E. Lagioia}
\affil{Dipartimento di Fisica e Astronomia `Galileo Galilei', Universit\`a
di Padova, Vicolo dell'Osservatorio 3, Padova I-35122, Italy}
\affil{Istituto Nazionale di Astrofisica\,--\,Osservatorio Astronomico di
Padova, Vicolo dell'Osservatorio 5, Padova I-35122, Italy}

\author[0000-0001-9673-7397]{M. Libralato}
\affil{Space Telescope Science Institute, 3700 San Martin Dr., Baltimore, MD
21218, USA}

\author[0000-0001-7506-930X]{A.~P. Milone}
\affil{Dipartimento di Fisica e Astronomia `Galileo Galilei', Universit\`a
di Padova, Vicolo dell'Osservatorio 3, Padova I-35122, Italy}
\affil{Istituto Nazionale di Astrofisica\,--\,Osservatorio Astronomico di
Padova, Vicolo dell'Osservatorio 5, Padova I-35122, Italy}

\author[0000-0003-2861-3995]{J. Anderson}        %%% Didn't find his ORCID
\affil{Space Telescope Science Institute, 3700 San Martin Dr., Baltimore, MD
21218, USA}

\author[0000-0002-6054-0004]{A. Aparicio}
\affil{Instituto de Astrof\`isica de Canarias, E-38200 La Laguna, Canary
Islands, Spain}
\affil{Department of Astrophysics, University of La Laguna, E-38200 La Laguna,
Tenerife, Canary Islands, Spain}
%\affil{Instituto de Astrof\'{i}sica de Canarias (IAC), calle V\'{i}a L\'{a}ctea
%s/n, E-38205, La Laguna, Tenerife, Spain}
%\affil{Universidad de La Laguna, Dpto. Astrof\'{i}sica, E-38206, Avenida
%Francisco S\'{a}nchez s/n, E-38206, La Laguna, Tenerife, Spain}

\author[0000-0003-4080-6466]{L.~R. Bedin}
\affil{Istituto Nazionale di Astrofisica\,--\,Osservatorio Astronomico di
Padova, Vicolo dell'Osservatorio 5, Padova I-35122, Italy}

% NARDIELLO: REMOVE?? Not active...
%\author[0000-0003-3858-637X]{A. Bellini}
%\affil{Space Telescope Science Institute, 3700 San Martin Dr., Baltimore, MD
%21218, USA}

\author[0000-0002-1793-9968]{T.~M. Brown}
\affil{Space Telescope Science Institute, 3700 San Martin Dr., Baltimore, MD
21218, USA}

%\author{A.~M. Cool}     % Orcid???
%\affil{Department of Physics and Astronomy, San Francisco State University,
%1600 Holloway Avenue, San Francisco, CA 94132, USA}

%\author[0000-0002-0002-9298]{S. Hidalgo}   % REMOVED: E-mail.
%\affil{Instituto de Astrof\`isica de Canarias, E-38200 La Laguna, Canary
%Islands, Spain}
%\affil{Department of Astrophysics, University of La Laguna, E-38200 La Laguna,
%Tenerife, Canary Islands, Spain}

\author{I.~R. King}     % Orcid???
\affil{Department of Astronomy, University of Washington, Box 351580, Seattle}

\author[0000-0002-1276-5487]{A.~F. Marino}
\affil{Dipartimento di Fisica e Astronomia `Galileo Galilei', Universit\`a
di Padova, Vicolo dell'Osservatorio 3, Padova I-35122, Italy}

%\author[0000-0001-5292-6380]{M. Monelli}   % REMOVED: E-mail.
%\affil{Instituto de Astrof\`isica de Canarias, E-38200 La Laguna, Canary
%Islands, Spain}
%\affil{Department of Astrophysics, University of La Laguna, E-38200 La Laguna,
%Tenerife, Canary Islands, Spain}

\author[0000-0003-3795-9031]{A. Pietrinferni}
\affil{INAF - Astronomical Observatory of Abruzzo, Via M. Maggini, sn, 64100
Teramo, Italy}

\author[0000-0002-7093-7355]{A. Renzini}
\affil{Istituto Nazionale di Astrofisica\,--\,Osservatorio Astronomico di
Padova, Vicolo dell'Osservatorio 5, Padova I-35122, Italy}

\author{A. Sarajedini}      % Orcid???
\affil{Charles E. Schmidt College of Science, Florida Atlantic University,
777 Glades Rd, Boca Raton,  FL 33431, USA}

\author[0000-0001-7827-7825]{R. van der Marel}
\affil{Space Telescope Science Institute, 3700 San Martin Dr., Baltimore,
MD 21218, USA}
\affil{Center for Astrophysical Sciences, Department of Physics \& Astronomy,
Johns Hopkins University, Baltimore, MD 21218, USA}

\author[0000-0003-2742-6872]{E. Vesperini}
\affil{Department of Astronomy, Indiana University, Bloomington, IN 47401,
USA}

\begin{abstract}
In the present work we analyzed seven globular clusters selected from
their location in the Galactic bulge and with metallicity values in the range
$-1.30 \lesssim \rm {[Fe/H]}\lesssim -0.50$. The aim of this work is first to
derive cluster ages assuming single stellar populations, and secondly, to
identify the stars from first (1G) and second generations (2G) from the main
sequence, subgiant and red giant branches, and to derive their age differences.
Based on a combination of UV and optical filters used in this project, we apply
the Gaussian mixture models to distinguish the multiple stellar populations.
Applying statistical isochrone fitting, we derive self-consistent ages,
distances, metallicities, and reddening values for the sample clusters.
An average age of $12.3\pm0.4$\,Gyr was obtained both using Dartmouth and BaSTI
(accounting atomic diffusion effects) isochrones, without a clear distinction
between the moderately metal-poor and the more metal-rich bulge clusters, except
for NGC\,6717 and the inner halo NGC\,6362 with $\sim 13.5$\,Gyr. We derived a
weighted mean age difference between the multiple populations hosted by each
globular cluster of $41\pm170$\,Myr adopting canonical He abundances; whereas
for higher He in 2G stars, this difference reduces to $17\pm170$\,Myr, but with
individual uncertainties of $500$\,Myr.
%The stellar generations are coeval within the total errors of $\pm 0.5$\,Gyr.
\end{abstract}


%%%%%%%%%%%%%%%%%%%%%%%%%%%%%%%%%%%%%%%%%%%%%%%%%%%%%%%%%%%%%%%%%%%%%%%%%%%%%%%%
%%%%%%%%%%%%%%%%%%%%%%%%%%%%%%%%% INTRODUCTION %%%%%%%%%%%%%%%%%%%%%%%%%%%%%%%%%
%%%%%%%%%%%%%%%%%%%%%%%%%%%%%%%%%%%%%%%%%%%%%%%%%%%%%%%%%%%%%%%%%%%%%%%%%%%%%%%%
\section{Introduction} \label{sec:intro}

The early formation and present configuration of the Galactic bulge is complex,
and studies of its stellar populations can give hints on its formation and
evolution processes \citep{2018ARA&A..56..223B}. Globular clusters (GCs) are
tracers of the formation and chemodynamical evolution of the Milky Way (MW).
In particular, the bulge GCs are witnesses of the earliest stages of the Galaxy
formation. As an example, the old age of $\sim13$\,Gyr derived for some of the
moderately metal-poor bulge GCs with a blue horizontal branch, in particular
NGC\,6522, NGC\,6626 and HP\,1 \citep{2018ApJ...853...15K,2019MNRAS.484.5530K},
suggest that they were formed before the present configuration of the bulge/bar
%\citep[][]{2018ApJ...863...16R},
%The Galactic bar was recently estimated to
%have an age of $8\pm2$\,Gyr by \citet{2018ApJ...861...88B} and \citet
%{2019MNRAS.490.4740B}.
component \citep[][]{2018ApJ...863...16R}, considering that the Galactic bar was
recently estimated to have an age of $8\pm2$\,Gyr \citep{2018ApJ...861...88B,
2019MNRAS.490.4740B}.

The \textit{Hubble Space Telescope} (\textit{HST}) UV Legacy Survey of Galactic
GCs \citep[GO-13297 program, PI G. Piotto;][hereafter \citetalias
{2015AJ....149...91P} of this series]{2015AJ....149...91P} allowed to obtain
photometry for 56 GCs with the UV/blue filters F275W, F336W and F438W of the
Ultraviolet and Visual Channel of the Wide Field Camera 3 (UVIS/WFC3). The main
goals of this survey are to identify and investigate the nature of the multiple
stellar populations (MPs) in GCs. The bandpasses of this ``magic trio'' of
filters include the OH, NH, CN and CH molecular bands \citepalias
{2015AJ....149...91P}, thus showing the C, N and O abundance variations, as
detected in spectroscopy \citep[e.g.][]{2012A&ARv..20...50G}. Previous data in
optical filters (F606W and F814W), obtained with the Advanced Camera for Surveys
(ACS), the Survey of Galactic GCs \citep[GO-10775 program, PI A. Sarajedini;][]
{2007AJ....133.1658S}, are also available.

In the present work, we analyze the seven bulge GCs observed in these \textit
{HST} programs that are within the selection of bulge GCs by \citet
{2016PASA...33...28B}: NGC\,6304, NGC~6352, NGC\,6624 and NGC\,6637 (M69), with
literature metallicities in the range $-0.75 \lesssim \rm{[Fe/H]} \lesssim
-0.35$; and NGC\,6652, NGC\,6717 (Pal\,9) and NGC\,6723, with $-1.25\lesssim
\rm{[Fe/H]} \lesssim-0.75$ (\citealp [][2010 edition]{1996AJ....112.1487H}%
\footnote{\url{http://www.physics.mcmaster.ca/~harris/mwgc.dat}}; \citealt
{2009A-A...508..695C}). They are located at Galactic latitudes $|b|\leq 17\fdeg
50$ and Galactocentric distances $R_{\rm GC}\leq 3.3$\,kpc, therefore within
the bulge volume, and they are a representative sample of the moderately
metal-rich and moderately metal-poor bulge GCs. In a recent classification from
orbital analysis by \citet{2019MNRAS.tmp.2773P}, the GCs NGC\,6304, NGC\,6624,
NGC\,6637 and NGC\,6717 are confirmed as bulge members. 
%NGC\,6352 could be bulge or thick disk, depending on the velocity pattern of
%the bar, and
NGC\,6352 has a higher probability to be part of the thick disk for a high bar
speed, whereas for the case of slow bar pattern speed it could also have a
significant probability to be part of the bulge;
NGC\,6723 and NGC\,6652 would belong to the thick disk. The old inner halo
cluster NGC\,6362, recently studied by \citet{2018ApJ...853...15K} in terms of
ages, and by \citet{2016ApJ...824...73M} and \citet{2017MNRAS.468.1249M} in
terms of spectroscopic abundances, was also selected for comparison purposes.

Accurate age determinations for GCs depend on high-precision photometry and
are indeed a great challenge, particularly for those in the Galactic bulge,
due to the combination of high field stellar contamination, strong total and
differential extinction, and also stellar crowding effects normally present in
any GC. Recent efforts have been made by using the \textit{HST} in the visible
and near-infrared (NIR), NIR detectors assisted by adaptive optics (AO) systems
in $8-10$\,m class telescopes and, when possible, proper-motion (PM) cleaning
techniques to overcome these obstacles \citep[e.g.][]{2014AJ....148...18C,
2018AJ....156...41C,2014ApJ...782...50L,2016ApJ...823...18C,2018ApJ...864..147C,
2016ApJ...828...75F,2016ApJ...832...48S,2019ApJ...874...86S,2018ApJ...853...15K,
2019MNRAS.484.5530K}.

The reliability of age derivations require a statistical and self-consistent
analysis to be applied to deep and multiband color-magnitude diagrams (CMDs),
such as those provided among  the more recent ones, by the \textit{HST} UV
Legacy Survey of Galactic GCs \citep{2015AJ....149...91P}, the ACS Survey of
Galactic GCs \citep{2007AJ....133.1658S} and the WFC3/NIR survey of GCs toward
the Galactic bulge \citep{2018AJ....156...41C}. 

Absolute and relative ages were determined for 69 GCs using the ACS Survey of
Galactic GCs data \citep{2007AJ....133.1658S}, applying different isochrone
fitting methods and theoretical models \citep{2009ApJ...694.1498M,
2010ApJ...708..698D,2013ApJ...775..134V,2017MNRAS.468.1038W}. The bulge GC
NGC\,6352 was analyzed in detail by \citet[\citetalias{2015MNRAS.451..312N}]
{2015MNRAS.451..312N}. In the present work we carry out a detailed analysis for
the seven sample bulge GCs, including NGC~6352.

%The GCs were originally thought to host a single stellar population (SSP),
%meaning that their stars would have formed in a single burst and share the same
%age, initial chemical composition, as well as distance and reddening. 
As concerns MP analyses,
the first detections of anomalous abundances of light elements from
proton-capture processes were presented in the 1970s \citep[e.g.][]
{1971Obs....91..223O}, but no theories about MPs were developed yet, because
the evidence was restricted to evolved giant stars. For a few clusters, in
particular M22 and $\omega$ Centauri, evidence on metallicity variations were
suggested \citep[e.g.][]{1982ApJ...263..187P,1977ApJS...33..471H} -- see also
photometric evidence from \citet{1999Natur.402...55L}, but these clusters are
considered to be special cases until today.
Since the first photometric evidence of MPs among unevolved stars of GCs with
\textit{HST} by \citet{2004ApJ...605L.125B} and \citet{2005ApJ...621..777P}, the
observations have been showing that the phenomenon is common to almost all GCs
so far studied. Reviews on the evidence revealing chemical abundance
anomalies can be found in \citet{1994PASP..106..553K} and \citet
{2012A&ARv..20...50G}. Reviews on photometric evidence of MPs were presented
by \citet{2009IAUS..258..233P}, and more recently by \citet[and references
therein]{2018ARA&A..56...83B}.

\citet[\citetalias{2015MNRAS.454.4197R}]{2015MNRAS.454.4197R} describe the
possible scenarios for 2G formation from the ejected material processed by 1G
stars, including massive asymptotic giant branch (AGB) stars, massive
interacting binaries and fast-rotating massive stars. These models predict
age differences between 1G and 2G going from about zero \citep[for the
supermassive star model, e.g.][]{2018MNRAS.478.2461G}, to $\lesssim30$\,Myr
\citep[for the massive rotating star model, see e.g.][]{2007A-A...464.1029D},
to about $50-100$\,Myr \citep[for the AGB model, e.g.][]{2008MNRAS.391..825D}
and possibly up to $\sim 150$\,Myr \citep{2016MNRAS.458.2122D} for the Type II
clusters \citepalias{2017MNRAS.464.3636M}.
Although none of the scenarios can reproduce all the observational evidence, a
reliable determination of the age differences between MPs might turn out to
favor one or some of them. Therefore, an extremely accurate determination
of age differences, of less than 1\% of the cluster ages, would be needed to
constrain the models. This level of precision cannot be reached
for old clusters, and it might be feasible by analysing younger clusters, 
although other problems (e.g. rotation)
may arise  to hamper accurate determinations \citep[e.g.][]{2017NatAs...1E.186D,
2018MNRAS.477.2640M}.

In this work, the ages, metallicities, distance moduli and reddening values
are derived via isochrone fitting following a Bayesian approach, assuming both
the single and multiple stellar populations. The MPs are identified from the
main sequence (MS) to the red giant branch (RGB). Theoretical stellar
evolutionary models with canonical He abundances and $\alpha$-enhancement, from
\textit{A Bag of Stellar Tracks and Isochrones} \citep[BaSTI,][]
{2006ApJ...642..797P} and \textit{Dartmouth Stellar Evolutionary Database}
\citep[DSED,][]{2008ApJS..178...89D} were adopted. The isochrone fitting was
carried out in membership probability cleaned CMDs with optical filters.

The high precision in the relative age derivations from our isochrone fitting
method allows us to investigate a possible age difference between the MPs,
otherwise impossible with higher uncertainties. In \citetalias
{2015MNRAS.451..312N} the relative ages between the MPs in NGC\,6352 were
derived, obtaining an age difference of $10\pm110$\,Myr and a He abundance
variation of $\Delta Y = 0.029\pm0.006$, assuming no difference in $\rm{[Fe/H]}
$ and $\rm{[\alpha/Fe]}$. However, adopting a small variation in $\rm{[Fe/H]}$
and $\rm{[\alpha/Fe]}$, the uncertainty in the age difference rises to 280\,Myr.

In Section~\ref{sec:Obs}, the GO-13297 observations are briefly described.
Section~\ref{sec:Isoc-Method} presents the statistical methods for isochrone
fitting, following a Bayesian approach. In Section~\ref{sec:Results}, the
isochrone fits considering the sample GCs as a single stellar population (SSP)
are shown. Section~\ref{sec:MPs} presents the separation of their MPs and the
age derivation for each generation. In Section~\ref{sec:Concl}, conclusions
are drawn.

\begin{figure}
 \centering
 \includegraphics[width=0.99\columnwidth]{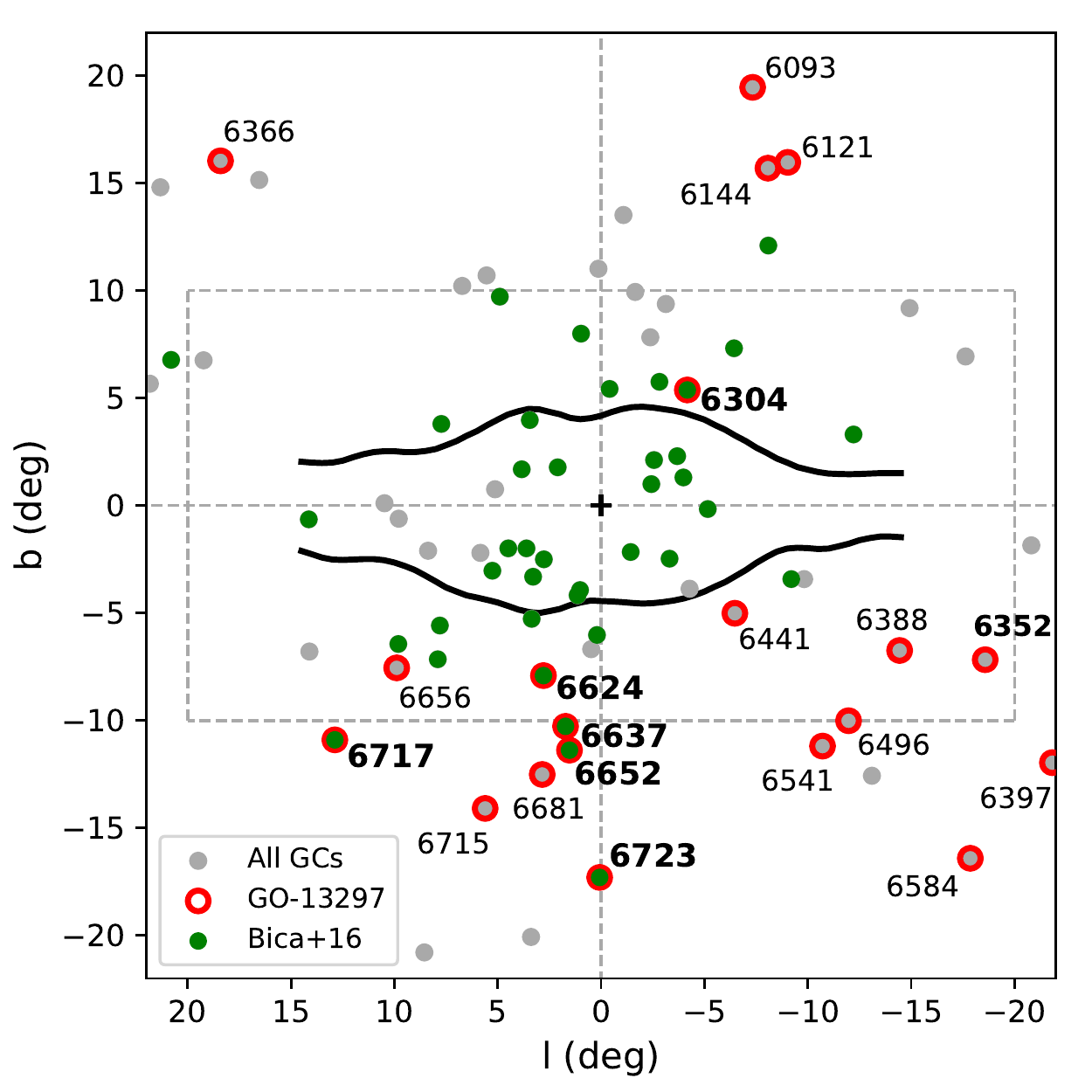}
 \caption{Location in Galactic coordinates of the GCs presented in \citet[2010
 edition]{1996AJ....112.1487H} catalog, within $|\ell|<22\arcdeg$ and $|b|<
 22\arcdeg$. The open red circles correspond to the 56 GCs observed in the
 GO-13297 program. The black contours correspond to the COBE/DIRBE outline of
 the peanut bulge \citep{1994ApJ...425L..81W,2017A&A...598A.101J}, and the
 dashed lines delimit the central region with $|\ell|<20\arcdeg$ and $|b|<10
 \arcdeg$. The only six clusters with $R_{\rm GC} < 3$\,kpc and $\rm{[Fe/H]} >
 -1.5$, classified as bulge GCs in \citet[green circles]{2016PASA...33...28B},
 were selected for this present analysis. NGC\,6352, which is classified as an
 outer bulge GC, was also selected.}
 \label{fig:piotto-GCs}
\end{figure}

\begin{deluxetable*}{lcccccccccc}[ht]
\tablenum{1}
\tablecaption{Cluster parameters and metallicity from \citet[2010 edition, H10]
{1996AJ....112.1487H}, other metallicity values from \citet[C09]
{2009A-A...508..695C} and \citet[V18]{2018A-A...619A..13V}, HB color differences
from \citet{2010ApJ...708..698D} and masses from \citet{2018MNRAS.478.1520B}. The
horizontal lines in the following tables divide the bulge GCs from the inner
halo one.}
\label{tab:Harris}
\tablewidth{0pt}
\tablehead{
\colhead{Cluster} & \colhead{$\ell$} & \colhead{$b$} &  \colhead{$d_{\odot}$} &
\colhead{$R_{\rm GC}$} & \colhead{$E(B-V)$} & $\rm{[Fe/H]}_{\rm{H10}}$ & $\rm
{[Fe/H]}_{\rm{C09}}$ & $\rm{[Fe/H]}_{\rm{V18}}$ & \colhead{$\Delta(V-I)$} &
\colhead{Mass} \\ & \colhead{(deg)} & \colhead{(deg)} & \colhead{(kpc)} &
\colhead{(kpc)} & \colhead{(mag)} & \colhead{(dex)} & \colhead{(dex)} &
\colhead{(dex)} & \colhead{(mag)} & \colhead{($10^5$\,M$_{\odot}$)}
}
%\decimalcolnumbers
\startdata
NGC\,6304 & 355.83 & 5.38 & 5.9 & 2.3 & 0.54 & $-0.45$ & $-0.37$ & $-0.43\pm0.05$
& $0.105$ & 2.61 \\ % \pm0.004
NGC\,6352 & 341.42 & $-7.17$ & 5.6 & 3.3 & 0.22 & $-0.64$ & $-0.62$ & $-0.54\pm
0.03$ & $0.123$ & 0.596 \\ % \pm0.012
% 17 25 29.11  -48 25 19.8   341.42   -7.17    5.6   3.3   5.3  -1.8  -0.7
NGC\,6624 & 2.79 & $-$7.91 & 7.9 & 1.2 & 0.28 & $-0.44$ & $-0.42$ & $-0.37\pm
0.01$ & $0.135$ & 0.930 \\ % \pm0.012
NGC\,6637 & 1.72 & $-$10.27 & 8.8 & 1.7 & 0.18 & $-0.64$ & $-0.59$ & $-0.48\pm
0.02$ & $0.138$ & 1.63 \\ % \pm0.003
%\hline
NGC\,6652 & 1.53 & $-11.38$ & 10.0 & 2.7 & 0.09 & $-0.81$ & $-0.76$ & $-0.82\pm
0.07$ & $0.141$ & 0.521 \\ % \pm0.006
NGC\,6717 & 12.88 & $-10.90$ & 7.1 & 2.4 & 0.22 & $-1.26$ & $-1.26$ & $-1.17\pm
0.09$ & $0.948$ & 0.181 \\ % \pm0.021
NGC\,6723 & 0.07 & $-17.30$ & 8.7 & 2.6 & 0.05 & $-1.10$ & $-1.10$ & $-1.01\pm
0.06$ & $0.371$ & 1.69 \\ % \pm0.039
\hline
NGC\,6362 & 25.55 & $-17.57$ & 7.6 & 5.1 & 0.09 & $-0.99$ & $-1.07$ & $-1.03\pm
0.06$ & $0.247$ & 1.13 \\ % \pm0.012
\enddata
%\tablecomments{The metallicity values are from H10: \citet[2010 edition]
%{1996AJ....112.1487H}; C09: Carretta et al. (2009) and Dias catalogue (Vasquez
%et al. 2018)}
%\citet[D19]{2018A&A...619A..13V}}
\end{deluxetable*}
\begin{deluxetable*}{lccccccccccc}
%\tabletypesize{\scriptsize}
\tablenum{2}
\tablecaption{Metallicity and chemical abundances derived from high-resolution
spectroscopy of individual stars from the literature.}
\label{tab:Met-Abund}
\tablewidth{0pt}
\tablehead{
\colhead{Cluster} & \colhead{[Fe/H]} & \colhead{[O/Fe]} &  \colhead{[Na/Fe]}
& \colhead{[Al/Fe]} & \colhead{[Mg/Fe]} & \colhead{[Si/Fe]} & \colhead
{[Ca/Fe]} & \colhead{[Ti/Fe]} & \colhead{[Ba/Fe]} & \colhead{[$\alpha$/Fe]} &
\colhead{Ref.}}
\startdata
%NGC\,6304 & $-0.61\pm0.02$ & --- & --- & --- & $+0.31$ & $+0.20$ & $+0.17$ &
%$+0.30$ & --- & $+0.18^{\dagger}$ & C18 \\
NGC\,6352 & $-0.55\pm0.03$ & --- & $+0.18$ & $+0.32$ & $+0.47$ & $+0.20$ &
$+0.19$ & $+0.15$ & --- & $+0.20$ & F09 \\
NGC\,6624 & $-0.69\pm0.02$ & $+0.41$ & --- & $+0.39$ & $+0.42$ & $+0.38$ &
$+0.40$ & $+0.37$ & --- & $+0.39$ & V11 \\
% & $-0.77\pm0.01$ & --- & --- & --- & $+0.36$ & %$+0.25$ & $+0.23$ &
% $+0.41$ & & C18 \\
NGC\,6637 & $-0.77\pm0.02$ & $+0.20$ & $+0.35$ & $+0.49$ & $+0.28$ & $+0.45$
& $+0.20$ & $+0.24$ & $+0.22$ & $+0.27^{\dagger}$ & L07 \\
% & $-0.90\pm0.01$ & --- & --- & --- & $+0.43$ & %$+0.20$ & $+0.30$ &
% $+0.54$ & & & C18 \\
%\hline
%NGC\,6652 & $-0.93\pm0.02$ & --- & --- & --- & $+0.45$ & $+0.23$ & $+0.25$ &
%$+0.33$ & --- & $+0.32^{\dagger}$ & C18 \\
NGC\,6723 & $-0.98\pm0.08$ & $+0.29$ & $+0.00$ & $+0.31$ & $+0.23$ & $+0.36$
& $+0.30$ & $+0.24$ & +0.22 & $+0.28^{\dagger}$ & R16 \\
 & --- & $+0.39$ & $+0.05$ & --- & $+0.52$ & --- & --- & --- & --- &
$+0.46^{\dagger}$ & G15$^{a}$ \\
 & $-1.22\pm0.01$ & $+0.53$ & $+0.13$ & --- & $+0.51$ & $+0.60$ &
$+0.81$ & --- & $+0.75$ & $+0.61^{\dagger}$ & G15$^{b}$ \\
 & $-0.93\pm0.05$ & $+0.39$ & $+0.14$ & $+0.32$ & $+0.47$ & $+0.52$ &
$+0.37$ & $+0.34$ & $+0.36$ & $+0.39$ & C19 \\
\hline
NGC\,6362 & $-1.09\pm0.01$ & --- & +0.00 & --- & --- & --- &
--- & --- & --- & --- & M16$^{a}$ \\
 & $-1.09\pm0.01$ & --- & +0.33 & --- & --- & --- & --- &
--- & --- & --- & M16$^{b}$ \\
 & $-1.07\pm0.01$ & --- & --- & $+0.51$ & $+0.54$ & $+0.45$ &
$+0.26$ & $+0.24$ & +0.61& $+0.32$ & M17 \\
\enddata
\tablecomments{F09: \citet{2009A-A...493..913F};%Feltzing et al. (2009);
V11: \citet{2011MNRAS.414.2690V};
L07: \citet{2007RMxAC..28..120L}; R16: \citet{RojasArriagada16}; G15: \citet
{Gratton15} for blue (G15$^a$) and red (G15$^b$) HB stars; C19: \citet
{2019MNRAS.tmp.1639C}; M16: \citet{2016ApJ...824...73M} for 1G (M16$^a$) and 2G
(M16$^b$) stars; M17: \citet{2017MNRAS.468.1249M}. $^{\dagger}$The
$\alpha$-element abundances which were not made explicit in the references were
computed here as the mean of O, Mg, Si, Ca and Ti abundances, if available.
NGC\,6304, NGC\,6652 and NGC\,6717 are not included in any high-resolution
spectroscopic study of individual stars.}
\end{deluxetable*}

%%%%%%%%%%%%%%%%%%%%%%%%%%%%%%%%%%%%%%%%%%%%%%%%%%%%%%%%%%%%%%%%%%%%%%%%%%%%%%%%
%%%%%%%%%%%%%%%%%%%%%%%%% OBSERVATIONS AND REDUCTIONS %%%%%%%%%%%%%%%%%%%%%%%%%%
%%%%%%%%%%%%%%%%%%%%%%%%%%%%%%%%%%%%%%%%%%%%%%%%%%%%%%%%%%%%%%%%%%%%%%%%%%%%%%%%
\section{Observations: GO-13297 program}\label{sec:Obs}

The objective of the GO-13297 program is the identification of MPs in a sample
of 56 GCs (the most central of them in the Galaxy are identified in Figure~\ref
{fig:piotto-GCs}), using the WFC3/UVIS UV and blue filters F275W, F336W and
F438W. The exposure times were set up to reach a color precision of $0.02$\,mag
in F275W just below the main-sequence turnoff (MSTO). \citetalias
{2015AJ....149...91P} presented the exposure times and observing strategies
adopted. Previous photometry with the ACS/WFC F606W  and F814W optical filters
\citep[GO-10775,][]{2007AJ....133.1658S} was also obtained for this sample.

The $2.6\arcmin\times2.6\arcmin$ field of view of WFC3 is slightly smaller
than that of ACS/WFC ($3.4\arcmin\times3.4\arcmin$) and therefore GO-13297 data
target a more central region of the GCs. The data reduction pipelines \citep
[adapted from][]{2008AJ....135.2055A} and the astro-photometric catalogs
used in this work are described in \citetalias{2015AJ....149...91P} and \citet
[\citetalias{2018MNRAS.481.3382N}]{2018MNRAS.481.3382N}. We adopted
the same procedure described in \citetalias{2018MNRAS.481.3382N} for selecting
the well-measured stars, based on the photometric errors and two quality
parameters for the five \textit{HST} filters.

\begin{figure*}
    \centering
    \includegraphics[width=0.96\textwidth]{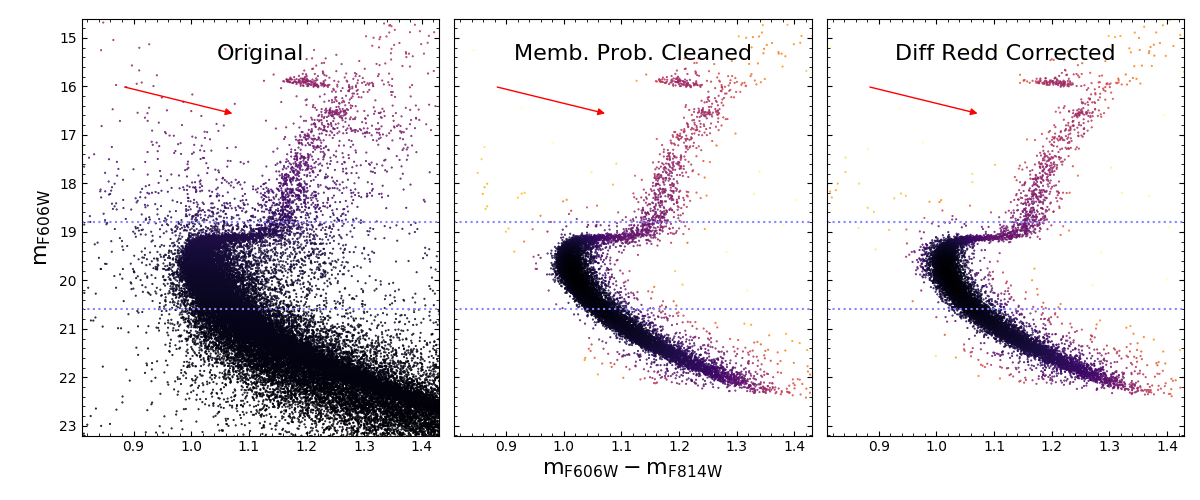}
    \caption{Example of the CMD cleaning process for NGC\,6304. The left
    panel shows the original $m_{\rm F606W}$ vs. $m_{\rm F606W}-m_{\rm F814W}$
    CMD, the middle panel shows the membership probability cleaned CMD, and the
    right panel shows the final CMD after applying also the differential
    reddening correction. The dotted-blue lines enhance the most age-sensitive
    region in the CMD, whose stars are actually used in our isochrone fitting
    method. The relative PMs used to calculate the membership were measured
    combining GO-10775 and GO-13297 photometry ($\Delta \mathrm{t} = 7-8$\,yr).
    A reddening vector of $E(B-V)=0.20$ is given by the red arrow.}
    \label{fig:cmds_ngc6304}
\end{figure*}

The photometry was also corrected for differential reddening (DR) as in \citet
{2012A&A...540A..16M} and decontaminated from field stars via
membership-probability cleaning \citepalias{2018MNRAS.481.3382N}. The DR
correction is crucial for age derivation, since the reddening vector is almost
perpendicular to the subgiant branch (SGB) and MSTO. The original CMD and those
resulting from the membership-probability and DR-cleaning processes are shown in
Figure~\ref{fig:cmds_ngc6304} for the moderately metal-rich cluster NGC\,6304,
which has the largest reddening in our sample. 

\subsection{The sample}

The GO-10775 and GO-13297 programs covered only a few outer bulge objects
(Figure~\ref{fig:piotto-GCs}), essentially because they focused on nearby
GCs with low reddening values, to ensure the feasibility of observations in a
few \textit{HST} orbits. The distribution of the central GCs in Galactic
coordinates presented in Figure~\ref{fig:piotto-GCs} justifies the sample
selection: among the 20 observed clusters within $|b|<22\arcdeg$ and $|\ell|<
22 \arcdeg$, only six of them have $R_{\rm GC}\lesssim 3 $\,kpc and $\rm{[Fe/H]}
>-1.5$, classified as bulge GCs, in \citet
{2016PASA...33...28B}. We also include NGC\,6352 which is slightly farther,
at 3.3 kpc from the Galactic center.

 The twin clusters NGC\,6388 and NGC\,6441 were not analyzed because
they are characterized by an extremely blue
horizontal branch morphology despite their quite high metallicity \citep
{1997ApJ...484L..25R,2007A&A...474..105B}.
 Besides, they are classified as type-II and type-I ambiguous respectively,
by \citet[][\citetalias{2017MNRAS.464.3636M}]{2017MNRAS.464.3636M}, making
their stellar populations even more difficult to be disentangled. \citet
{2013ApJ...765...32B} carried out a multicolor analysis with \textit{HST}
filters to detect their MPs. 

\begin{figure*}
    \centering
    \includegraphics[width=\textwidth]{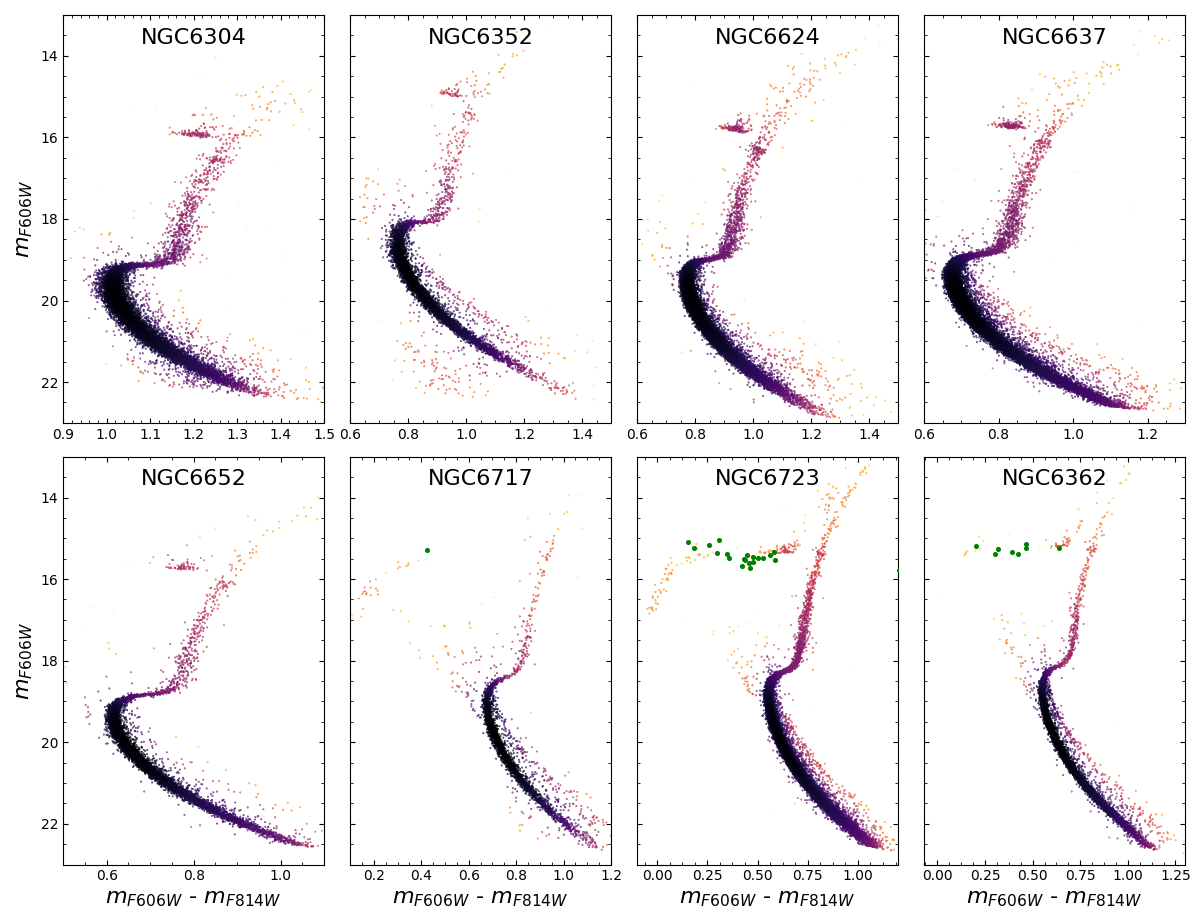}
    \caption{Membership probability and differential reddening cleaned CMDs of the seven
    bulge GCs and the reference halo one NGC\,6362. The upper panels show the
    moderately metal-rich, and the lower panels show the moderately metal-poor
    GCs. Green points in the last panels show the cataloged RR Lyrae stars that
    returned a cross-match with this \textit{HST} photometry (Section~\ref
    {subsec:priors}).}
    \label{fig:cmds}
\end{figure*}

Table~\ref{tab:Harris} presents the coordinates and photometric parameters
of the sample clusters from \citet[2010 edition]{1996AJ....112.1487H} and
mass from \citet{2018MNRAS.478.1520B}. Literature metallicity values from
\citet{2009A-A...508..695C} and \citet{2018A-A...619A..13V}, according to the
updated compilation\footnote
{\url{www.sc.eso.org/~bdias/catalogues.html}} by \citet{2015A&A...573A..13D,
2016A&A...590A...9D}, are included. In Table~\ref{tab:Met-Abund} are reported
the metallicities and chemical abundances derived from high-resolution
spectroscopy, available in the literature. Figures~\ref{fig:cmds_ngc6304} and
\ref{fig:cmds} show the $m_{\rm F606W}$ vs. $m_{\rm F606W}-m_{\rm F814W}$
membership and DR-cleaned CMDs of the sample clusters. Appendix~\ref{sec:ApxA}
presents an overview of previous literature work for these GCs. 

In a previous work in this series, \citet[\citetalias{2017MNRAS.464.3636M}]
{2017MNRAS.464.3636M} provided an atlas of MPs in all the 56 GCs, extracting
the so-called ``chromosome maps'' to perform a uniform analysis and determine
the fraction of 1G stars. We have followed the same steps to disentangle the
RGB and MS stellar populations, together with the method described in
\citetalias{2015MNRAS.451..312N} to separate the SGB stellar populations. The
analysis of the multiple populations in this paper is carried out in
Section~\ref{sec:MPs}.

\citet[\citetalias{2016ApJ...822...44B}]{2016ApJ...822...44B} analyzed the HB
morphology of the sample clusters, showing that the four moderately metal-rich
GCs are dominated by red clump stars (NGC\,6304, NGC\,6352, NGC\,6624 and
NGC\,6637), as well as the slightly more metal-poor NGC\,6652; the most
metal-poor has an extended blue horizontal branch (NGC\,6717); and the two
remaining have intermediate morphologies (NGC\,6362 and NGC\,6723). The color
difference $\Delta(V-I)$ between the HB and RGB, from \citet
{2010ApJ...708..698D}, is listed in Table~\ref{tab:Harris} and agrees with
these HB morphologies. \citet{2014ApJ...785...21M} also carried out a similar
analysis, but with the F606W and F814W \textit{HST} filters.

\citet[\citetalias{2018MNRAS.481.5098M}]{2018MNRAS.481.5098M} have shown that
the average He difference between the 2G and 1G stars does not exceed
0.010 in mass fraction for the sample GCs (and maximum internal variation
below 0.030); considering all the 56 GCs, the average enhancement is also
of $\sim 0.010$. It is consistent with \citet[\citetalias
{2018MNRAS.475.4088L}] {2018MNRAS.475.4088L}, where an average He enhancement
of $0.011\pm0.002$ in 2G stars was derived from an analysis of the RGB bump in
18 GCs.

%%%%%%%%%%%%%%%%%%%%%%%%%%%%%%%%%%%%%%%%%%%%%%%%%%%%%%%%%%%%%%%%%%%%%%%%%%%%%%%%
%%%%%%%%%%%%%%%%%%%%%%%%%%% ISOCHRONE FITTING METHOD %%%%%%%%%%%%%%%%%%%%%%%%%%%
%%%%%%%%%%%%%%%%%%%%%%%%%%%%%%%%%%%%%%%%%%%%%%%%%%%%%%%%%%%%%%%%%%%%%%%%%%%%%%%%
\section{Isochrone fitting method}\label{sec:Isoc-Method}

The physical parameters of the GCs were determined by statistical comparisons
between the observed stars (in a limited magnitude range from the MS to the
RGB) and theoretical models. We have followed the basis of analysis previously
presented by \citet{2007A&A...462..139K}. Here, we use the \texttt{SIRIUS} code
\citep[Statistical Inference of physical paRameters of sIngle and mUltiple
populations in Stellar clusters -- ][]{2020arXiv200102697S}, with the aim of
having a uniform and self-consistent method for age derivation via statistical
isochrone fitting. It follows a Bayesian approach with the Markov chain Monte
Carlo (MCMC) sampling method, in order to derive ages, distances, metallicity
and reddening values for stellar clusters. In this work, \texttt{SIRIUS} is
applied to the sample clusters as a single stellar population, as well as to
their 1G and 2G stars separately.

The likelihood function is computed for each star~$i$, relative to the
$j^{th}$ closest point of each isochrone defined as the combination of four
free parameters (age, $\rm{[Fe/H]}$, $E(B-V)$ and $(m-M)_0$), through:
\begin{equation}
 \displaystyle L \propto \prod_{i=1}^{\rm{N}} \max \left \{ \prod_{j=1}^{\rm{M}}
 \exp{\left(- \chi_{\rm{mag,j,i}}^2 - \chi_{\rm{col,j,i}}^2
 \right)} \right \}\;,
\end{equation}
where $mag$ refers to $m_{\rm F606W}$, $col$ is the color $m_{\rm F606W}
-m_{\rm F814W}$ and $\sigma$ correspond to the uncertainties. The
chi-squares $\chi_{\rm{mag,j,i}}^2$ and $\chi_{\rm{col,j,i}}^2$ are given by:
\begin{eqnarray}
 \ \chi_{\rm{mag,j,i}}^2 = \frac{\left(mag_i - mag_j \right)^2}{\sigma_
    {\rm{mag,i}}^2} \\
 \ \chi_{\rm{col,j,i}}^2 = \frac{\left(col_i - col_j \right)^2}{\sigma_
    {\rm{col,i}}^2} \;.
\end{eqnarray}

In order to take into account the number of stars in each evolutionary
stage, we apply a factor inversely proportional to the star count
inside a small region around each star, in the likelihood function.
The contribution of all $N$ stars are combined computing the natural logarithm
($\mathcal{L} =\ln L$), relatively to the $j$-th closest point of the
isochrone with M points:
\begin{equation}
  \mathcal{L} \propto \sum_{i=1}^{\rm{N}} \max \left\{-\sum_{j=1}^{\rm{M}}\left(\chi_
  {\rm{col,j,i}}^2 + \chi_{\rm{mag,j,i}}^2\right)\right\}\;.
\end{equation}

Thus, Gaussian distributions are assumed for the magnitude and color
distributions of stars, following the evolutionary path given by the isochrone.
The higher is the $\mathcal{L}$ value, the greater is the plausibility that a
model represents the observation. The MCMC sampling is carried out using the
\texttt{emcee} package \citep{2013PASP..125..306F} to simulate stochastic chains
according to a number of walkers and steps. Some similar isochrone fitting
methods have been applied recently. For instance, \citet{2018ApJ...853...15K}
applied a similar likelihood function, but comparing synthetic and observed
fiducial lines instead of considering all stars. \texttt{SIRIUS} was already
applied in \citet{2019MNRAS.484.5530K} and \citet{2019A&A...627A.145O}.

\subsection{Stellar Evolutionary Models}

Two sets of $\alpha$-enhanced
isochrones are compared with the membership probability cleaned CMDs of the
sample clusters:
\begin{itemize}
\item \textit{Dartmouth Stellar Evolutionary Database} \citep[DSED\footnote
{\url{http://stellar.dartmouth.edu/models/grid.html}},][]{2008ApJS..178...89D};
\item \textit{A Bag of Stellar Tracks and Isochrones} \citep[BaSTI\footnote
{\url{http://basti.oa-abruzzo.inaf.it/}},][]{2006ApJ...642..797P}.
\end{itemize}

All available ages (from 10.0 to 15.0 Gyr, in steps of 0.5\,Gyr) and
metallicity values were considered.~Besides, we downloaded interpolated DSED
isochrones in the online interpolator\footnote
{\url{http://stellar.dartmouth.edu/models/isolf_new.html}} and performed linear
interpolations in the original BaSTI isochrones, to build a more complete grid
in [Fe/H] (steps of $0.01$\,dex) and ages (0.1\,Gyr), for the sake of a higher 
stability in the simulated chains.

Considering the results of \citetalias{2018MNRAS.475.4088L} and \citetalias
{2018MNRAS.481.5098M}, we assume here that the typical helium enhancement
between MPs should not exceed $\sim0.01$. Therefore, in the analysis of CMDs
as SSPs, we adopt isochrones with a standard He abundance of $Y\sim
0.25$. See Section~\ref{subsec:MPsAges} for calculations with
an ad hoc helium content for the MPs hosted by the globular clusters.

The BaSTI $\alpha$-enhanced models adopted in the present investigation do not
account for atomic diffusion. Recently the BaSTI database has been updated\footnote{\url{http://basti-iac.oa-abruzzo.inaf.it/}} by
also accounting for the effects of atomic diffusion \citep{2018ApJ...856..125H}.
However, being this updated library available only for scaled-solar heavy
elements distribution and being the updated $\alpha$-enhanced sets of models
still under-construction, we have decided for the present investigations to
rely on the previous BaSTI database. This notwithstanding, since the updated
BaSTI models for the scaled-solar case have been computed for different
assumption about the atomic diffusion efficiency \citep[see][for details]
{2018ApJ...856..125H}, we have adopted a subset of these new BaSTI models in
order to properly estimate the impact of GC age dating of alternatively using
model predictions accounting or not accounting for diffusive processes.

By comparing suitable, self-consistent\footnote{Stellar models computed by
adopting exactly the same physical framework and stellar evolution code, but
two different assumptions about atomic diffusion efficiency: no diffusion and
full efficient atomic diffusion process.} isochrones we have estimated that
using non-diffusive theoretical isochrones implies an overestimate of the
cluster age by 0.80\,Gyr in the metallicity range of our sample GCs. 
Therefore, the isochrone fits were carried out with the original BaSTI
isochrones \citep{2006ApJ...642..797P}, and for the sake of clarity the offset
of $0.80$\,Gyr was included in the BaSTI solutions, represented in the text
and tables hereafter by BaSTI*.

We have adopted the UV/optical photometric data (F275W, F336W and F438W filters)
to properly tag and separate the individual MPs in the sample GCs, and
subsequently we use only the optical bands F606W and F814W to derive ages. This
is because UV filters are sensitive to the peculiar abundances of light elements
characteristic of 2G stars, while optical bands are only sensitive to
the He enhancement via its effect on the stellar effective temperature scale
\citep[see][]{2011A&A...534A...9S,2013A&A...554A..19C,2012ApJ...744...58M,
2018MNRAS.481.5098M}. It would be extremely difficult and computing demanding
to compute model atmospheres and, hence, suitable bolometric corrections for
the relevant photometric passbands, accounting for each individual chemical
patterns observed in 2G stars in the selected GC sample.

\subsection{T$_{\rm eff}$-dependent reddening corrections}

Given that the extinction in the sample bulge GCs~is rather high ($A_V
\lesssim 1.60$), a second-order reddening correction must be applied to the
isochrones, due to the extinction dependency on the effective temperature
\citep{2005MNRAS.357.1038B}. \citet{2005PASP..117.1049S} described in detail
the corrections for \textit{HST}/ACS CCD detectors, increasingly important for
wider and bluer filters.

\begin{figure}
    \centering
    \includegraphics[width=0.85\columnwidth]{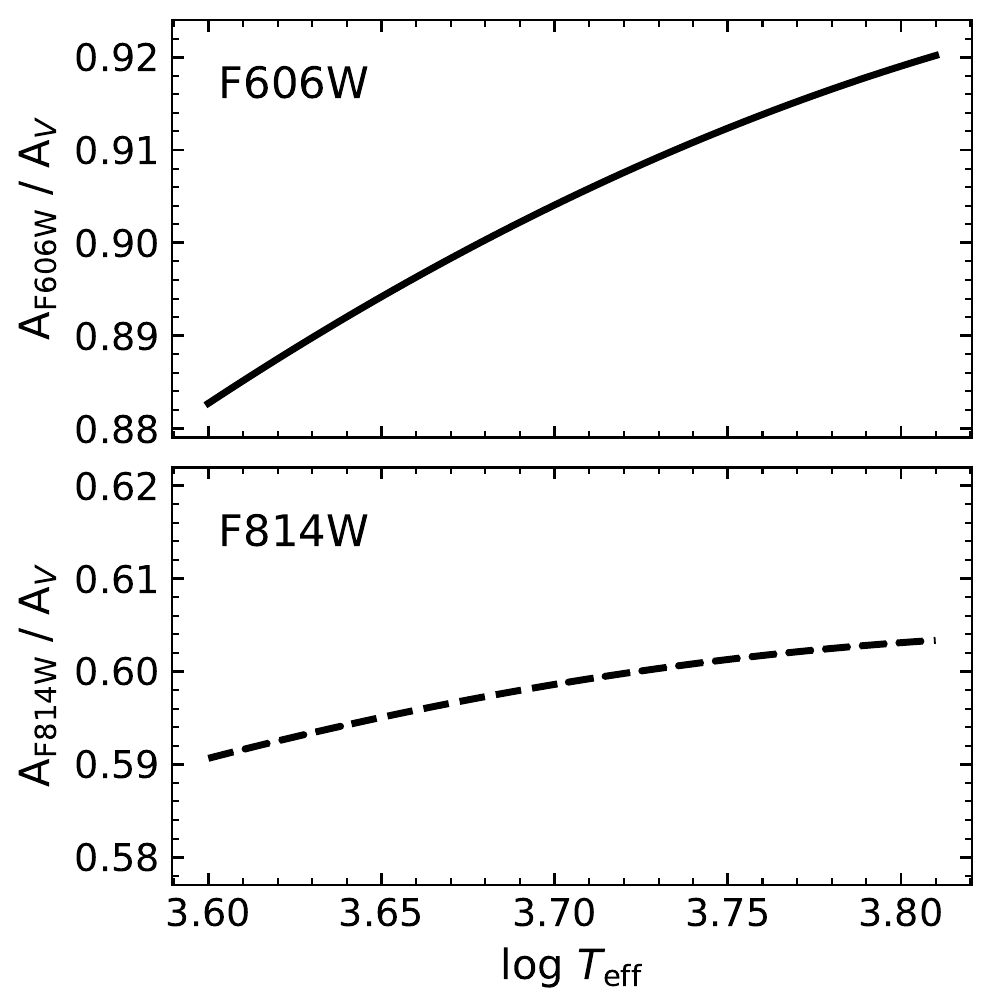}
    \caption{$A_{\rm F606W}/A_{V}$ and $A_{\rm F814W}/A_{V}$ variations
    as a function of the stellar effective temperature.}
    \label{fig:Teff-corr}
\end{figure}

We apply this correction along the DSED and BaSTI models, by using \textit
{PAdova and TRieste Stellar Evolution Code} \citep[PARSEC,][]
{2012MNRAS.427..127B} isochrones including interstellar extinction\footnote
{\url{http://stev.oapd.inaf.it/cgi-bin/cmd_2.8}}. Isochrones with $A_{\rm
V}=0.00$ and $1.55$ are compared for each value of $T_{\rm eff}$, and the
differences in magnitude between them are fitted to a quadratic function as a
function of $T_{\rm eff}$. Figure~\ref{fig:Teff-corr} shows the derived
$A_{\lambda}/A_{V}$ variation for the filters
F606W ($\sim 5\%$) and F814W ($\sim 3\%$), used in
the isochrone fitting. For the sake of completeness, we present the derived functions for the five \textit{HST} bands, where $x=\log T_{\rm eff}$:
\begin{eqnarray}
 \label{eq:teff1}
 & A_{\rm F275W}/A_{V} = -9.302 x^2 + 70.336 x - 131.021 \\
 & A_{\rm F336W}/A_{V} = -0.895 x^2 + 6.777 x - 11.165 \\
 & A_{\rm F438W}/A_{V} = -0.065 x^2 + 0.590 x + 0.037 \\
 \label{eq:teff2}
 & A_{\rm F606W}/A_{V} = -0.320 x^2 + 2.552 x - 4.152 \\
 & A_{\rm F814W}/A_{V} = -0.174 x^2 + 1.347 x - 2.008
 \label{eq:teff3}
\end{eqnarray}

In the $m_{\rm F606W}$ versus $m_{\rm F606W}-m_{\rm F814W}$ CMD, these
corrections have the effect of steepening the slope of the RGB and MS  by
$\Delta(m_{\rm F606W}-m_{\rm F814W})\sim 0.02$ and $0.03$ respectively. \citet
{2018ApJ...853...15K} derived a similar value of $\Delta(m_{\rm F435W}-m_{\rm
F625W})\sim0.05$. Assuming this correction is a linear function of the
reddening, the coefficients of the derived functions for $A_{V} = 1.55$
(Equations~\ref{eq:teff1} to \ref{eq:teff3}) are weighted for the desired
$A_{V}$ value.

\subsection{Prior distributions}
\label{subsec:priors}

In order to better constrain the free parameters during the isochrone fitting
processes, we adopted the following set of prior distributions: canonical He
content, non-negative $E(B-V)$, $\rm{[Fe/H]}$ from spectroscopic studies, and
apparent distance modulus $(m-M)_V$ from RR Lyrae mean magnitudes, when
available. The age was the only free parameter with a flat prior distribution.

Table \ref{tab:Priors} provides the Gaussian prior distributions employed in 
the metallicity for each cluster. For NGC\,6352, NGC\,6624, NGC\,6637 and
NGC\,6362, the central values come from high-resolution spectroscopy and we
assumed 3$\sigma$ as the uncertainty (see Table \ref{tab:Met-Abund}). Although
NGC\,6723 has high-resolution spectroscopic studies, the metallicity
determinations are discrepant and an average value was adopted as the prior.
For NGC\,6304 and NGC\,6652, the metallicity was derived only from integrated
spectra in the literature \citep{2018ApJ...854..139C}, therefore we used the
values given in Table \ref{tab:Harris} with $\pm 0.15$\,dex. There are no
high-resolution spectroscopic studies for individual stars for NGC\,6717, and
we adopted the value from \citet{2009A-A...508..695C} with a $\sim10 \%$
uncertainty.

The luminosity-metallicity ($M_V-\rm{[Fe/H]}$) relation derived from RR Lyrae
stars (RRLs) by \citet{2018MNRAS.481.1195M}, together with the mean magnitudes
of the cluster RRLs, allow us to obtain a reliable and independent constraint
on the apparent distance modulus $(m-M)_{V}$. For each random set of
parameters, the $E(B-V)$ value corresponds to a respective prior in the absolute distance
modulus $(m-M)_0$.

\begin{figure}
\centering
\includegraphics[width=0.96\columnwidth]{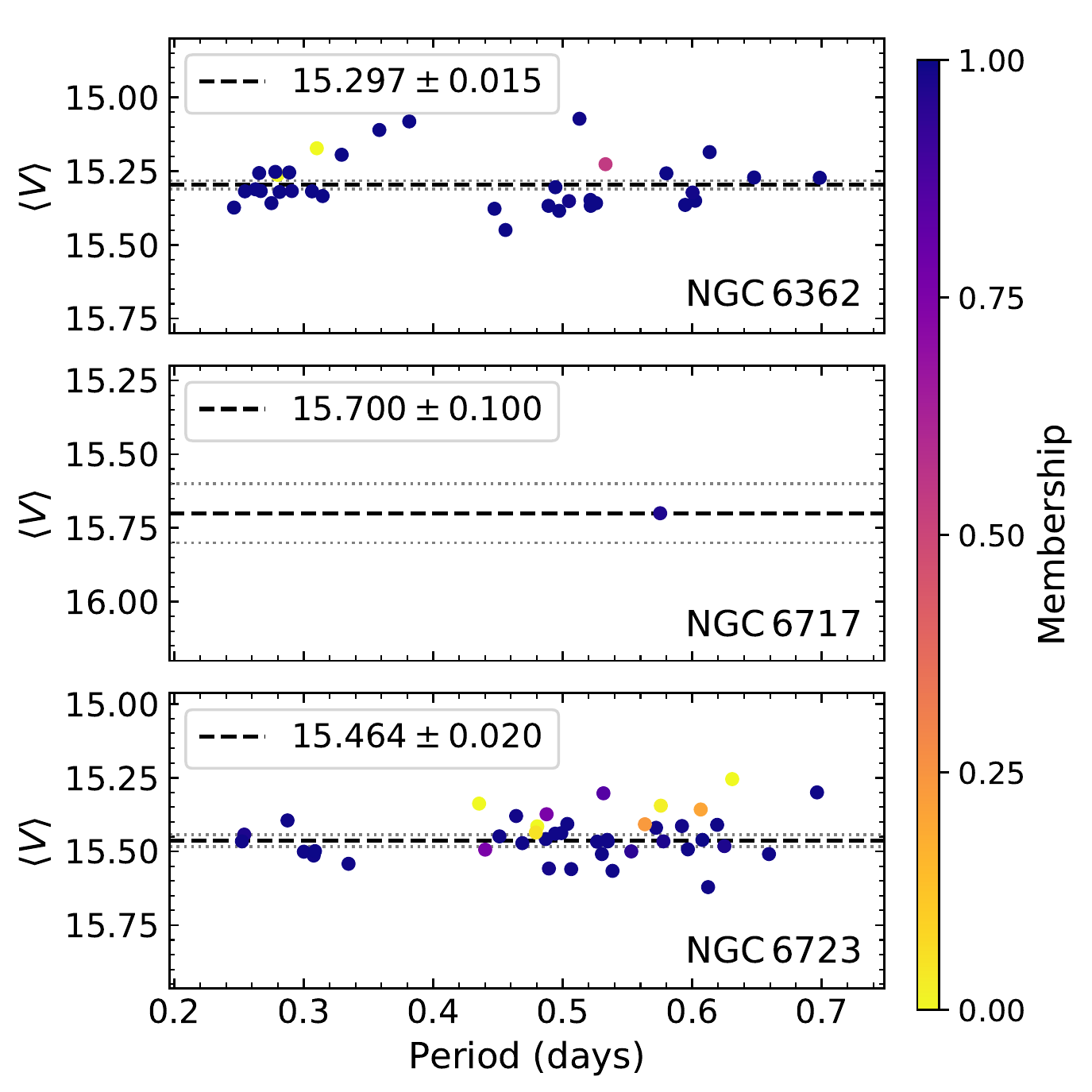}
\caption{Mean $V$ magnitudes vs. period of the RRLs in the fields that contain
NGC\,6362, NGC\,6717 and NGC\,6723. The derived membership probabilities are
represented by the color bar and the weighted averages are shown in each panel.}
\label{fig:RRL-mag}
\end{figure}

Using the catalogs from \citet[2017 edition\footnote
{\url{http://vizier.u-strasbg.fr/viz-bin/VizieR?-source=V/150}}]
{2001AJ....122.2587C} and OGLE Collection of Variable Stars \citep[OCVS\footnote
{\url{http://ogledb.astrouw.edu.pl/~ogle/OCVS}},][]{2014AcA....64..177S}, we
retrieved the coordinates and magnitude of the RRLs located in a radius of
$10\arcmin$ around the cluster center. NGC\,6304 was the only cluster inside
the OGLE covered area in the Galactic bulge. The catalogs contain 21 RRLs for
NGC\,6304, 4 for NGC\,6652, 43 for NGC\,6723, 35 for NGC\,6362 and one RRL
for each of the other GCs. Moreover, 3 red variables with unknown type and
period are listed for NGC\,6352. The \textit{Gaia} DR2 catalog of RR
Lyrae \citep{2018A&A...616A...1G} was checked in the fields around the sample
GCs, but no new RR Lyrae were found.

The proper motions of all these RRLs were retrieved from \textit{Gaia} DR2
\citep{2016A&A...595A...1G, 2018A&A...616A...1G} data, by cross-matching
coordinates. The Gaussian mixture models (GMM) were implemented in order to
identify cluster and field stars, adjusting two Gaussian both for the right
ascension and declination PMs. A membership probability was computed for each
RRL using the equations from \citet{2009A&A...493..959B}, which consider the
measured PMs of this RRL, the cluster and the field, and their respective
uncertainties.

The membership information given in the \citetalias{2018MNRAS.481.3382N}
catalogs cannot be applied to the selected RRLs, since many of them are outside the
$2.6\arcmin\times2.6\arcmin$ WFC3 field of view. To
illustrate this fact, Figure~\ref{fig:cmds} presents the RRLs that returned a
cross-match with the adopted \textit{HST} catalogs, showing that around half
of the original RRLs are located inside the WFC3 covered area. In turn, the RRLs
detected in the cross-match populate exactly the instability strip region.

\begin{figure}
\centering
\includegraphics[width=0.98\columnwidth]{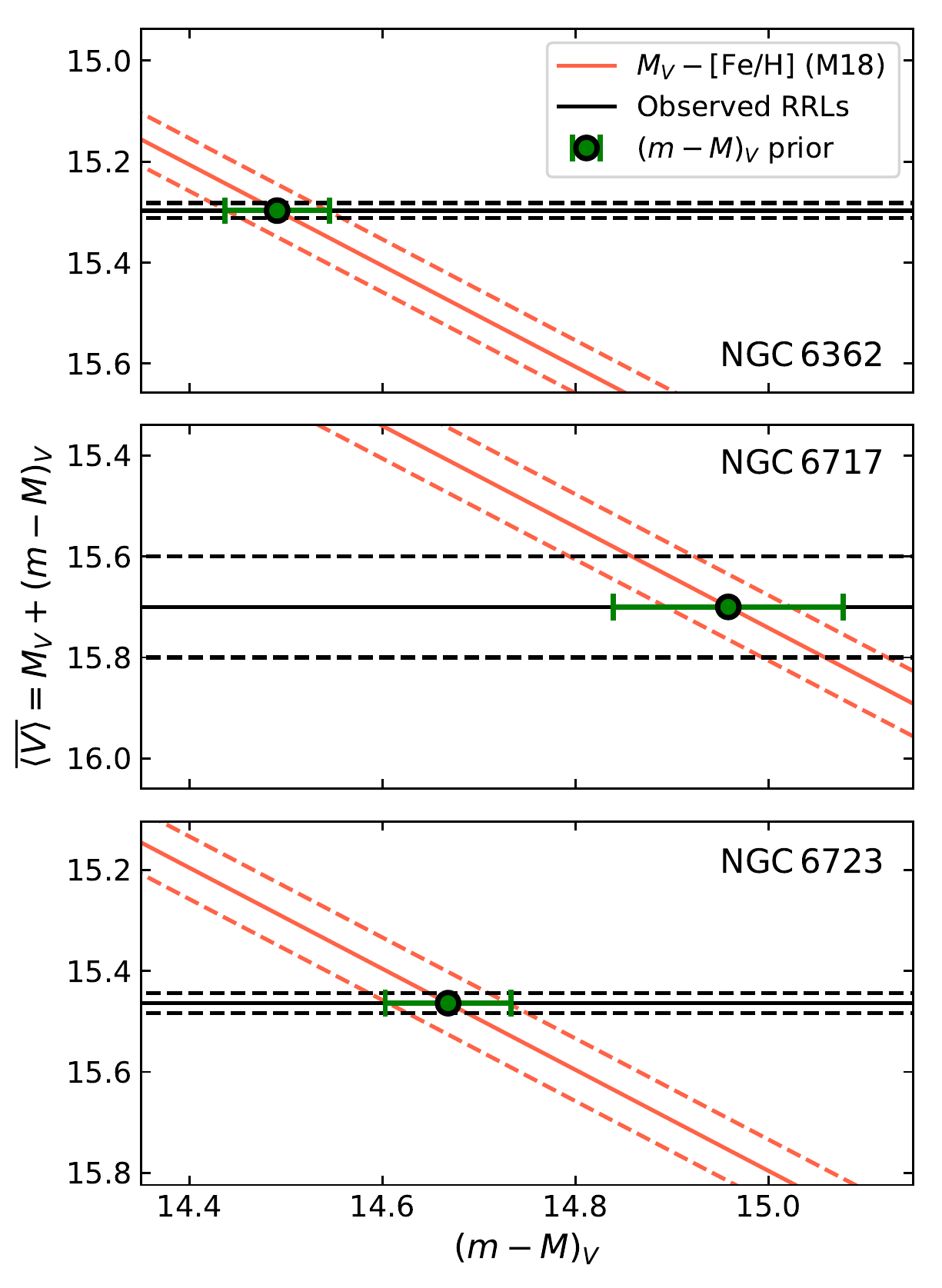}
\caption{Mean magnitudes versus $(m-M)_{\rm{V}}$ of the RRLs in NGC\,6362, 
NGC\,6717 and NGC\,6723. The apparent distance moduli (green circles) are given
by the intersection of the $M_{\rm{V}}$$-$$\rm{[Fe/H]}$ relation \citep[from]
[M18, red diagonal lines]{2018MNRAS.481.1195M}, evaluated for the
respective $\rm{[Fe/H]}$, and the average of the RRL mean magnitudes (black
horizontal lines). The central values and uncertainties are represented
by the solid and the dashed lines. The panels cover the same interval in
$x$-axis and have the same scale in $y$-axis.}
\label{fig:gaiarrlyrae}
\end{figure}

The membership analysis shows that only NGC\,6362, NGC\,6717 and NGC\,6723
have RR Lyrae member stars. If a limit is set in $P_{i} \geq 75\%$, NGC\,6362
contains 32 member RRLs, NGC\,6723 contains 35 RRLs and NGC\,6717 remains with
its unique RRL. \citet{2019AJ....157...47S} report the discovery of new variable
stars in NGC\,6652, but no new RRLs; furthermore, one of the four cataloged
RRLs was selected by them as a member, but our membership analysis concluded
that its PM is not compatible with the derived values for the cluster stars.

In Figure~\ref{fig:RRL-mag} are shown the RRL mean magnitudes $\langle V
\rangle$ versus period for NGC\,6362, NGC\,6717 and NGC\,6723. The dashed line
indicates the weighted average and standard deviation of $\langle V \rangle$
and the colors indicate the membership value of the RRLs, according to the
color bar. The RRLs with low membership values ($P_i\leq75\%$) were not
excluded from the weighted average and standard deviation calculations, since
its membership probability was used as the
weight ($\overline{\langle V\rangle}
= \sum \langle V \rangle_i P_i/\sum P_i$).

Relations $M_V-\rm{[Fe/H]}$  with slightly
different slopes are found in the literature \citep[e.g.][]{1993AJ....106..703S, 2003AJ....125.1309C, 2017A&A...605A..79G}. In this work we adopted 
the recent Bayesian
calibration derived by \citet{2018MNRAS.481.1195M}, by using 381 local RRLs with
accurate \textit{Gaia} DR2 parallaxes \citep{2018A&A...616A...1G} together with
photometric data \citep{2013MNRAS.435.3206D}:
\begin{equation}
    \ M_V = (0.34\pm0.03)\cdot\rm{[Fe/H]} + (1.17 \pm 0.04) \; .
\end{equation}

Therefore, the absolute magnitude was computed according to the metallicity in
Table~\ref{tab:Priors} and, combined with the mean apparent magnitude of the
sample RRLs, the apparent distance modulus was estimated and used as a prior.
Figure~\ref{fig:gaiarrlyrae} presents the intersection between the average of
the observed mean magnitudes and the empirical $M_V-\rm{[Fe/H]}$ calibration,
providing the expected $(m-M)_V$ (green circles, Table~\ref{tab:Priors}).

\begin{deluxetable}{lccc}
\tablenum{3}
\tablecaption{Priors on metallicity, based on Table \ref{tab:Met-Abund}, and apparent distance modulus,
applied for isochrone fitting.}
\label{tab:Priors}
\tablewidth{0pt}
\tablehead{
\colhead{Cluster} & \colhead{$\rm{[Fe/H]}$} & $\langle V \rangle$&
\colhead{$(m-M)_{V}$}
}
%\decimalcolnumbers 
\startdata
 NGC\,6304 & $-0.45\pm0.15$ & --- & --- \\
 NGC\,6352 & $-0.55\pm0.09^\dagger$ & --- & --- \\
 NGC\,6624 & $-0.69\pm0.06^\dagger$ & --- & --- \\
 NGC\,6637 & $-0.77\pm0.06^\dagger$ & --- & --- \\
 NGC\,6652 & $-0.85\pm0.15$ & --- & --- \\
 NGC\,6717 & $-1.26\pm0.10$ & $15.7\pm0.1$ & $14.96\pm0.12$ \\
 NGC\,6723 & $-1.10\pm0.10$ & $15.464\pm0.020$ & $14.67\pm0.07$ \\
 \hline
 NGC\,6362 & $-1.07\pm0.03^\dagger$ & $15.297\pm0.015$ & $14.49\pm0.05$ \\
\enddata
\tablecomments{ $^\dagger$A 3-$\sigma$ deviation from high-resolution
measurement uncertainties was adopted in the prior.}
\end{deluxetable}

Since only three GCs have member RRLs, the prior in $(m-M)_V$ was applied only
for them. For the other five GCs, the distance modulus $(m-M)_0$ was
left to vary uniformly between 12.0 and 16.0, within a flat prior distribution.
Figure~\ref{fig:gaiarrlyrae} also shows that the He abundances must be
canonical, otherwise the RRLs would be much brighter, implying a higher
$(m-M)_V$, as shown in \citet[Figure~15]{2018ApJ...853...15K}. Note that the RR
Lyrae are 1G stars (see Section~\ref{subsec:MPsAges}).

\begin{deluxetable*}{lccccccc}
\tablenum{4}
\tablecaption{Parameters derived from the isochrone fitting to the observed
$m_{\rm F606W}$ vs. $m_{\rm F606W}-m_{\rm F814W}$ CMD, for each model (DSED and
BaSTI; where BaSTI* refers to BaSTI models corrected by an offset of 0.80\,Gyr,
due to atomic diffusion effects) and a mean value. The uncertainties of the mean
value are a combination of the error propagation and the systematic errors due
to the models.}
\label{tab:Isoc-Results}
\tablewidth{0pt}
\tablehead{
\colhead{Cluster} & \colhead{Model} & \colhead{Age (Gyr)} & \colhead{[Fe/H]} & \colhead{$E(B-V)$} &
\colhead{$(m-M)_0$} &  \colhead{$(m-M)_{V}$} & \colhead{$d_{\odot}$ (kpc)}
}
%\decimalcolnumbers
\startdata
%\multicolumn{8}{c}{F606W vs. F438W$-$F606W // $m_{\rm F606W}$ vs. $m_{\rm F438W}-m_{\rm F606W}$} \\
%\hline
 \multirow{3}{*}{NGC\,6304$^\dagger$} &   DSED & $11.60^{+0.90}_{-0.63}$ & $-0.49^{+0.03}_{-0.02}$ & $0.49^{+0.02}_{-0.02}$ & $14.00^{+0.05}_{-0.05}$ &
 $15.42^{+0.07}_{-0.07}$ & $6.28^{+0.09}_{-0.09}$ \\
                                      & BaSTI* & $12.90^{+0.74}_{-0.75}$ & $-0.46^{+0.02}_{-0.02}$ & $0.50^{+0.02}_{-0.02}$ & $13.84^{+0.04}_{-0.04}$ &
 $15.39^{+0.07}_{-0.07}$ & $5.86^{+0.08}_{-0.08}$ \\
                    & \textbf{Mean} & $\mathbf{12.30^{+0.82}_{-0.69}}$ & $\mathbf{-0.48^{+0.03}_{-0.02}}$ & $\mathbf{0.50^{+0.02}_{-0.02}}$ & $\mathbf{13.92^{+0.05}_{-0.05}}$ &
 $\mathbf{15.40^{+0.07}_{-0.07}}$ & $\mathbf{6.07^{+0.09}_{-0.09}}$ \\
\hline
 \multirow{3}{*}{NGC\,6352$^\dagger$} &   DSED & $12.10^{+0.85}_{-0.73}$ & $-0.58^{+0.06}_{-0.06}$ & $0.24^{+0.02}_{-0.02}$ & $13.60^{+0.05}_{-0.04}$ &
 $14.38^{+0.07}_{-0.07}$ & $5.28^{+0.15}_{-0.15}$ \\
                                      & BaSTI* & $12.20^{+0.63}_{-0.66}$ & $-0.56^{+0.03}_{-0.03}$ & $0.26^{+0.02}_{-0.02}$ & $13.58^{+0.04}_{-0.04}$ &
 $14.39^{+0.07}_{-0.07}$ & $5.20^{+0.15}_{-0.15}$ \\
                            & \textbf{Mean}   & $\mathbf{12.20^{+0.75}_{-0.70}}$ & $\mathbf{-0.57^{+0.05}_{-0.05}}$ & $\mathbf{0.25^{+0.02}_{-0.02}}$ & $\mathbf{13.59^{+0.05}_{-0.04}}$ &
 $\mathbf{14.39^{+0.07}_{-0.07}}$ & $\mathbf{5.24^{+0.15}_{-0.15}}$ \\
\hline
 \multirow{3}{*}{NGC\,6624} &   DSED & $12.00^{+0.51}_{-0.56}$ & $-0.74^{+0.02}_{-0.02}$ & $0.25^{+0.02}_{-0.02}$ & $14.56^{+0.04}_{-0.04}$ &
 $15.34^{+0.05}_{-0.05}$ & $8.17^{+0.15}_{-0.15}$ \\
                            & BaSTI* & $11.30^{+0.64}_{-0.73}$ & $-0.72^{+0.02}_{-0.02}$ & $0.26^{+0.02}_{-0.02}$ & $14.57^{+0.04}_{-0.04}$ &
 $15.38^{+0.05}_{-0.05}$ & $8.20^{+0.15}_{-0.15}$ \\
                            & \textbf{Mean}   & $\mathbf{11.70^{+0.58}_{-0.65}}$ & $\mathbf{-0.73^{+0.02}_{-0.02}}$ & $\mathbf{0.26^{+0.02}_{-0.02}}$ & $\mathbf{14.57^{+0.04}_{-0.04}}$ &
 $\mathbf{15.36^{+0.05}_{-0.05}}$ & $\mathbf{8.19^{+0.15}_{-0.15}}$ \\
\hline
 \multirow{3}{*}{NGC\,6637} &   DSED & $12.30^{+0.78}_{-0.36}$ & $-0.82^{+0.02}_{-0.02}$ & $0.16^{+0.01}_{-0.02}$ & $14.71^{+0.03}_{-0.04}$ &
 $15.21^{+0.03}_{-0.05}$ & $8.75^{+0.12}_{-0.16}$ \\
                            & BaSTI* & $12.20^{+0.51}_{-0.45}$ & $-0.81^{+0.02}_{-0.02}$ & $0.16^{+0.01}_{-0.01}$ & $14.71^{+0.03}_{-0.03}$ &
 $15.21^{+0.03}_{-0.03}$ & $8.75^{+0.12}_{-0.12}$ \\
                            & \textbf{Mean}  & $\mathbf{12.30^{+0.67}_{-0.41}}$ & $\mathbf{-0.82^{+0.02}_{-0.02}}$ & $\mathbf{0.16^{+0.01}_{-0.02}}$ & $\mathbf{14.71^{+0.03}_{-0.04}}$ &
 $\mathbf{15.21^{+0.03}_{-0.04}}$ & $\mathbf{8.75^{+0.12}_{-0.12}}$ \\
\hline
%\hline
 \multirow{3}{*}{NGC\,6652} &   DSED & $12.80^{+0.60}_{-0.52}$ & $-0.94^{+0.05}_{-0.05}$ & $0.11^{+0.02}_{-0.02}$ & $14.87^{+0.04}_{-0.04}$ &
 $15.21^{+0.05}_{-0.05}$ & $9.42^{+0.18}_{-0.18}$ \\
                            & BaSTI* & $12.50^{+0.69}_{-0.64}$ & $-0.90^{+0.04}_{-0.04}$ & $0.10^{+0.02}_{-0.02}$ & $14.83^{+0.04}_{-0.04}$ & 
 $15.14^{+0.05}_{-0.05}$ & $9.25^{+0.17}_{-0.17}$ \\
                            & \textbf{Mean}   & $\mathbf{12.70^{+0.65}_{-0.58}}$ & $\mathbf{-0.92^{+0.05}_{-0.05}}$ & $\mathbf{0.11^{+0.02}_{-0.02}}$ & $\mathbf{14.85^{+0.04}_{-0.04}}$ & 
 $\mathbf{15.18^{+0.05}_{-0.05}}$ & $\mathbf{9.34^{+0.18}_{-0.18}}$ \\
\hline
 \multirow{3}{*}{NGC\,6717} &   DSED & $13.70^{+0.73}_{-0.84}$ & $-1.29^{+0.07}_{-0.06}$ & $0.18^{+0.02}_{-0.02}$ & $14.32^{+0.04}_{-0.04}$ &
 $14.88^{+0.05}_{-0.05}$ & $7.31^{+0.14}_{-0.14}$ \\
                            & BaSTI* & $13.20^{+0.60}_{-0.68}$ & $-1.28^{+0.04}_{-0.04}$ & $0.19^{+0.02}_{-0.01}$ & $14.33^{+0.03}_{-0.03}$ & 
 $14.92^{+0.05}_{-0.03}$ & $7.35^{+0.10}_{-0.10}$ \\
                            & \textbf{Mean}   & $\mathbf{13.50^{+0.67}_{-0.76}}$ & $\mathbf{-1.29^{+0.07}_{-0.06}}$ & $\mathbf{0.19^{+0.02}_{-0.02}}$ & $\mathbf{14.33^{+0.04}_{-0.04}}$ & 
 $\mathbf{14.90^{+0.05}_{-0.05}}$ & $\mathbf{7.33^{+0.12}_{-0.12}}$ \\
\hline
 \multirow{3}{*}{NGC\,6723} &   DSED & $12.50^{+0.69}_{-0.55}$ & $-1.09^{+0.06}_{-0.07}$ & $0.05^{+0.01}_{-0.01}$ & $14.54^{+0.03}_{-0.03}$ &
 $14.69^{+0.03}_{-0.03}$ & $8.09^{+0.11}_{-0.11}$ \\
                            & BaSTI* & $12.60^{+0.48}_{-0.50}$ & $-1.15^{+0.04}_{-0.03}$ & $0.06^{+0.01}_{-0.01}$ & $14.54^{+0.03}_{-0.03}$ & 
 $14.73^{+0.03}_{-0.03}$ & $8.09^{+0.11}_{-0.11}$ \\
                            & \textbf{Mean}   & $\mathbf{12.60^{+0.59}_{-0.53}}$ & $\mathbf{-1.12^{+0.06}_{-0.07}}$ & $\mathbf{0.06^{+0.01}_{-0.01}}$ &  $\mathbf{14.54^{+0.03}_{-0.03}}$ &
 $\mathbf{14.71^{+0.03}_{-0.03}}$ & $\mathbf{8.09^{+0.11}_{-0.11}}$ \\
\hline
\hline
 \multirow{3}{*}{NGC\,6362} &   DSED & $13.80^{+0.51}_{-0.56}$ & $-1.07^{+0.03}_{-0.03}$ & $0.04^{+0.01}_{-0.01}$ & $14.38^{+0.03}_{-0.03}$ &
 $14.50^{+0.03}_{-0.03}$ & $7.52^{+0.10}_{-0.10}$ \\
                            & BaSTI* & $13.40^{+0.43}_{-0.48}$ & $-1.09^{+0.03}_{-0.02}$ & $0.04^{+0.01}_{-0.01}$ & $14.36^{+0.03}_{-0.03}$ & 
 $14.48^{+0.03}_{-0.03}$ & $7.45^{+0.10}_{-0.10}$ \\
                            & \textbf{Mean}   & $\mathbf{13.60^{+0.47}_{-0.52}}$ & $\mathbf{-1.08^{+0.03}_{-0.03}}$ & $\mathbf{0.04^{+0.01}_{-0.01}}$ & $\mathbf{14.37^{+0.03}_{-0.03}}$ &
 $\mathbf{14.49^{+0.03}_{-0.03}}$ & $\mathbf{7.49^{+0.10}_{-0.10}}$ \\
\enddata
\tablecomments{$^\dagger$ Isochrones with $\rm{[\alpha/Fe]} = +0.2$ were applied, instead of $\rm{[\alpha/Fe]} = +0.4$.}
\end{deluxetable*}

 %Previous ages:
 %NGC\,6304 & DSED & $12.50^{+0.44}_{-0.46}$ & $-0.53^{+0.05}_{-0.05}$ &
%$14.17^{+0.03}_{-0.03}$ & --- & $0.41^{+0.02}_{-0.03}$ & --- \\
% NGC\,6362 & DSED & $12.45^{+0.48}_{-0.52}$ & $-1.08^{+0.06}_{-0.06}$ &
%$14.64^{+0.04}_{-0.05}$ & --- & $0.00^{+0.02}_{-0.02}$ & --- \\
% NGC\,6624 & DSED & $13.11^{+0.40}_{-0.53}$ & $-0.58^{+0.06}_{-0.06}$ &
%$14.84^{+0.03}_{-0.04}$ & --- & $0.12^{+0.02}_{-0.02}$ & --- \\
% NGC\,6637 & DSED & $12.28^{+0.36}_{-0.48}$ & $-0.70^{+0.05}_{-0.06}$ &
%$15.01^{+0.04}_{-0.04}$ & --- & $0.06^{+0.03}_{-0.02}$ & --- \\
% NGC\,6652 & DSED & $12.35^{+0.60}_{-0.65}$ & $-0.88^{+0.07}_{-0.05}$ &
%$15.07^{+0.04}_{-0.05}$ & --- & $0.06^{+0.03}_{-0.03}$ & --- \\
% NGC\,6723 & DSED & $12.69^{+0.37}_{-0.42}$ & $-1.11^{+0.04}_{-0.05}$ &
%$14.75^{+0.04}_{-0.04}$ & --- & $0.01^{+0.02}_{-0.02}$ & --- \\

%%%%%%%%%%%%%%%%%%%%%%%%%%%%%%%%%%%%%%%%%%%%%%%%%%%%%%%%%%%%%%%%%%%%%%%%%%%%%%%%
%%%%%%%%%%%%%%%%%%%%%%%%% RESULTS FROM ISOCHRONE FITS %%%%%%%%%%%%%%%%%%%%%%%%%%
%%%%%%%%%%%%%%%%%%%%%%%%%%%%%%%%%%%%%%%%%%%%%%%%%%%%%%%%%%%%%%%%%%%%%%%%%%%%%%%%
\section{Single stellar population analysis}\label{sec:Results}

In this work, the ages were derived with the membership probability cleaned
$m_{\rm F606W}$ vs. $m_{\rm F606W}-m_{\rm F814W}$ CMDs, first considering the
GCs as single stellar populations. These optical filters were chosen due to
their low sensitivity to extinction and to variation of C, N, O abundances
compared to bluer filters.

We adopted a parameter space with: \textit{(i)} ages in the $10-15$\,Gyr range,
with steps of 0.1\,Gyr; \textit{(ii)} [Fe/H] between $-2.00$ and $-0.05$\,dex,
with steps of 0.01\,dex; \textit{(iii)} reddening $E(B-V)$ between 0.00 and
1.00\,mag; and \textit{(iv)} distance modulus in the $13.0-16.0$ mag range.
Although the MCMC sample is composed of continuous values, the parameter space
is discrete in ages and [Fe/H]. In these cases, the random value is changed by
the nearest one. Two Gaussian prior distributions were applied: one in 
$\rm{[Fe/H]}$ centered on the literature values and another in the apparent
distance modulus $(m-M)_{V}$ derived from the RR Lyrae analysis for
three GCs. The adopted uncertainties in these priors are shown in Table~\ref
{tab:Priors}.

For the isochrone fitting, we computed a fiducial line and, in each magnitude
bin, only stars within  $3\sigma$ from this fiducial line are selected. Thus,
binaries and blue straggler stars are identified and discarded. Besides, a
magnitude threshold is selected for the isochrone fitting: stars with magnitude
between $0.8$\,mag above the MSTO and the completeness limit are considered in
the fitting. This is because the CMD region most sensitive to different ages
goes from the MSTO to the lower SGB \citep[e.g.][]{2016ApJ...832...48S}.

We do not include the RGB stars in the isochrone fitting.
The reason is that the shape of the isochrone depends on the precise value that
the builder of the stellar models has chosen to treat convection \citep
{2018NatAs...2..270D}. In other words, the color distance between the MSTO and
the RGB cannot be trusted to derive a precise age. 

As a rule of thumb, a more efficient convection model will provide bluer RGBs,
therefore at a fixed age, the distance between MSTO and RGB colors will be
smaller. Consequently, a fit taking into consideration the full morphology of
the isochrones will tend to attribute a smaller age from models with efficient convection models and a larger age for less efficient convection. 

\begin{figure*}
\centering
\includegraphics[width=7.65cm]{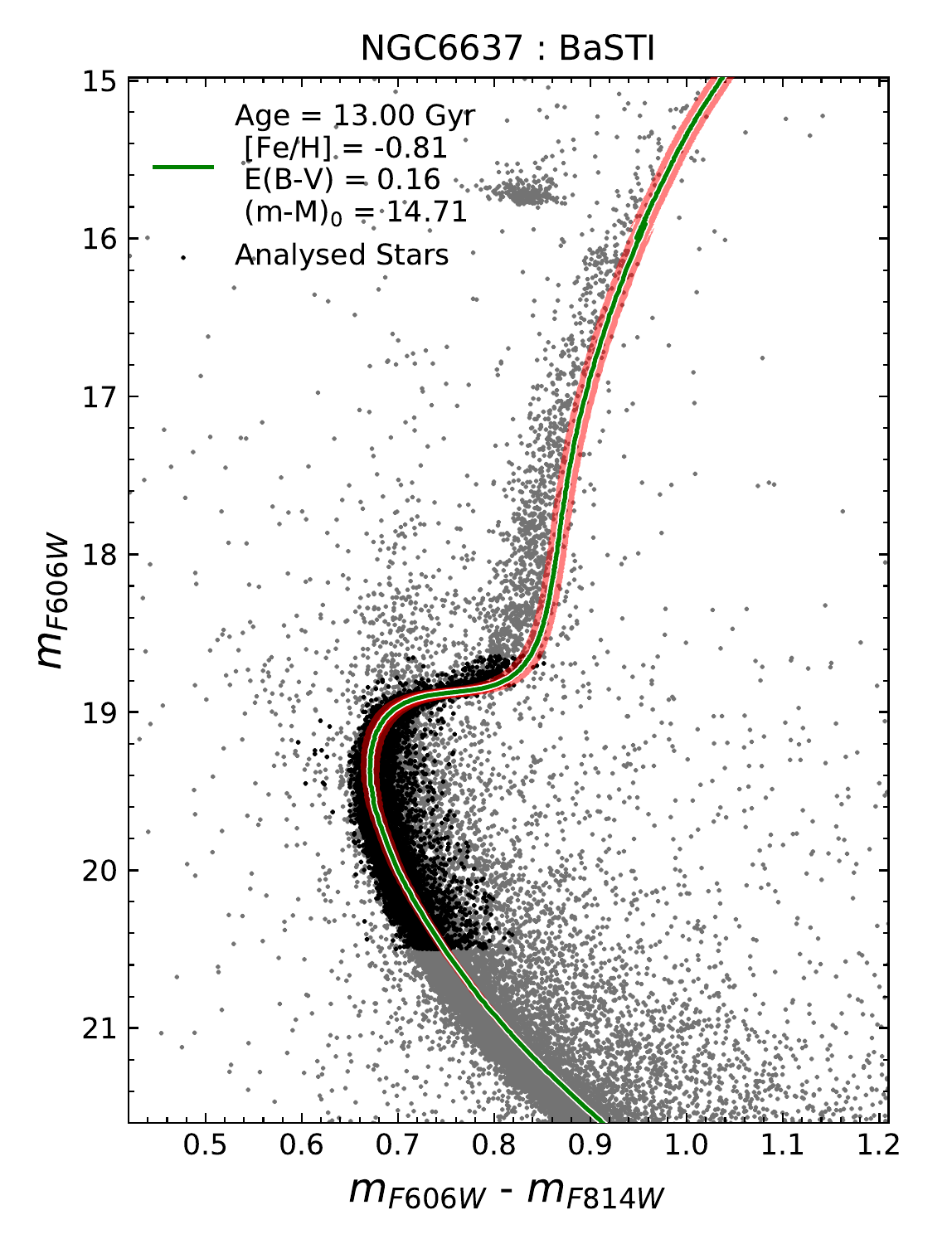}
\hspace{2mm}\includegraphics[width=9.7cm]{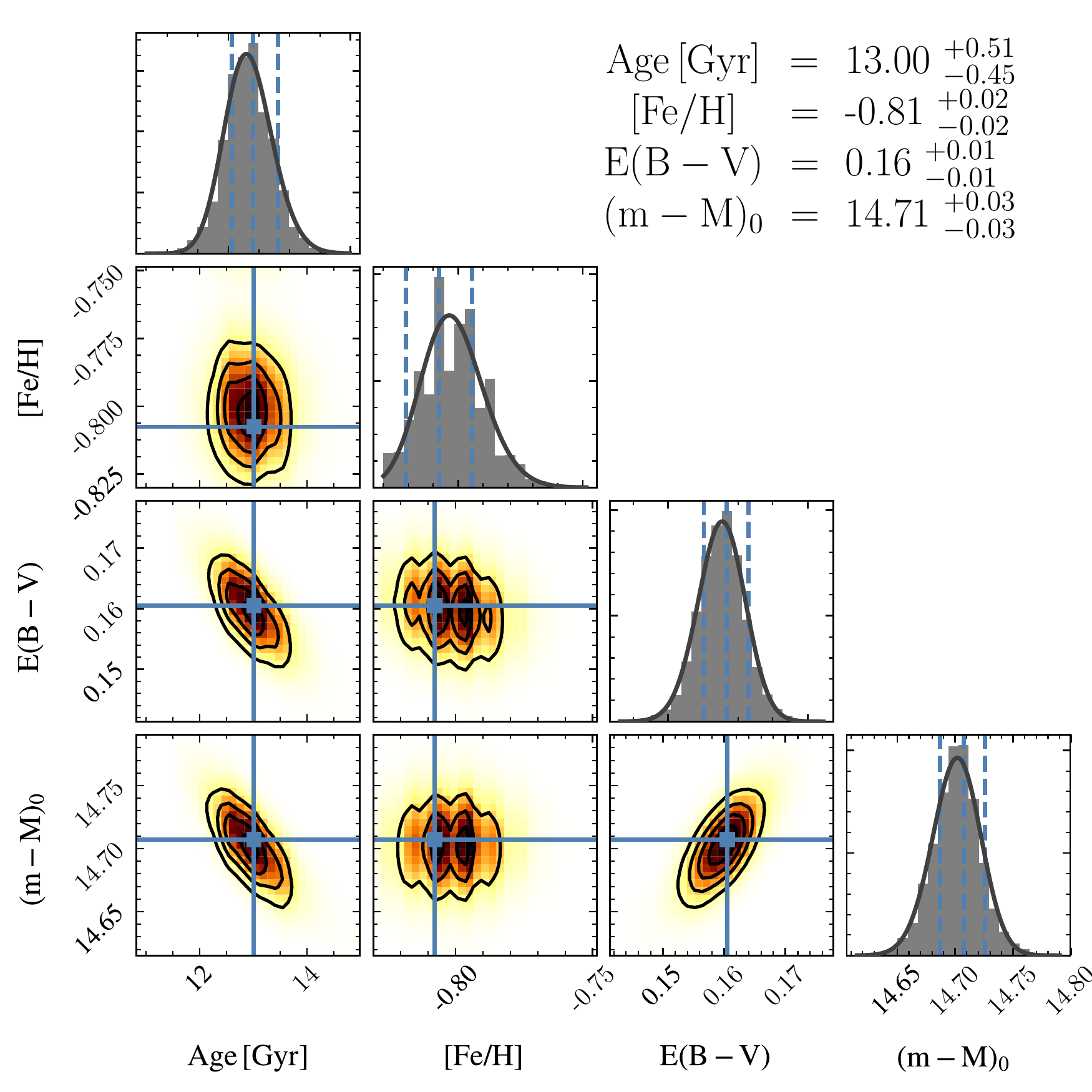}
\caption{Results of the isochrone fits for NGC\,6637 (M69) with BaSTI isochrones.
(\textit{Left:}) CMD with the isochrone representing the parameters of the
central solution (green line). The black dots represent the stars used in the
isochrone fitting, whereas the red region highlight the $1\sigma$ deviation
around the solution (\textit{Right:}) Corner plots showing the probability
distribution function in all the free parameters. The parameters of the solution
are shown along with the uncertainties in the upper right.}
\label{fig:Fig9}
\end{figure*}

\begin{figure*}
\centering
\includegraphics[width=7.65cm]{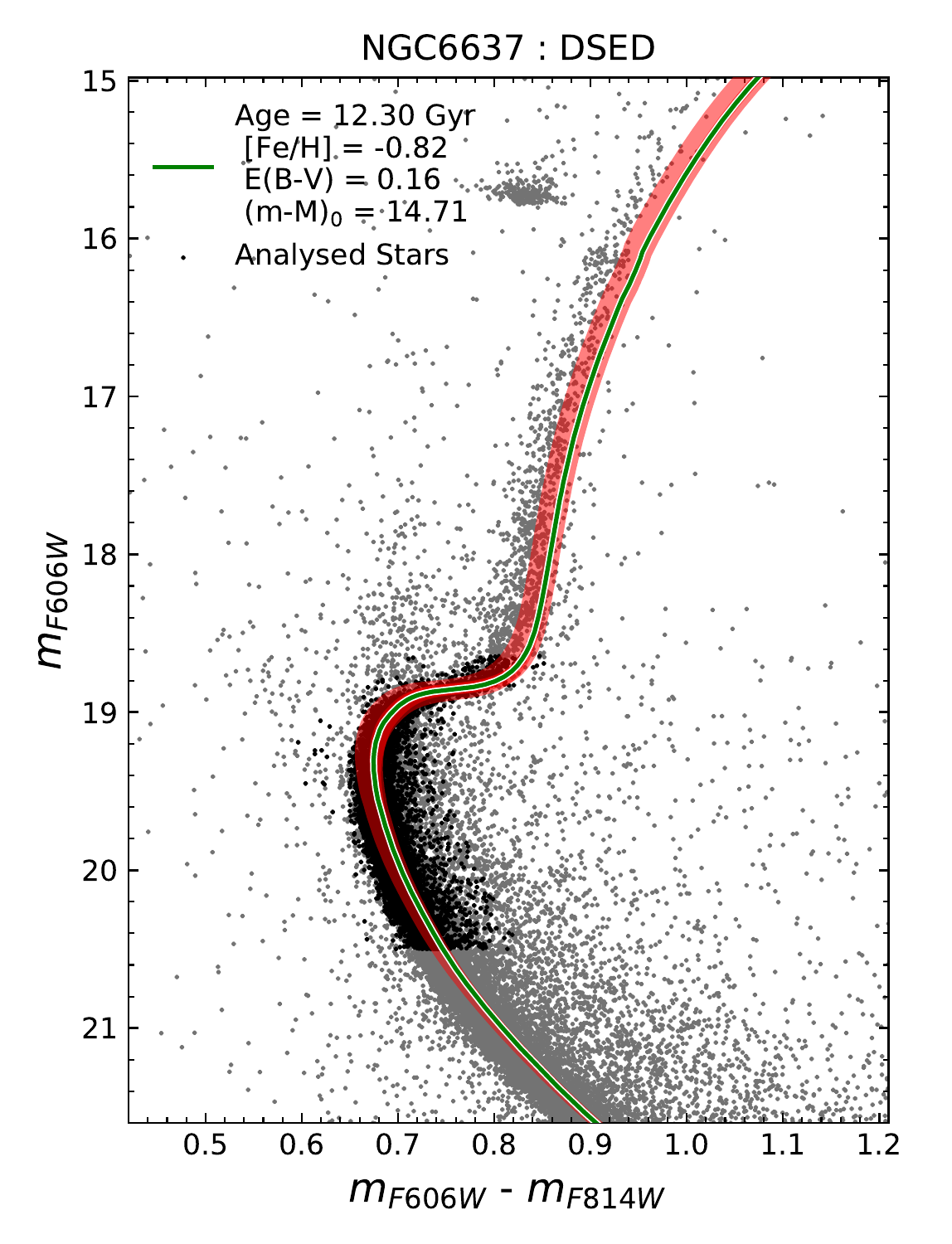}
\hspace{2mm}\includegraphics[width=9.7cm]{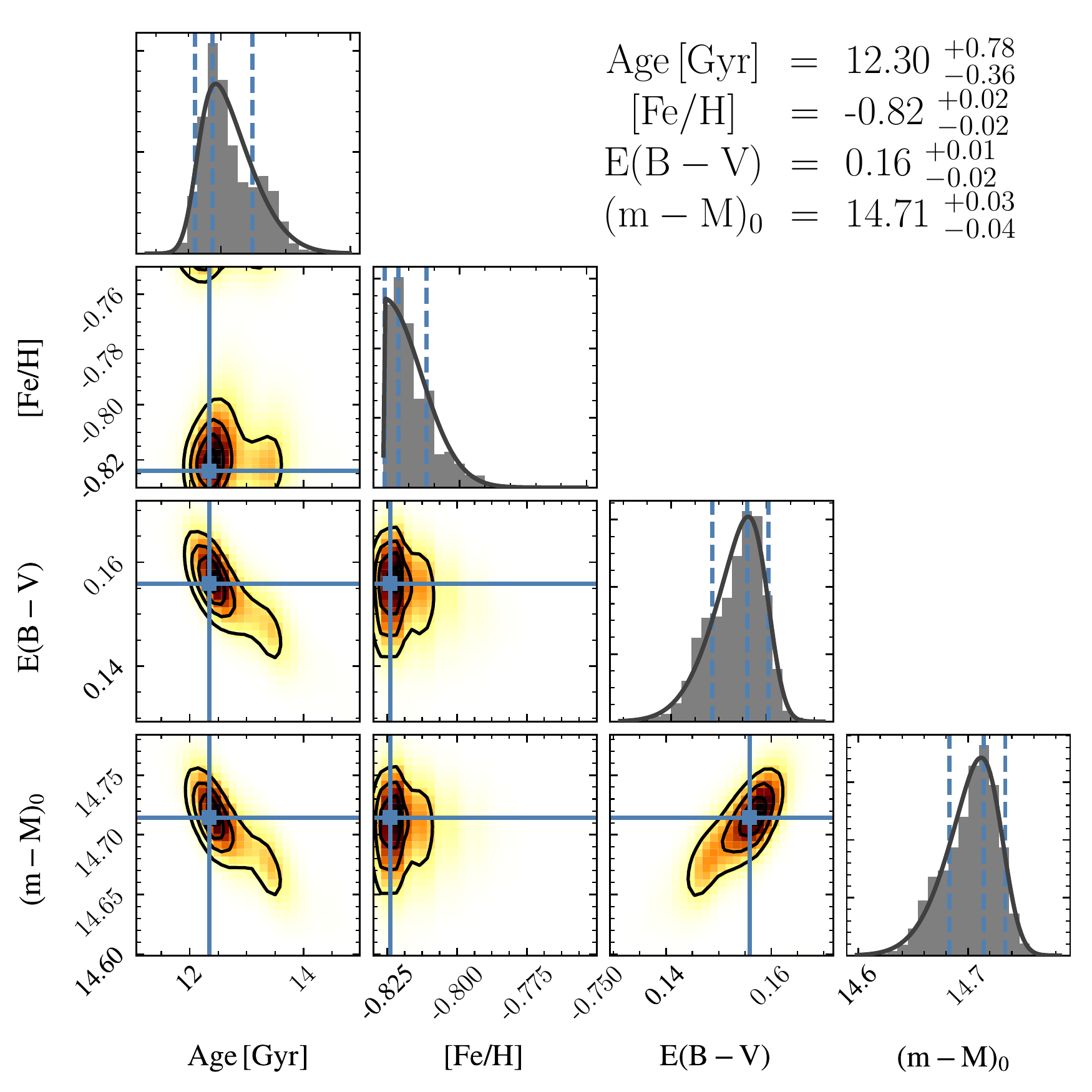}
\caption{Same as Figure~\ref{fig:Fig9} for DSED isochrones.}
\label{fig:Fig12}
\end{figure*}

\begin{figure*}
    \centering
    \includegraphics[width=0.80\textwidth]{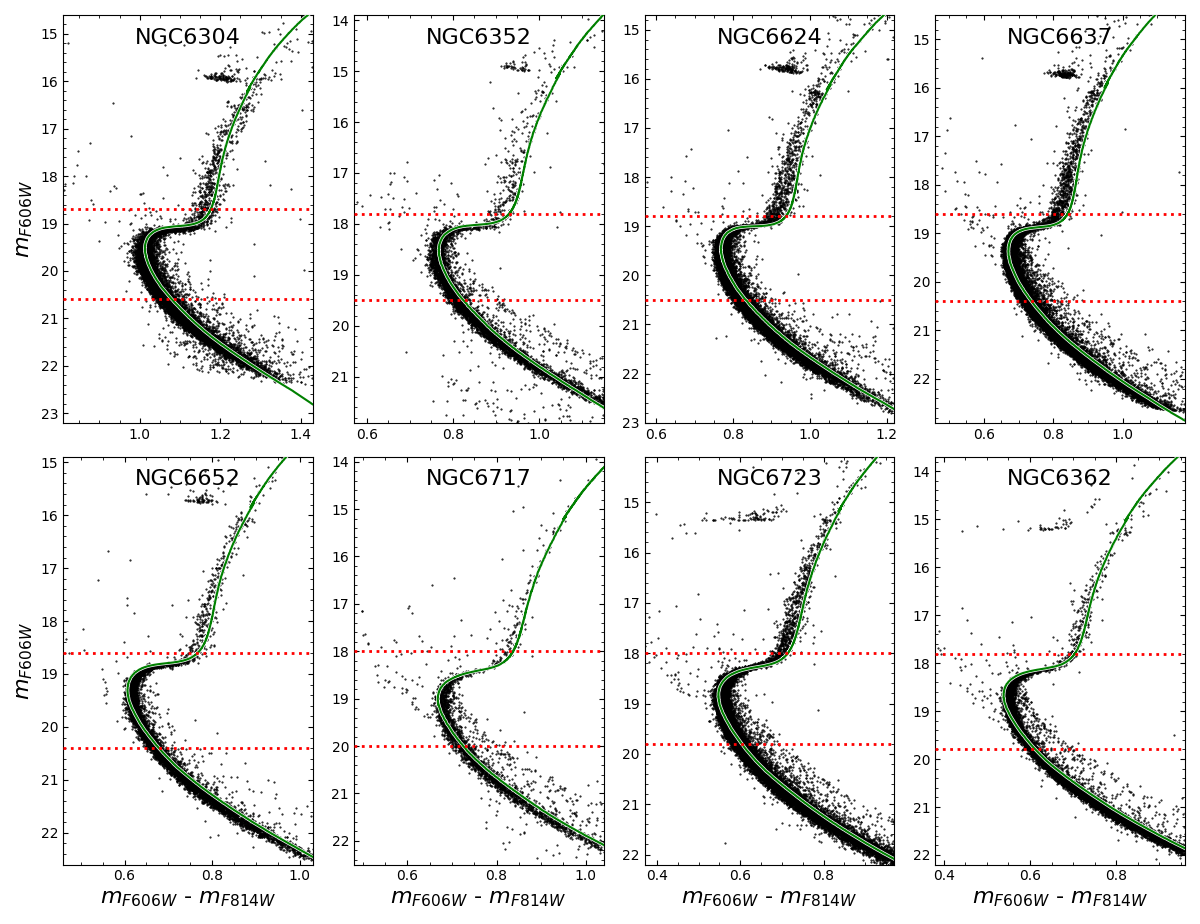}
    \caption{Results of the isochrone fits for all eight clusters with BaSTI isochrones. The red dotted lines represent the magnitude range adopted for the fitting.}
    \label{fig:results-basti}
\end{figure*}

\begin{figure*}
    \centering
    \includegraphics[width=0.80\textwidth]{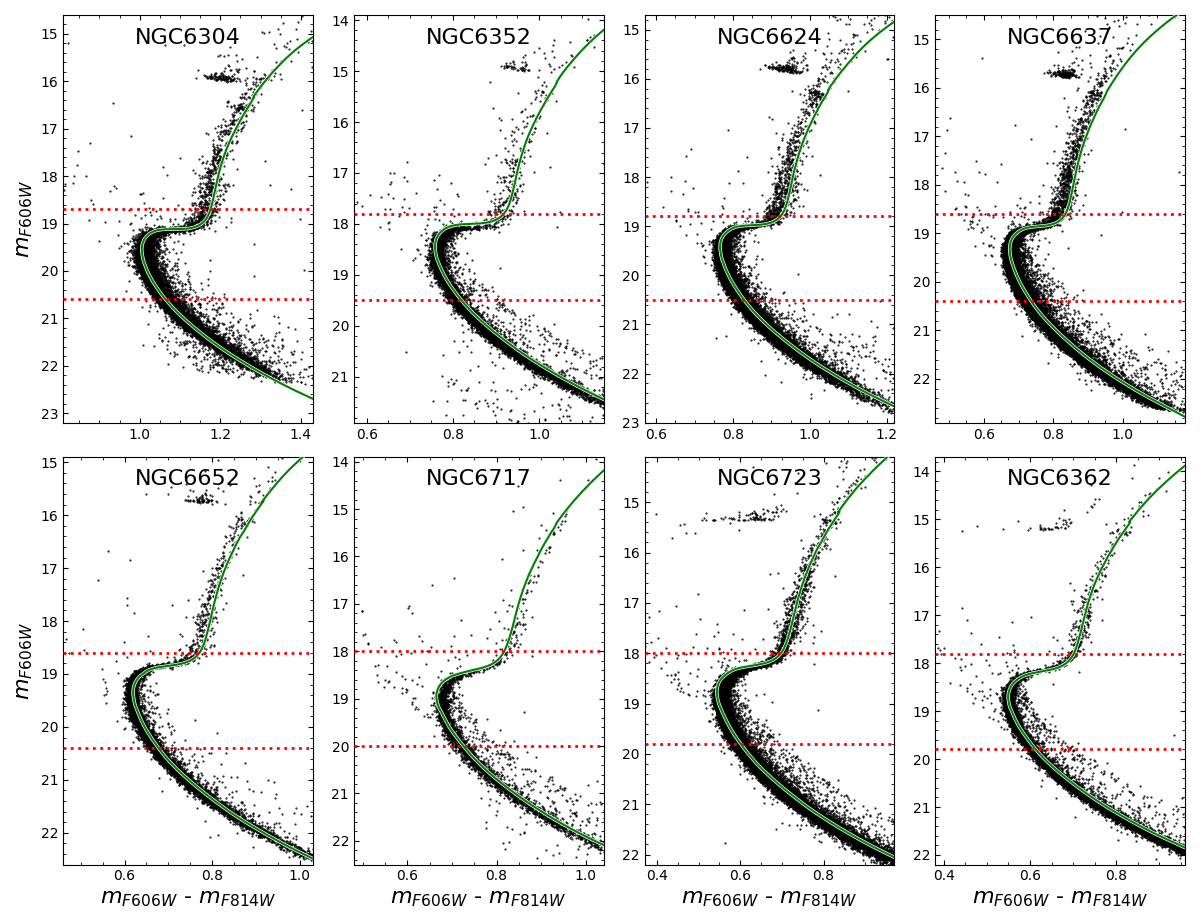}
    \caption{Same as Figure \ref{fig:results-basti} for DSED isochrones.}
    \label{fig:results-dsed}
\end{figure*}

\subsection{Isochrone fitting: DSED and BaSTI}

The results from the isochrone fitting considering each GC as a SSP and
adopting two sets of isochrones (DSED and BaSTI) are shown in Table~\ref
{tab:Isoc-Results}. The average result for each cluster is calculated as a
simple average of the center values for BaSTI and DSED, and the non-symmetric
uncertainties are calculated as the quadrature sum of the two model
uncertainties.

The adopted free parameters are age, $\rm{[Fe/H]}$, $(m-M)_0$ and $E(B-V)$,
but the apparent distance modulus $(m-M)_{V}$ and distance are also shown
in Table~\ref{tab:Isoc-Results}. The ages and distance results are discussed
in Section~\ref{subsec:disc}, and compared with previous literature results. The
derived ages present a mean absolute error between $\sim0.4$ and 0.8\,Gyr, where
the uncertainty propagation and the systematic errors from the comparison of
DSED and BaSTI models are considered.

For the most metal-rich GCs, an important ingredient to be considered is the
 $\alpha$-enhancement. For NGC\,6304 and NGC\,6352 ($\rm
{[Fe/H]}\sim-0.50$), we adopted $\rm{[\alpha/Fe]}=+0.2$, more consistent with
the values from Table~\ref{tab:Met-Abund}. The reason for this comes from the
[$\alpha$/Fe] decrease with increasing metallicity \citep{2018ARA&A..56..223B},
such that it would correspond to $\rm{[\alpha/Fe]}\sim+0.2$ for these GCs. In
this case, the isochrones were interpolated. Considering $\rm{[\alpha/Fe]}=+0.2$
leads to slightly older ages (relative to adopting $\rm{[\alpha/Fe]}=+0.4$). It
suggests that other metal-rich clusters, where enhanced [$\alpha$/Fe] values
were considered, may have to be reassessed \citep[e.g.][]{2014ApJ...782...50L}.
Further spectroscopic derivations of accurate $\alpha$-element abundances are
greatly needed.

\begin{figure}
    \centering
    \includegraphics[width=\columnwidth]{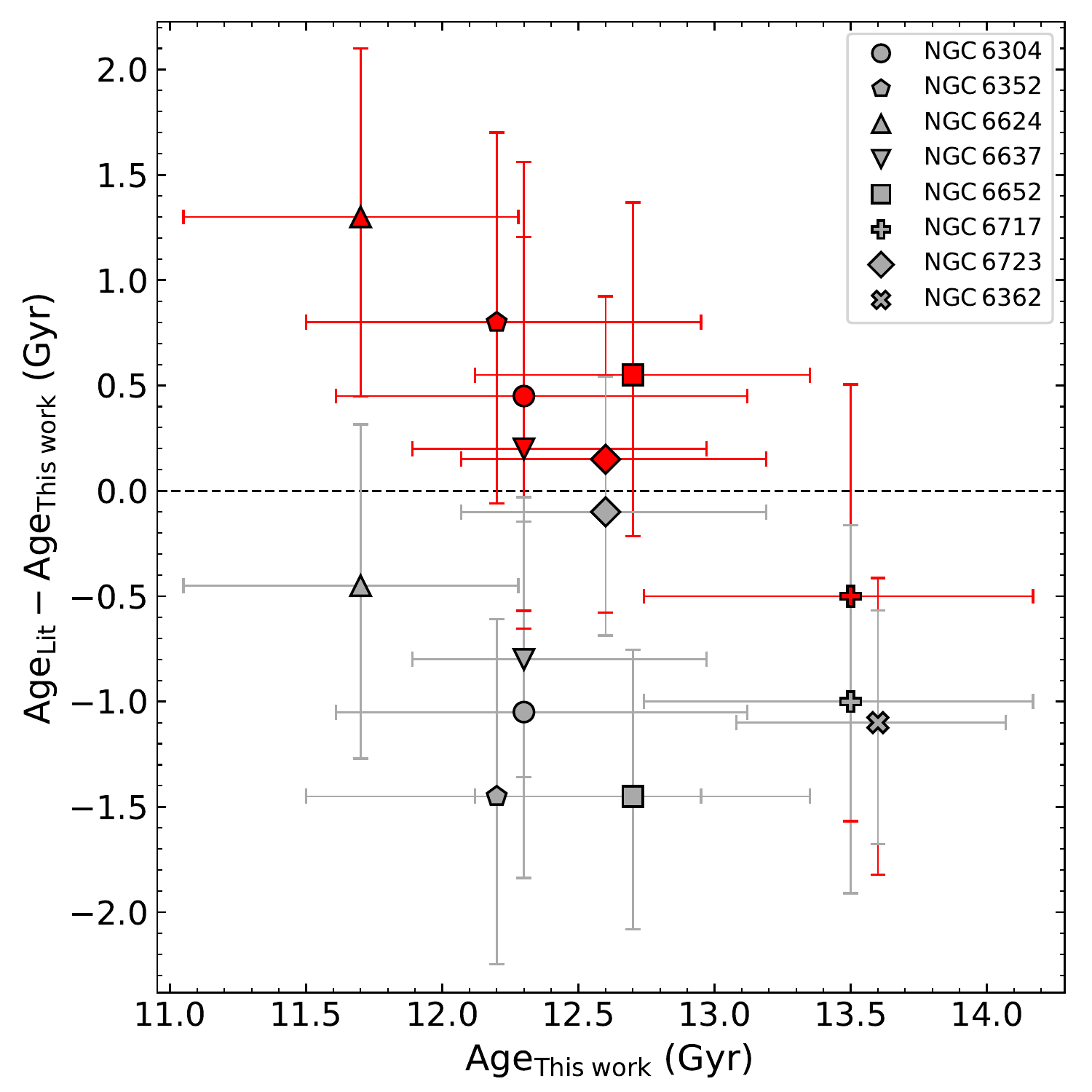}
    \caption{Comparison between the ages derived in \citet[red symbols]
    {2010ApJ...708..698D} and \citet[gray symbols]{2013ApJ...775..134V}, and
    our derived ages. The result for NGC\,6362 is the same in
    both literature works, then its symbol and the error bars overlap.}
    \label{fig:Lit-Comp}
\end{figure}

The best isochrone fits to CMDs and corresponding corner plots for NGC\,6637 are
presented in Figures~\ref{fig:Fig9} (BaSTI isochrones before the correction)
and \ref {fig:Fig12} (DSED isochrones). The best fit CMD can be seen in the
central solution identified with the solid line, whereas the red area shows the
region around the isochrone corresponding to the derived parameters within
1$\sigma$. The black dots correspond to the stars used in the isochrone
fitting and the gray ones to all the observed stars. The corner plots present
the probability distribution function derived for the free parameters in the
diagonal panels and the correlations between two parameters in the other panels.

Figures \ref{fig:results-basti} and \ref{fig:results-dsed} show the best
isochrone fit to all the sample clusters, employing respectively the BaSTI and
DSED isochrones. The fit is carried out to stars between the lower RGB and 1.0
magnitude below the MSTO, as shown by red dotted lines. RGB stars are avoided
due to convection issues, as explained above, but even so the fits to the RGBs
are very good in most cases.

\subsection{Discussion on ages and distances} \label{subsec:disc}

\begin{figure}
    \centering
    \includegraphics[width=0.98\columnwidth]{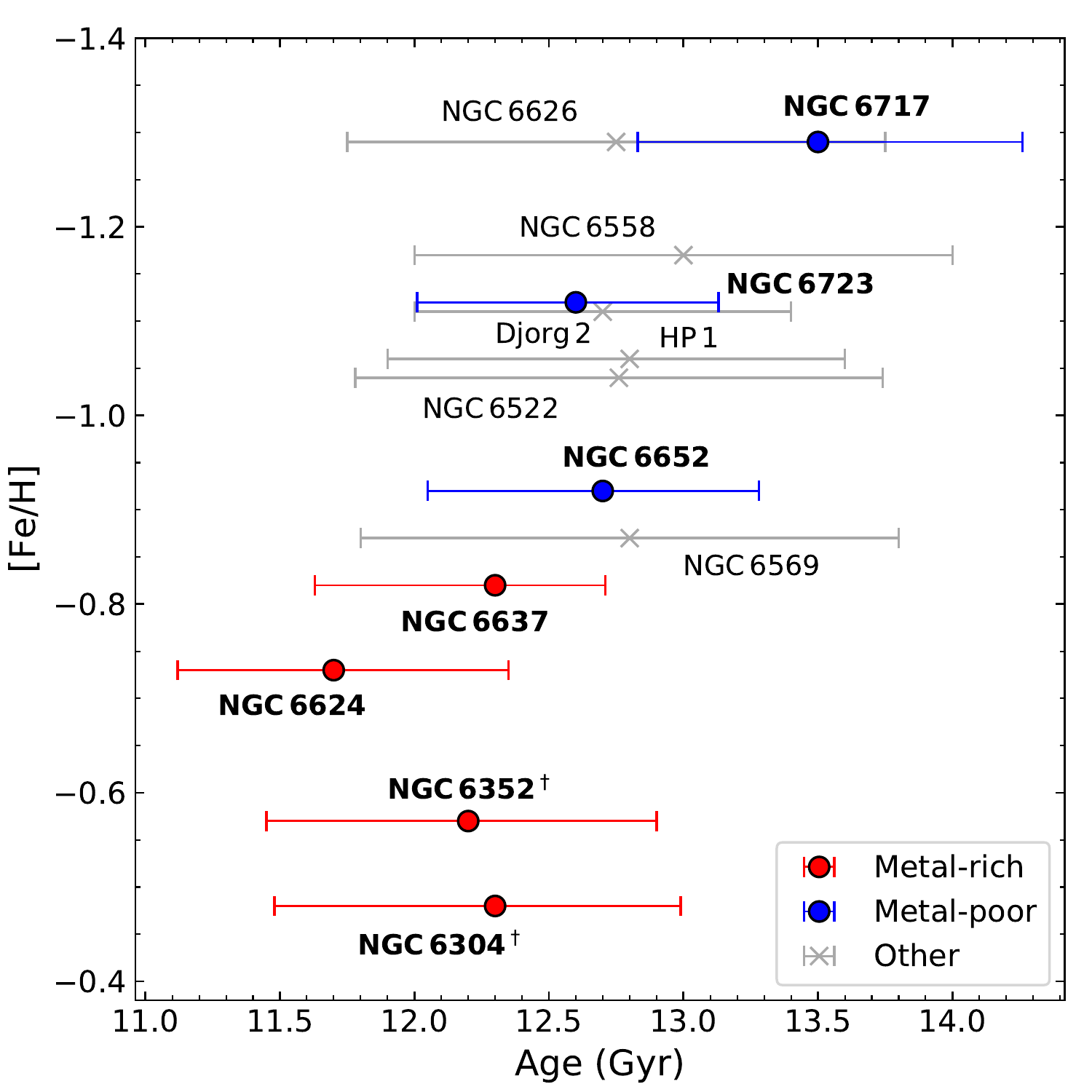}
    \caption{Distribution of the derived ages vs. metallicities in the
    range $-1.3 < \rm{[Fe/H]} < -0.4$. The circles represent the seven
    sample bulge GCs (four moderately metal-rich in red and three moderately
    metal-poor in blue). The gray markers correspond to the other bulge GCs that
    have accurate age measurements in the literature: NGC\,6626 and NGC\,6522
    \citep{2018ApJ...853...15K}; NGC\,6558 \citep{2018A&A...619A.178B}; HP\,1
    \citep{2019MNRAS.484.5530K}; NGC\,6569 \citep{2019ApJ...874...86S}; and
    Djorg\,2 \citep{2019A&A...627A.145O}. Those marked with $\dagger$ correspond
    to $[\alpha\rm{/Fe}] = +0.2$.}
    \label{fig:FEHvsAge}
\end{figure}

Previous determinations of ages, distances and reddening values through
isochrone fitting methods for the sample GCs are reported in Table~\ref
{tab:LitAges}, namely from: \citet{2010ApJ...708..698D}, \citet{2013ApJ...775..134V}
and \citet{2017ApJ...838..162O} adopting ACS photometry (F606W and F814W
filters) from the GO-10775 program; \citet{2018ApJ...853...15K} adopting ACS
and WFC3 photometry (F625W and F438W filters) from GO-12008 and GO-13297
programs; and \citet{2016ApJ...832...48S} with Gemini near-infrared data
(GeMS+GSAOI).

\citet{2010ApJ...708..698D} used an isochrone fitting method
that consists in measuring the MSTO absolute magnitude and then interpolating
an isochrone grid of the MSTO as a function of age and metallicity, with fixed
$E(B-V)$ and absolute distance modulus $(m-M)_0$, applying DSED isochrones.

\citet{2013ApJ...775..134V} derived the ages for 55 GCs from $\Delta_{\rm
TO}^{\rm HB}$ measurement, with the Victoria-Regina models. They report that
the ages derived for NGC\,6304, NGC\,6624 and NGC\,6637 are the least reliable
ones because their CMDs are strongly affected by differential reddening and
field-star contamination.

\begin{deluxetable*}{lcccccccc}
\tablenum{5}
\tablecaption{Literature ages, reddening and distance modulus derived for a
given $\rm{[Fe/H]}$ and $\rm{[\alpha/Fe]}$.}
%  except for \citet{2016MNRAS.463.3768W}, which leaves this parameter
% free between $0.23$ and $0.40$
\label{tab:LitAges}
\tablewidth{0pt}
\tablehead{
\colhead{Cluster} & \colhead{[Fe/H]} & \colhead{[$\alpha$/Fe]} & \colhead{$Y$} & \colhead
{Age (Gyr)} & \colhead{$E(B-V)$} & $d_{\odot}$ (kpc) & Model & \colhead{Ref.}}% & Data & Filters}
%\decimalcolnumbers
\startdata
\multirow{2}{*}{NGC\,6304}
& $-0.50$ & $+0.20$ & 0.25 & $12.75\pm0.75$ & $0.482$ & $6.21$ & DSED & D10 \\
& $-0.37$ & $+0.22$ & 0.264 & $11.25\pm0.38$ & $0.500$ & $6.28$ & V-R & VdB13 \\
% & $-0.45$ & $+0.00$ & $0.311^{+0.009}_{-0.006}$ & $12.950^{+0.199}_{-0.186}$
% & DSED & WK17 \\
\hline
\multirow{2}{*}{NGC\,6352}
 & $-0.80$ & $+0.40$ & 0.25 & $13.00\pm0.50$ & $0.258$ & $5.29$ & DSED & D10 \\
 & $-0.62$ & $+0.37$ & 0.259 & $10.75\pm0.38$ & $0.27$ & $5.44$ & V-R & VdB13 \\
\hline
\multirow{3}{*}{NGC\,6624}
 & $-0.50$ & $+0.00$ & 0.25 & $13.00\pm0.55$ & $0.258$ & $8.19$ & DSED & D10 \\
 & $-0.42$ & $+0.25$ & 0.263 & $11.25\pm0.50$ & $0.268$ & $7.99$ & V-R & VdB13 \\
 & $-0.60$ & $ +0.40$ & $\sim 0.26$ & $12.50\pm0.50$ & $0.28$ & $7.91$ & DSED,
 BaSTI, V-R & S16 \\
% & $-0.44$ & $+0.00$ & $0.303^{+0.004}_{-0.003}$ & $12.608^{+0.101}_{-0.086}$ & %DSED & WK17 \\
\hline
\multirow{2}{*}{NGC\,6637}
 & $-0.70$ & $+0.20$ & 0.25 & $12.50\pm0.75$ & $0.166$ & $8.86$ & DSED & D10 \\
 & $-0.59$ & $+0.35$ & 0.260 & $11.00\pm0.38$ & $0.163$ & $8.97$ & V-R & VdB13 \\
% & $-0.64$ & $+0.00$ & $0.279^{+0.003}_{-0.003}$ & $13.488^{+0.011}_{-0.035}$ &
% DSED & WK17 \\
\hline
\multirow{3}{*}{NGC\,6652}
 & $-0.75$ & $+0.20$ & 0.25 & $13.25\pm0.50$ & $0.115$ & $9.26$ & DSED & D10 \\
 & $-0.76$ & $+0.46$ & 0.257 & $11.25\pm0.25$ & $0.116$ & $9.59$ & V-R & VdB13 \\
 %& $-1.10$ & $+0.00$ & $0.235^{+0.005}_{-0.003}$ & $13.497^{+0.003}_{-0.018}$ &
 %DSED & WK17 \\
 & $-0.76$ & $+0.40$ & --- & $11.4\pm2.0$ & $0.11$ & $9.68$ & DSED & OM17 \\
\hline
\multirow{2}{*}{NGC\,6717}
 & $-1.10$ & $+0.20$ & 0.25 & $13.00\pm0.75$ & $0.207$ & $7.55$ & DSED & D10 \\
 & $-1.26$ & $+0.46$ & 0.250 & $12.50\pm0.50$ & $0.225$ & $7.27$ & V-R & VdB13 \\
% & $-1.26$ & $+0.00$ & $0.330^{+0.002}_{-0.010}$ & $13.426^{+0.070}_{-0.139}$ &
% DSED & WK17 \\
\hline
\multirow{3}{*}{NGC\,6723}
 & $-1.00$ & $+0.20$ & 0.25 & $12.75\pm0.50$ & $0.074$ & $8.06$ & DSED & D10 \\
 & $-1.10$ & $+0.46$ & 0.250 & $12.50\pm0.25$ & $0.07$ & $8.05$ & V-R & VdB13 \\
% & $-1.10$ & $+0.00$ & $0.275^{+0.003}_{-0.003}$ & $13.493^{+0.007}_{-0.026}$ &
% DSED & WK17 \\
 & $-1.10$ & $+0.40$ & --- & $11.9\pm2.0$ & $0.05$ & $8.75$ & DSED & OM17 \\
\hline\hline
\multirow{4}{*}{NGC\,6362}
 & $-1.10$ & $+0.40$ & 0.25 & $12.50\pm0.50$ & $0.071$ & $7.64$ & DSED & D10 \\
 & $-1.07$ & $+0.46$ & 0.25 & $12.50\pm0.25$ & $0.076$ & $7.46$ &  V-R & VdB13 \\
% & $-0.99$ & $+0.00$ & $0.272^{+0.003}_{-0.003}$ & $13.495^{+0.005}_{-0.018}$ &
% DSED & WK17 \\
 & $-1.07$ & $+0.40$ & --- & $11.4\pm2.0$ & $0.07$ & $8.09$ & DSED & OM17 \\
 & $-1.08$ & $+0.40$ & 0.25 & $13.7\pm1.0$ & $0.053$ & $7.69$ &  DSED, BaSTI & K18
 \enddata
\tablecomments{D10: \citet{2010ApJ...708..698D}; VdB13: \citet[where V-R
refers to the Victoria-Regina models]{2013ApJ...775..134V}; S16: \citet
{2016ApJ...832...48S}; OM17: \citet{2017ApJ...838..162O}; and K18: \citet
{2018ApJ...853...15K}. The color excess $E(m_{\rm{F606W}}-m_{\rm{F814W}})$ and
apparent distance moduli were transformed using the extinction coefficients
from the PARSEC database and the transformation given in \citet
{2013MNRAS.433..243C}.}
% WK17: \citet{2017MNRAS.468.1038W};
\end{deluxetable*}

\citet{2016ApJ...832...48S} and \citet{2018ApJ...853...15K} applied $\chi^2$
calculation in the isochrone fitting for NGC\,6624 and NGC\,6362 respectively.
\citet{2016ApJ...832...48S} kept the age as the only free parameter and assumed
the minimum value of $\chi^2$ as the solution. On the other hand, \citet
{2018ApJ...853...15K} considered the age, distance modulus and reddening as
the free parameters and also applied the \texttt{emcee} code to sample their
posterior probabilities. \citet{2017ApJ...838..162O} applied a Monte Carlo
analysis to carry out a MS-fitting to derive the ages and distances for 22 GCs,
resulting in higher uncertainties.

Figure~\ref{fig:Lit-Comp} compares the ages from \citet[red symbols]
{2010ApJ...708..698D} and \citet[gray symbols]{2013ApJ...775..134V} with our
results (Table~\ref{tab:Isoc-Results}). In general, the results from \citet
{2010ApJ...708..698D} are around 0.5\,Gyr older than our results, whereas
those from \citet{2013ApJ...775..134V} point systematically toward $1.0-1.5
$\,Gyr younger ages, showing a disagreement larger than the error bars.
Most of the  \citet{2010ApJ...708..698D} derived ages are compatible within 1$\sigma$
with the present results.

Figure \ref{fig:FEHvsAge} shows the distribution of derived ages vs.
metallicities, updating an interesting plot presented in \citet[their
Figure~16]{2019ApJ...874...86S}. The moderately metal-poor GCs ($\rm{[Fe/H]}
\lesssim-0.85$) appear to be slightly older than the metal-rich ones, with average
ages of $12.86\pm0.36$\,Gyr and $12.12\pm0.32$\,Gyr respectively. The
very low statistics of objects, the individual age uncertainties of $\sim
0.50$\,Gyr and the probability that several of the metal-rich ones might
be assigned to a thick disk population,
\citep{2019MNRAS.tmp.2773P}, prevent any further conclusion on systematic age
difference as a function of $\rm{[Fe/H]}$ for bulge clusters.
%Despite the trend of younger ages for the
%more metal-rich GCs, when considering the uncertainties there is no clear
%distinction between the moderately metal-rich and metal-poor GCs. 
It is worth noting that some clusters analyzed here (NGC\,6717 and NGC\,6362)
and in previous works \citep[e.g.][]{2018ApJ...853...15K,2019MNRAS.484.5530K}
are revealed to be among the oldest GCs in the Galaxy. 

For all clusters the ages from DSED and BaSTI are compatible within
1$\sigma$. The clusters NGC\,6304 and NGC\,6624 show the largest age differences
between the derivations based on DSED and BaSTI, amounting to 1.3\,Gyr and
0.7\,Gyr respectively. This is due to their larger reddening, and consequently
a larger uncertainty in the distances -- see discussion in \citet
{2019MNRAS.tmp.2773P}. As a matter of fact, the uncertainty in distances is a
major difficulty for establishing precise ages and orbits.

The derived metallicities are in very good agreement within 1$\sigma$ with
the adopted values in the Gaussian priors (Table~\ref{tab:Priors}). 
For the GCs that presented at least one
member RRL (NGC\,6717, NGC\,6723 and NGC\,6362), the derived apparent distance
moduli $(m-M)_{V}$ are also compatible with those given by RRL mean
magnitudes (Figure~\ref{fig:gaiarrlyrae}).
The derived distances and reddening values (Table~\ref{tab:Isoc-Results}) are
also consistent with those from \citet{2010ApJ...708..698D}, \citet
{2013ApJ...775..134V}, \citet{2016ApJ...832...48S} and \citet
{2018ApJ...853...15K}, as shown in Table~\ref{tab:LitAges}. 

Comparing our derived distances with those obtained from the \textit
{Gaia} DR2 parallaxes \citep{2018A&A...616A..12G}, given in \citet
{2019MNRAS.tmp.2773P} for bulge GCs, a rather large discrepancy 
from $20$ to $65$ per cent is observed for NGC\,6304,
NGC\,6352, NGC\,6637 and NGC\,6723. For NGC\,6624, NGC\,6652 and NGC\,6717,
this discrepancy remains below $10\%$. For these large distances
the {Gaia} DR2 parallaxes are often not suitable \citep[see][]
{2019MNRAS.tmp.2773P}.

%%%%%%%%%%%%%%%%%%%%%%%%%%%%%%%%%%%%%%%%%%%%%%%%%%%%%%%%%%%%%%%%%%%%%%%%%%%%%%%%
%%%%%%%%%%%%%%%%%%%%%%%% MULTIPLE POPULATIONS ANALYSIS %%%%%%%%%%%%%%%%%%%%%%%%%
%%%%%%%%%%%%%%%%%%%%%%%%%%%%%%%%%%%%%%%%%%%%%%%%%%%%%%%%%%%%%%%%%%%%%%%%%%%%%%%%
\section{Multiple Stellar Population Analysis}\label{sec:MPs}

In \citet[\citetalias{2017MNRAS.464.3636M}]{2017MNRAS.464.3636M}, the
identification of the multiple stellar populations, with the ``chromosome
map'' diagram, was described in detail and applied to the RGB stars for the 56
GCs from the GO-13297 program. Previously, \citet[\citetalias
{2015ApJ...808...51M}]{2015ApJ...808...51M} performed the identification of
the MPs in both the MS and RGB of NGC\,2808, also using this diagram.

In this work, we distinguish the MPs from the MS to the RGB, including the SGB,
to carry out the isochrone fitting both to the first
(1G) and second generation (2G) stars  simultaneously, and to check whether there
occurs any detectable age difference.

Note that, in the pseudo-color $C_{\rm F275W,F336W,F438W} =
(m_{\rm F275W}-m_{\rm F336W})-(m_{\rm F336W}-m_{\rm F438W})$,
 F275W, F336W, and F438W are dominated
by OH, NH and CN bands respectively. C, O are anticorrelated with
respect to N in 2G stars. In particular, a 2G star is fainter than 1G stars
in F336W because it is N and Na-richer, and brighter than 1G stars in F275W
and F438W filters because they are O- and C-poorer, compared to 1G stars. The
filter F336W is counted twice in the pseudo-color, thus enhancing the
contrast \citepalias{2015AJ....149...91P}.

In some combinations of colors and magnitudes in different CMDs, the MPs seem
to be entangled in the SGB region, as they are not horizontally separated
from each other (Section~\ref{subsec:TCD}). The chromosome map basically
rectifies a distribution of stars, and shows whether it follows a bimodal (or
multimodal) horizontal distribution. Therefore, it cannot be applied to SGB
stars, since this region is typically horizontal in CMDs.

\subsection{Stellar population separation in RGB and MS: Chromosome maps}
\label{subsec:CMap}

The original chromosome map analysis was presented in \citetalias
{2017MNRAS.464.3636M}. Here we apply the same method, but improved in terms of
a more accurate separation of MPs. This is obtained by applying GMM algorithms
\citep[for details see][]
{2020arXiv200102697S}. We also compute the fractions of stars from first
(N$_{1}$/N$_{\rm TOT}$) and second generations and compare our results to those
from \citetalias{2017MNRAS.464.3636M} in Table~\ref{tab:BonaFide}, showing a
good agreement.

\begin{figure}
    \centering
    \includegraphics[width=0.9\columnwidth]{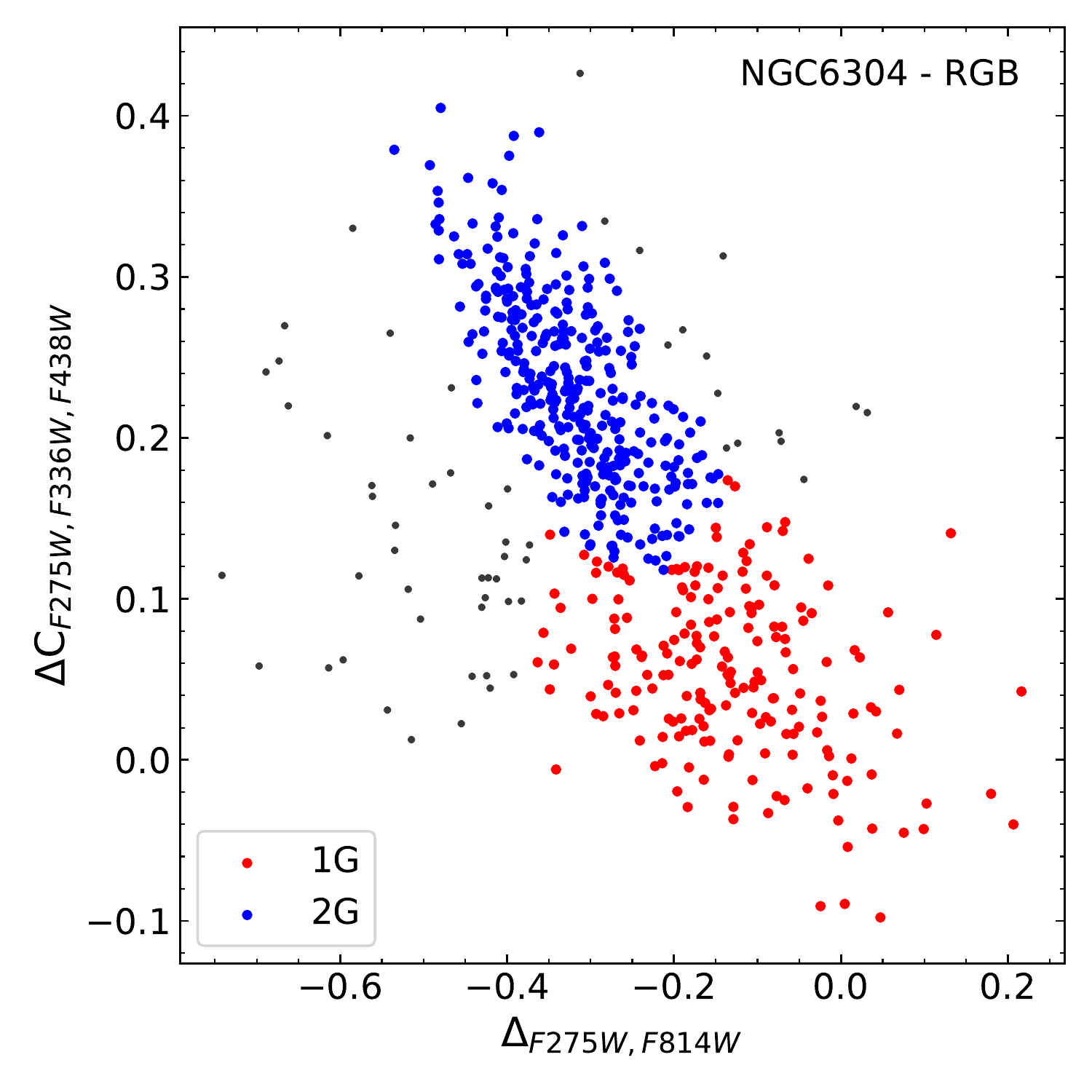}
    \caption{Chromosome map of RGB stars in NGC\,6304. The 1G and 2G were
    separated using GMM techniques. }
    \label{fig:6304_RGB_ChMap}
\end{figure}

\citetalias{2017MNRAS.464.3636M} concluded that the seven GCs of the present
sample are type-I clusters, meaning that their chromosome maps do not present
additional sequences and that their 1G and 2G stars can be separated more
clearly for some clusters such as NGC\,6352, and less so for cases as NGC\,6304.
These patterns are also observed in the present analysis, where we use an
automatic approach allowing to separate satisfactorily the MPs, even for
NGC\,6304 (Figure~\ref{fig:6304_RGB_ChMap}).

\begin{deluxetable}{lcc}
\tablenum{6}
\tablecaption{Comparison between the fractions of 1G stars over the total number of
RGB stars from \citet[M17]{2017MNRAS.464.3636M} and the values derived in the
present work.}
\label{tab:BonaFide}
\tablewidth{10pt}
\tablehead{
\colhead{Cluster} & \colhead{($N_1$/$N_{\rm TOT}$)$_{\rm M17}$} & \colhead{$N_1$/$N_{\rm TOT}$}%$_{\rm TOT}$}
%& \colhead{($N_1$/$N_{\rm TOT}$)$_{\rm BF}$}
}
%\decimalcolnumbers 
\startdata
NGC\,6304 &         ---       &  $0.362\pm0.059$ \\%& $0.333\pm0.046$ \\
NGC\,6352 &  $0.474\pm0.035$  &  $0.426\pm0.041$ \\%& $0.438\pm0.029$ \\
NGC\,6624 &	$0.279\pm0.020$	 &  $0.462\pm0.103$ \\%& $0.281\pm0.042$ \\
NGC\,6637 &	$0.425\pm0.017$	 &  $0.481\pm0.036$ \\%& $0.529\pm0.033$ \\
NGC\,6652 &	$0.344\pm0.026$	 &  $0.371\pm0.041$ \\%& $0.467\pm0.042$ \\
NGC\,6717 &	$0.637\pm0.039$	 &  $0.635\pm0.052$ \\%& $0.611\pm0.026$ \\
NGC\,6723 &	$0.363\pm0.017$	 &  $0.377\pm0.029$ \\%& $0.607\pm0.033$ \\
%NGC~6752 &	$0.294\pm0.023$	 &  $0.322\pm0.033$ & $0.479\pm0.037$ \\
\hline
NGC\,6362 &	$0.574\pm0.035$	 &  $0.584\pm0.041$	\\%& $0.576\pm0.029$
\enddata
%\tablecomments{$^\dagger$ represents that a 3-$\sigma$ from the high-resolution
%measurements uncertainties was adopted.}
\end{deluxetable}

%{\bf For the RGB stars, in order to retrieve a bulk sample of the MPs} (by
%neglecting stars in the intersection between 1G and 2G), we 
%%proceeded as in \citetalias{2018MNRAS.481.5098M}, by
%%increased the number of GMM components, taking into account only the two
%extremes, as exemplified in Figure \ref{fig:bonafide} for NGC\,6637.
%The fraction of bona-fide 1G stars is also shown in Table~\ref{tab:BonaFide}.
%{\bf For clusters with no variation of Fe, when selecting these bulk stars, we
%are selecting part of the 1G and part of the 2G populations. These
%sub-groups of 1G and 2G stars represent the two different populations, whose
%stars have different chemical composition \citep[e.g. \citetalias
%{2015ApJ...808...51M}, \citetalias{2018MNRAS.481.5098M} and][]
%{2019MNRAS.487.3815M}. For more complex clusters, such as $\omega$\,Cen,
%NGC\,2808 or M22 \citep{2020ApJ...888L...6L}, this separation of sub-populations
%would not apply. The fractions of 1G stars remain compatible within 3$\sigma$
%with those using all stars, except for NGC\,6723, where the bulk of 1G stars is
%more concentrated.}

\begin{figure}
\centering
\includegraphics[width=\columnwidth]{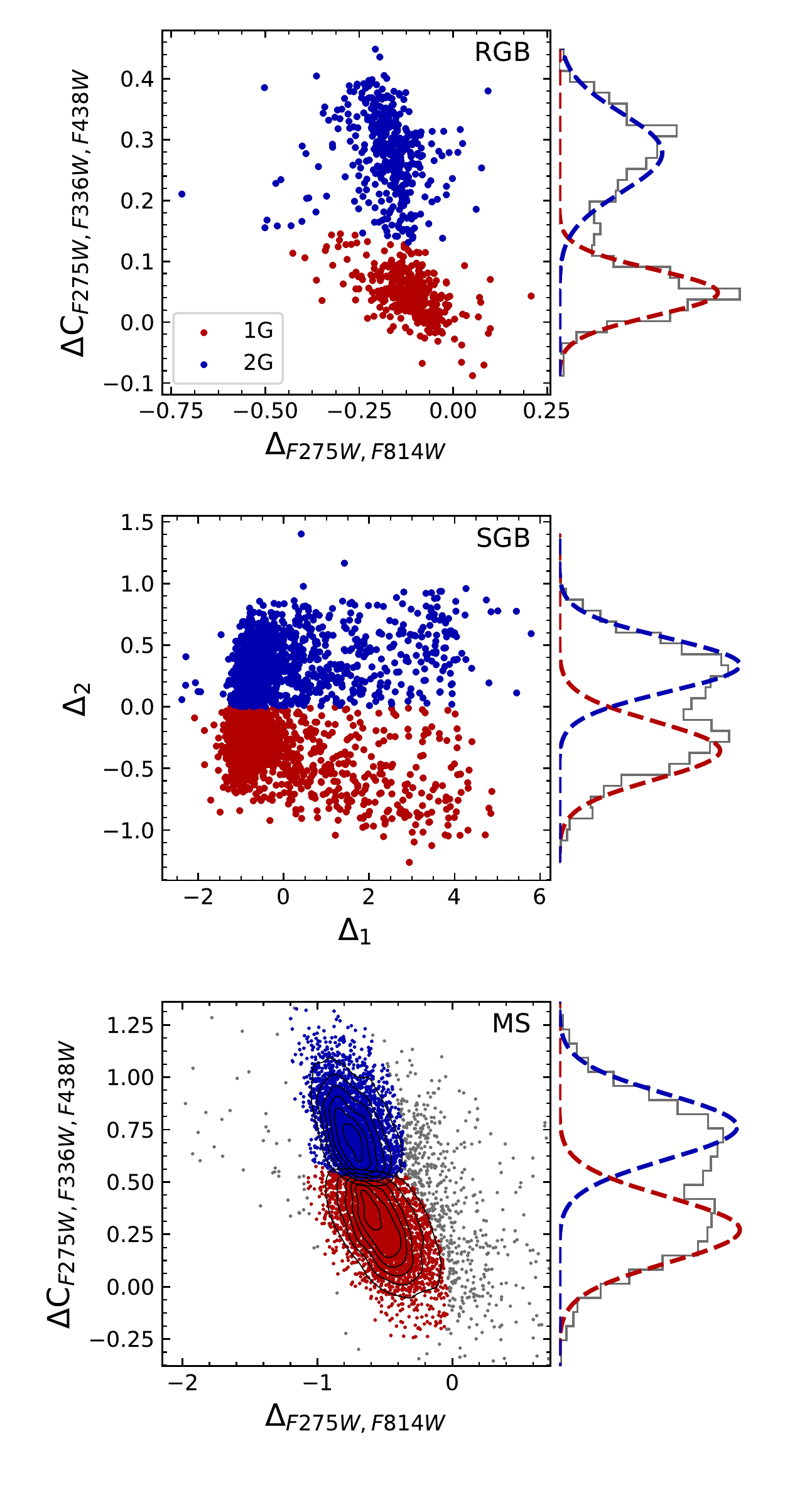}
\caption{Chromosome maps with the derived Gaussian distributions showing the MPs
separation for NGC\,6637, considering the RGB (top panel), MS (bottom panel),
and the two-color diagram applied to SGB (middle panel) stars. In the
middle panel, $\Delta_1$ and $\Delta_2$ are the coordinates obtained by rotating
the original two-color diagram counterclockwise by an angle $\theta=45^{\circ}
$. Since the MS is much more populated, we overplot the contour lines relative
to the two stellar generations. }
\label{fig:CMaps}
\end{figure}

Note that the fraction of 1G stars for NGC\,6717 ([Fe/H]\,$=-1.26$) is 0.635,
which is among the three largest 1G fractions presented in \citetalias
{2017MNRAS.464.3636M}, together with NGC\,6101 and NGC\,6496. The chromosome
maps applied to the RGB (top panel) and MS (bottom panel) stars of NGC\,6637
are presented in Figure~\ref{fig:CMaps}. For the MS stars, where the
chromosome map is much more populated, we have excluded from the analysis
the stars (grey points) outside the $3\sigma$ level of the fitted Gaussian
models.

%\begin{figure}
%    \centering
%    \includegraphics[width=0.85\columnwidth]{figures/NGC6637_split_msp_BF.pdf}
%    \caption{Chromosome map of RGB stars in NGC\,6637 to exemplify {\bf the
%    selection of the bulk sample both for 1G (green points over the red ones)
%    and 2G stars (red points over the blue ones).}}
%    \label{fig:bonafide}
%\end{figure}

\subsection{Stellar population separation in the SGB: Two-color diagrams}
\label{subsec:TCD}

The SGB morphology is a function of metallicity and of the choice of magnitude
in the CMD. In general terms, the chromosome map is not effective to separate
the stellar generations among these stars. Here, we apply the conventional
two-color diagram $m_{\rm F336W}-m_{\rm F438W}$ vs. $m_{\rm F275W}-m_{\rm
F336W}$, as described in \citet[\citetalias{2015MNRAS.451..312N}]
{2015MNRAS.451..312N}, with the implementation of the GMM algorithm \citep
{2020arXiv200102697S}.

The result for the SGB stars of NGC\,6637 is presented in the middle panel
of Figure~\ref{fig:CMaps} with a counterclockwise rotation of $45^{\circ}$,
showing an evident separation of the MPs in the horizontal direction and a
clear bimodality in the histogram.

\begin{figure*}
\centering
\includegraphics[width=0.80\textwidth]{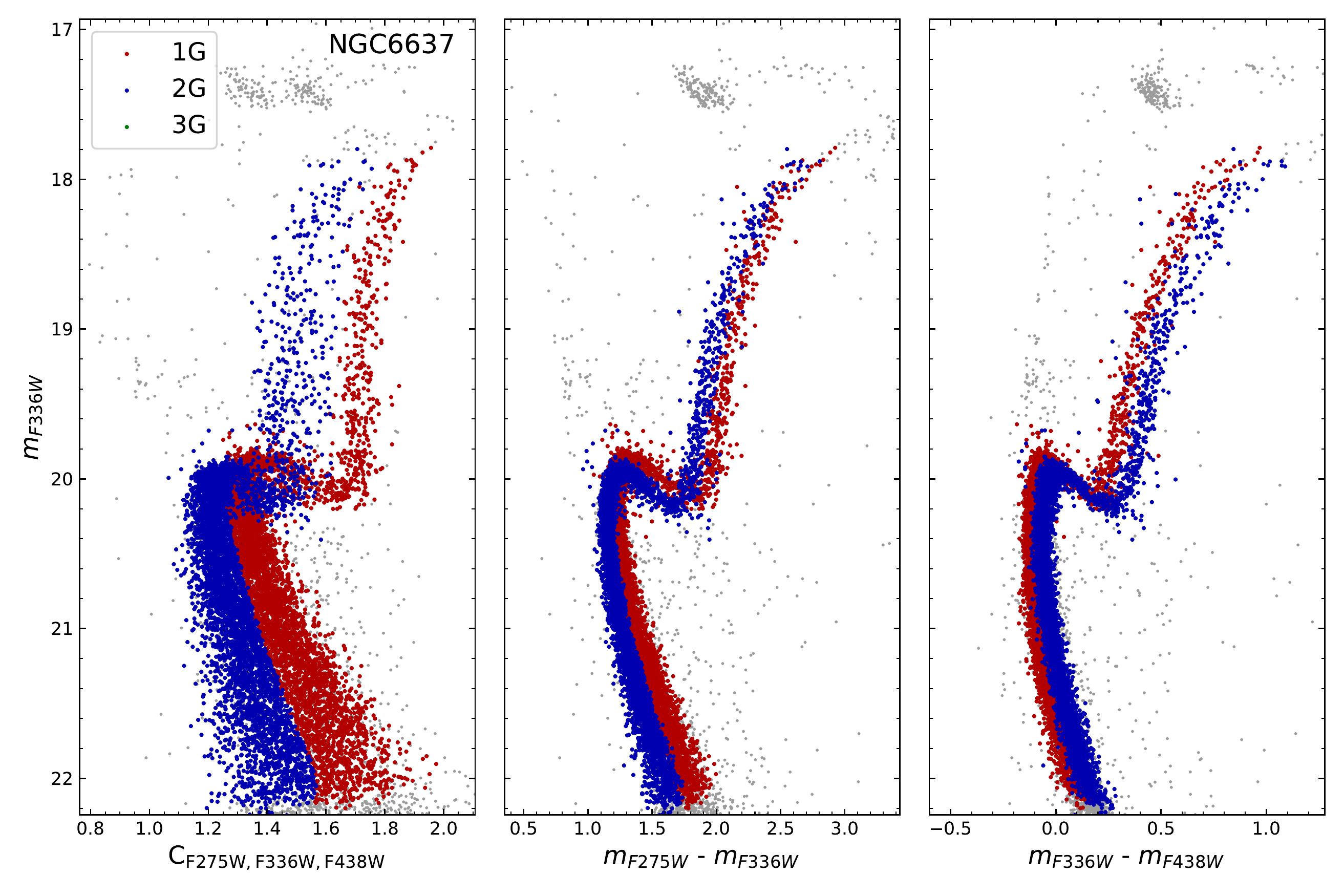}
\caption{MS-to-RGB MPs separation in CMD of NGC\,6637, from UV magnitudes,
colors, and pseudo-colors.}
\label{fig:3CMDs_N15}
\end{figure*}

Gathering the results from the MS to the RGB, several combinations of CMDs can
be tested. An example is shown in Figure~\ref{fig:3CMDs_N15} (similar to
Figure~2 from \citetalias{2015AJ....149...91P}), where the panels have the same
magnitude $m_{\rm F336W}$, but adopting colors such that the pseudo-color (left
panel) is defined by the subtraction of the two subsequent colors. As shown in
\citetalias{2015AJ....149...91P}, the 2G stars (N-rich and C-poor) are bluer in
the color $m_{\rm F275W}-m_{\rm F336W}$ than the 1G stars, but redder in $m_{\rm
F336W}-m_{\rm F438W}$. This is the reason why $C_{\rm F275,F336W,F438W}$
maximizes the MP separation, combining the ``magic trio'' of WFC3 filters.

\subsection{Multiple stellar populations: age differences} \label{subsec:MPsAges}

Derivations of age differences between stellar generations in a cluster were
carried out previously by \citet
{2008ApJ...673..241M} and \citet{2008ApJ...672L.115C} for NGC\,1851; \citet
{2011ApJ...733L..45R} for NGC\,288; \citet{2012ApJ...746...14M} and \cite
{2014ApJ...791..107V} for $\omega$\,Cen; \citet{2012A&A...541A..15M} for
NGC\,6656; \citet{2013ApJ...762...36J} for $\omega$\,Cen, M22 and NGC\,1851;
\citet{2013ApJ...778L..13L} for NGC\,2419; and \citet{2020arXiv200102697S} for NGC\,6752. In this collaboration, \citet[\citetalias{2015MNRAS.451..312N}]
{2015MNRAS.451..312N} obtained age differences for NGC\,6352, applying $\chi^2$
calculations for the isochrone fitting over synthetic CMDs. They infer a helium
abundance variation of $\Delta Y = 0.029\pm0.006$, and estimate an age
difference of $10\pm110$\,Myr assuming no difference in $\rm{[Fe/H]}$ and
$\rm{[\alpha/Fe]}$, with this uncertainty rising to $\sim300$\,Myr if a
difference of 0.02 is considered in both.

In this work, the age derivation was achieved from the 1G and 2G
($m_{\rm F606W}$ vs. $m_{\rm F606W}-m_{\rm F814W}$) CMDs simultaneously.
The analysis was carried out first considering the canonical helium for both
populations. As described in \citet{2020arXiv200102697S}, the likelihood function
for MPs assumes that the two stellar populations have the same distributions
of $E(B-V)$ and $(m-M)_0$ to perform the isochrone fitting, whereas [Fe/H] is
fixed. In that way, the best fit represents the best result for 1G and 2G at
the same time. After that, both generations are fitted assuming for 2G the
values of helium enhancement derived by \citet[\citetalias{2018MNRAS.481.5098M}]
{2018MNRAS.481.5098M}.

\begin{figure*}
    \centering
    \includegraphics[width=0.80\textwidth]{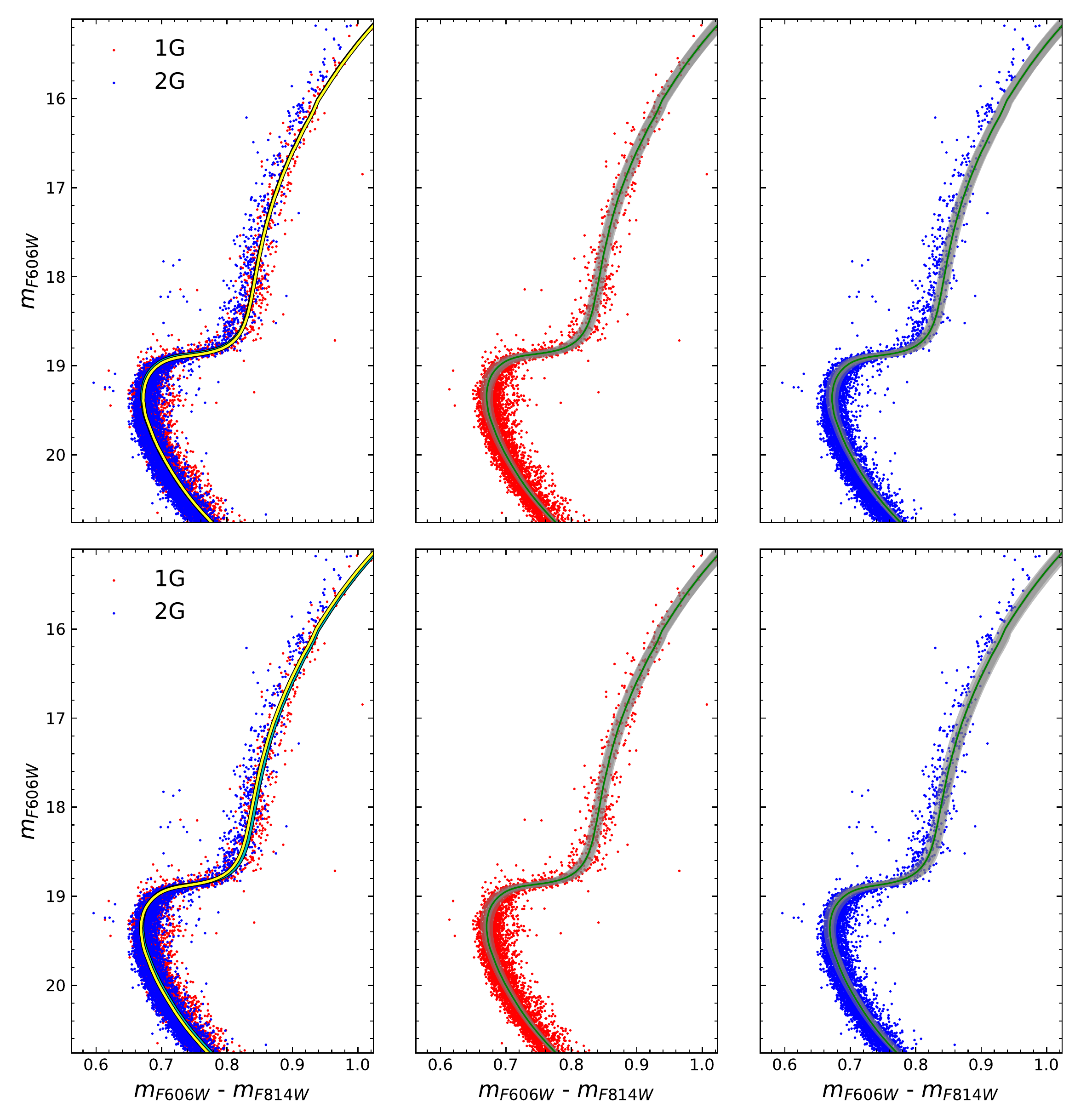}
    \caption{Isochrone fitting of MPs for NGC\,6637, considering canonical helium for both stellar populations (top panels), and considering helium enhancement for 2G stars (bottom panels). The dark region represents the result within 3$\sigma$.}
    \label{fig:ngc6637mps}
\end{figure*}

We derive a weighted mean age difference ($\left\langle\delta \tau_{\rm{1G,2G}}
\right\rangle$) of $41\pm 170$\,Myr, for the eight sample
clusters considering the canonical helium for both populations, as shown in
Table~\ref{tab:MPs-Ages}. This value reduces to $17\pm170$\,Myr
when the helium enhancement is taken into account. Recalling that the
individual age differences present an uncertainty of $\sim500$\,Myr, 
these smaller uncertainties were obtained by weighting the uncertainties of the
eight individual measurements.
Figure \ref{fig:ngc6637mps} shows the isochrone fitting for the MPs of
NGC\,6637. In the case of canonical He for 2G stars (top panels), a negative age
difference was derived for this cluster. The derived age difference vanishes
with the He enhancement in 2G (bottom panels), and shows a better fit for
2G stars.

 It is interesting to compare the age difference obtained for NGC\,6352, with
the results by \citet{2015MNRAS.451..312N}. We adopted $\rm{[Fe/H]}=-0.59$,
$[\alpha\rm{/Fe]}=+0.2$, canonical He abundance and $\Delta$Y = 0.027 for the
2G. \citet{2015MNRAS.451..312N} adopted $\rm{[Fe/H]}=-0.67$, $[\alpha\rm
{/Fe]}=+0.4$, He abundance and $\Delta$Y = 0.029 for the 2G. An age
difference of $10\pm110$ Myr was derived by \citet{2015MNRAS.451..312N}, which
is compatible within errors with the age difference derived by us, of $500\pm
480$\,Myr.

\begin{deluxetable}{lccc}
\tablenum{7}
\tablecaption{Age differences derived for the MPs, adopting both primordial and enhanced helium abundances.}
\label{tab:MPs-Ages}
\tablewidth{0pt}
\tablehead{
\colhead{Cluster} & \colhead{$\delta Y_{\rm max}$} & \colhead{$\delta \tau_{\rm 1G,2G}$} & \colhead{$\delta \tau_{\rm 1G,2G}^{\dagger}$} 
\\
  &  & (Gyr) & (Gyr) 
}
%\decimalcolnumbers 
\startdata
NGC\,6304$^\ddagger$ & $0.025\pm0.006$ & $-0.10^{+0.42}_{-0.46}$ & $-0.10^{+0.49}_{-0.45}$  \\
NGC\,6352$^\ddagger$ & $0.027\pm0.006$ & $ 0.40^{+0.49}_{-0.48}$ & $ 0.50^{+0.47}_{-0.48}$  \\
NGC~6624 & $0.023\pm0.003$ & $-0.10^{+0.46}_{-0.45}$ & $ 0.00^{+0.43}_{-0.50}$  \\
NGC~6637 & $0.012\pm0.005$ & $-0.20^{+0.58}_{-0.41}$ & $ 0.00^{+0.41}_{-0.36}$  \\
NGC~6652 & $0.017\pm0.011$ & $-0.10^{+0.53}_{-0.41}$ & $ 0.00^{+0.55}_{-0.49}$  \\
NGC~6717 & $0.003\pm0.009$ & $ 0.70^{+0.68}_{-0.70}$ & $ 0.30^{+0.70}_{-0.72}$  \\
NGC~6723 & $0.025\pm0.007$ & $ 0.00^{+0.51}_{-0.39}$ & $-0.10^{+0.51}_{-0.39}$  \\
\hline
NGC~6362 & $0.004\pm0.011$ & $0.10^{+0.44}_{-0.42}$  & $-0.20^{+0.40}_{-0.41}$ 
\enddata
\tablecomments{$^\dagger$ Results with helium enhancement in 2G stars, according
with \citetalias{2018MNRAS.481.5098M}. $^\ddagger$ $\rm{[\alpha/Fe]}=+0.2$.}
\end{deluxetable}

Systematic uncertainties in low-mass stellar models, affecting the cluster age
dating process, are mainly dominated by the treatment of superadiabatic
convection, diffusive process efficiency and low-temperature opacity; to these
sources of uncertainty one has to also add the error associated with the
still-present shortcomings in the bolometric corrections and effective
temperature - color relations. To firmly assess how much these systematics
contribute in the error budget of the GC age determination is difficult because
it depends on the metallicity range and adopted photometric systems. Data
listed in Table~\ref{tab:Isoc-Results} show that two independent, recent sets
of isochrones predict GC ages with a difference ranging between 0.1 and
1.0\,Gyr; when also accounting for the errors in the photometry and binary
star contamination in the MSTO region, it is safe to assume a realistic error
on the derived age of $\sim0.5-0.8$~Gyr.

In the case of relative ages such as the comparison between 1G and 2G, most of
these error sources are cancelled (all the zero points both of the photometry
and the models), but an additional source is added due to the effects of the
individual element abundance variation, in particular C, N, O and He. It is
well known that the helium, carbon, nitrogen and oxygen play an important role. The chemical
abundance uncertainties have an effect on the models and on the opacities. In
conclusion, a conservative uncertainty of $\pm$0.5 Gyr in the ages can be
adopted, and therefore, although the 2G generation is, as expected, younger
than 1G, the age difference between 1G and 2G are within errors, and the
quantitative difference cannot be specified.

\section{Conclusions}\label{sec:Concl}

The primary aim of the {\it HST} UV Legacy Survey of Galactic GCs collaboration
was the identification of multiple stellar populations in globular clusters.
In \citet[\citetalias{2017MNRAS.464.3636M}]{2017MNRAS.464.3636M} and \citet
[\citetalias{2015ApJ...808...51M}]{2015ApJ...808...51M}, a method of stellar
population separation was applied to the RGB and MS of the sample clusters.
\citet{2012ApJ...744...58M} and \citet[\citetalias{2015MNRAS.451..312N}]
{2015MNRAS.451..312N} analyzed the SGB stars based on two-color diagrams.
We applied the methods
described in the above cited papers to separate the stellar populations from
the MS to the RGB, with the fraction of 1G and 2G stars given in Table~\ref
{tab:BonaFide}.

In the present work we derive ages for the seven bulge globular clusters
included in \citet{2015AJ....149...91P}, both for a single stellar population
as well as for the first and second stellar generations. For the derivation of
age, distance, and reddening, we employed our new code \texttt{SIRIUS} \citep
{2020arXiv200102697S}, that uses Markov chain Monte Carlo algorithm for the
fitting of the isochrones to the observed CMDs. The $\alpha$-enhanced Dartmouth
and BaSTI isochrones were used for all clusters, for the analysis of single
stellar populations (Table \ref{tab:Isoc-Results}).

As shown in Figure~\ref{fig:FEHvsAge} \citep[updated from][]
{2019ApJ...874...86S}, we derived a weighted average age of $12.86\pm0.36
$\,Gyr and $12.12\pm0.32$\,Gyr for the moderately metal-poor and for the more
metal-rich bulge clusters respectively. We cannot conclude that this corresponds
to systematic age
difference as a function of $\rm{[Fe/H]}$ among bulge clusters, 
due to a low statistics of objects, the individual age uncertainties of
500\,Myr, and moreover, given that 3 of the metal-rich clusters appear
to belong to a thick disk (and not bulge) population. The clusters NGC\,6717 and NGC\,6362
(the latter being a reference halo cluster) are revealed to be among the oldest MW clusters with
$\sim 13.5$\,Gyr \citep{2018ApJ...853...15K,2019MNRAS.484.5530K}. Adopting lower
$\rm{[\alpha/Fe]}$ values for the more metal-rich GCs can influence their ages,
as applied here for NGC\,6304 and NGC\,6352, and further detailed studies using
original (not interpolated) sets of isochrones are needed. Also, accurate
$\alpha$-element abundances from high-resolution spectroscopy are of great
interest for further improving the precision in the age derivation.

We also derived age differences between 1G and 2G populations. A weighted
mean difference produced by our statistical fitting procedure, results to be of
$41\pm170$\,Myr adopting canonical He for both and smaller for enhanced
helium, reaching $17\pm170$\,Myr. For each individual cluster, the
typical uncertainty of the age difference between 1G and 2G is around $500$\,Myr
(Table~\ref{tab:MPs-Ages}), being of the same order as the assumed error on
ages of $\pm$0.5 Gyr.

In the derivation of relative ages of 1G and 2G stellar populations, most
of the sources of  errors on absolute ages are cancelled (zero points in the
photometry and the models), but the effects of the element abundance variations
in C, N, O and in particular in He are added instead. Therefore, adopting a
conservative uncertainty of $\pm$0.5\,Gyr in the individual age
differences, these quantitative values cannot be used to constrain models on
the formation of 2G stars \citep[see][]{2015MNRAS.454.4197R,2016MNRAS.458.2122D,
2018ARA&A..56...83B}.

Models for the formation of multiple populations predict age differences
between first and second generations, from about zero up to 150\,Myr \citep
{2018MNRAS.478.2461G,2007A-A...464.1029D,2008MNRAS.391..825D,
2016MNRAS.458.2122D,2017MNRAS.464.3636M}. An uncertainty of 0.5\,Gyr is typical
for the age of each individual cluster, whereas the weighted uncertainty from
the eight studied clusters results to be of $0.17$\,Gyr. Within these
uncertainties in the present age derivations, we cannot discriminate
between %these different predictions
 the different formation scenarios. This is due to the intrinsic uncertainties
inherent to detailed physics processes taken into account in stellar evolution
models.

%Therefore, an extremely accurate determination of age
%difference, of less than 1\% of the cluster ages, would be needed to constrain
%the models. For such an aim, it would be better to rely on age differences in
%clusters of younger ages, although other problems (e.g. rotation) arise there to
%hamper accurate determinations \citep[e.g.][]{2018MNRAS.477.2640M,
%2017NatAs...1E.186D}.

The present results add to the previous literature regarding the age of the
moderately metal-poor bulge globular clusters. This has an impact on the epoch
of bulge and globular cluster formation \citep[e.~g.][]{2018ARA&A..56..223B}.
The verification that the  bulge clusters are older than 11\,Gyr is an
important information with respect to the time of bar formation. According to
\citet{2018ApJ...861...88B}, from cosmological simulations, the bar should have
formed at about $8\pm2$\,Gyr ago. Also \citet{2019MNRAS.490.4740B} estimated
that the Galactic bar formed $\sim 8$\,Gyr ago, from an analysis of chemical
abundances of field stars, from the {\it Apache Point Observatory Galactic
Evolution Experiment} (APOGEE) survey, combined with kinematical information
from the Gaia collaboration. From this age difference, we can conclude that the
globular clusters were formed early in the Galaxy, before the bar formation, and
were later trapped by the bar \citep[see also][]{2018ApJ...863...16R}.
Therefore, scenarios of bar/bulge formation have to take into account the old
ages of the bulge globular clusters. It would be interesting to extend the
analysis to the other clusters of the UV Legacy Survey of Globular Clusters.

\acknowledgments

We thank an anonymous referee for suggestions that have improved the quality
of this paper.
RAPO and SOS acknowledge the FAPESP PhD fellowships nos. 2018/22181-0 and
2018/22044-3 respectively. RAPO, LOK, BB and EB acknowledge partial financial
support from FAPESP, CNPq, and CAPES - Finance Code 001. SO, GP and DN
acknowledge partial support by the Universit\`a degli Studi di Padova Progetto
di Ateneo BIRD178590. APV acknowledges the FAPESP postdoctoral fellowship no.
2017/15893-1. AA, SC, SH and GP acknowledge partial financial support from the
``Ministerio de Ciencia, Innovaci\'on y Universidades'' of Spain, under grant
AYA2017-89841-P. AA and SH acknowledge partial financial support to the
Instituto de Astrof\'isica de Canarias, under grant 309403.
Based on observations with the NASA/ESA Hubble Space Telescope, obtained at the
Space Telescope Science Institute, which is operated by AURA, Inc., under NASA
contract NAS 5-26555.

\appendix

\section{Literature information on the sample clusters}\label{sec:ApxA}

This appendix contains a bibliographic review on the sample GCs, regarding
the photometry and age derivations by isochrone fitting. We follow here the examples and references
given in \citet{2012AJ....143...70A} and \citet{2014ApJS..210...10R}.

\subsection{NGC 6304}

\citet{2000A&A...357..495O} obtained B and V photometry of NGC\,6304 and
derived $E(B-V) \approx 0.5$, $d_{\odot} \approx 6$ kpc and [Fe/H] $\approx
-0.6$, and provides a review of previous results. \citet{2005MNRAS.361..272V}
observed this GC and NGC\,6637 in the JHK filters from ESO/NTT telescope,
and estimated $\rm{[Fe/H]}=-0.70$, $E(B-V)=0.58$ and absolute distance modulus
of $(m-M)_0 = 13.88$, i.e. a distance of $\sim 6.0$\,kpc.

NGC\,6304 contains X-ray sources, and has been extensively observed in the
study of quiescent low-mass X-ray binaries and radio pulsars \citep[and
references therein]{2009ApJ...699.1418G,2013ApJ...772....7G}.

\subsection{NGC 6352}

This cluster has several photometric studies on the literature. The most recent
is \citet{2003A&A...399..121P}, with the F606W and F814W \textit{HST} filters
(WFPC2), estimating an age of 14\,Gyr, a distance modulus of $(m-M)_0 = 13.6$
and $E(B-V)=0.25$. \citet{2002ASPC..274..373F} analyzed older \textit{HST} data
in the F555W and F814W filters, and derived an age of 12.6\,Gyr, $(m-M)_0=13.58$
and $E(B-V)=0.26$. They also compared this cluster with NGC\,6624 and NGC\,6637
\citep[studied by][]{2000AJ....120..879H}, presenting similar ages and
metallicities. Previous photometry from ground-based \citep{2000A&AS..144....5R,
1994ApJS...93..161S} and space-based \citep{1995AJ....110..652F} telescopes were obtained, but no precise age derivation was carried out.

\citet{2010A&A...524A..44P} first showed the presence of MPs in this outer
bulge/disk cluster, detecting a bimodal CN and CH anticorrelation analyzing
low-resolution spectra of MS stars. \citet{2009A-A...493..913F} derived the
metallicity ($\rm{[Fe/H]} =-0.55\pm0.03$) and abundances for $\alpha$- and
iron-peak elements, from high-resolution UVES/VLT spectra of HB stars, as
reported in Table~\ref{tab:Met-Abund}. Previously, \citet{1987A-A...177..177G}
derived $\rm{[Fe/H]}=-0.79\pm0.06$ and the other abundances, but we adopted
the \citet{2009A-A...493..913F} results in the prior.

Inside this \textit{HST} collaboration, two works were focused on the NGC\,6352
MPs: \citet[\citetalias{2015MNRAS.451..312N}]{2015MNRAS.451..312N} study the age
and helium differences between them, and \citet[\citetalias
{2019ApJ...873..109L}]{2019ApJ...873..109L} analyze their radial distribution
and kinematics.

\subsection{NGC 6624}

\citet{2000AJ....120..879H} present the $V$ vs. $V-I$ observed CMD
 based on WFPC2/\textit{HST} photometry, deriving an apparent distance
modulus of $(m-M)_V = 15.42$ and $E(V-I) = 0.42$. \citet{2011MNRAS.414.2690V}
obtained infrared spectra for 5 stars with NIRSPEC at Keck II, deriving a
metallicity [Fe/H] $=-0.69$ and abundance ratios as shown in Table~\ref
{tab:Met-Abund}, with a mean $\alpha$-element enhancement of [$\alpha$/Fe]
$=+0.39$. They also determine a radial velocity of $-$47~km~s$^{-1}$.

This cluster hosts the low-mass X-ray binary 4U/1820-30, source of X-rays and gamma-rays, that is extensively observed \citep{2014ApJ...795..116P}. It also
contains 6 millisecond pulsars \citep{1994MNRAS.267..125B, 2012ApJ...745..109L}.

\subsection{NGC 6637 (M69)}

\citet{2000AJ....120..879H} derive an apparent distance modulus $(m-M)_{V}=
15.29$ and reddening $E(V-I)= 0.24$ from WFPC2/\textit{HST} data for NGC\,6637. \citet
{2005MNRAS.361..272V} obtained near-infrared photometry with SOFI@NTT/ESO,
deriving a metallicity [Fe/H] $=-0.68$, $E(B-V)=0.14$ and
$(m-M)_0=14.87$, i.e. an heliocentric distance of $d_{\odot}=9.42$ kpc.

\subsection{NGC 6652}

\citet{1994A&A...286..444O} obtained CMDs in BVRIz, and estimated a metallicity
of $\rm{[Fe/H]}=-0.9$, and a distance to the Sun of d$_{\odot}$ = 9.3\,kpc. From
\textit{HST} CMDs in filters F555W and F814W, \citet{2000AJ....120.3102C}
obtained a distance to the Galactic center of  $R_{\rm GC} = 2$\,kpc, and
$\rm{[Fe/H]}=-0.85$. This cluster has several X-ray binaries, and pulsars
\citep[and references therein]{2012ApJ...751...62S,2015ApJ...807L..23D}.

\subsection{NGC 6717 (Palomar 9)}
 
\citet{1996A&A...311..778B} observed the first deep $V$ vs. $B-V$ CMD for this
cluster, identifying a moderately blue extended HB and estimated $E(B-V)=0.22$
and $\rm{[Fe/H]}=-1.26\pm0.10$. \cite{1999A&AS..136..237O} obtained B,V CMDs,
deriving $E(B-V)=0.23$, and a distance of d$_{\odot}=7.1$\,kpc, placing the
cluster in the outskirts of the Galactic bulge. A metallicity of $\rm{[Fe/H]}
\sim -1.3$ was estimated, and a blue horizontal branch was also identified.
\citet{1979SvA....23..284G} identified the only catalogued RR Lyrae in the 
cluster field so far, obtaining $V = 15.7$, which was used in the present
RR Lyrae analysis (Figure~\ref{fig:RRL-mag}). X-ray sources were recently
detected in this cluster \citep{2015AAS...22534523M}.

\subsection{NGC 6723}

\cite{1999A&AS..136..461A} obtained CCD CMDs in B, V, I, and deduced $E(B-V)=
0.11$, and estimated a metallicity of $\rm{[Fe/H]}=-1.22$. \cite{Gratton15}
obtained a metallicity of $\rm{[Fe/H]}=-1.22$, and derived abundances of O, N,
Na, Mg, Ca, Ni, and Ba. These authors studied in detail the Na-O anticorrelation,
indicative of a second generation stellar population, and were able to
identify their location in the horizontal branch. \citet{RojasArriagada16}
carried out a detailed abundance analysis of 7 red giants, yielding $\rm{[Fe/H]}
=-0.98$, and radial velocity of $-96\pm3.6$\,km.s$^{-1}$. Abundances of O, Mg,
Si, Ca, Ti, Na, Al and Ba are reported in Table \ref{tab:Met-Abund}.

It is interesting to note that NGC\,6717 and NGC\,6723 are similar to other
bulge clusters with blue HBs: HP\,1, NGC\,6522, NGC\,6558, AL\,3, Terzan\,10
within the inner $6^{\circ}$, and NGC\,6325, NGC\,6355, NGC\,6453, NGC\,6626,
NGC\,6642 in the outer bulge within $12^{\circ}$ of the Galactic center.

\subsection{NGC 6362}

NGC 6362 is an inner halo globular cluster, located at $\ell=-34.45^{\circ}$
and $b=-17.57^{\circ}
$, with a reddening of $E(B-V)=0.11$. High-resolution abundance analyses by
\citet{2016ApJ...824...73M} and \citet{2017MNRAS.468.1249M} indicate
metallicity values of $\rm{[Fe/H]}=-1.09 \pm 0.01$ and $\rm{[Fe/H]}=-1.07 \pm 0.01$,
respectively. \cite{2016ApJ...824...73M} found evidence of multiple stellar
populations, through a Na-O anticorrelation, and suggest that this is the
lowest mass globular cluster with multiple populations. The age of NGC~6362
has been extensively analyzed in the literature, e.g. \citet
{2005AJ....130..116D}, \citet{2006A&A...456.1085M}, \citet
{2009ApJ...694.1498M}, \citet{2010ApJ...708..698D}, \citet
{2010AJ....139..476P}, \citet{2013ApJ...775..134V}, \citet
{2017MNRAS.468.1038W} and \citet{2018ApJ...853...15K}. The latter authors
deduce an age of $\sim13$\,Gyr for this cluster.

\bibliography{HST-PaperXX}

\begin{thebibliography}{}
\expandafter\ifx\csname natexlab\endcsname\relax\def\natexlab#1{#1}\fi
\providecommand{\url}[1]{\href{#1}{#1}}

\bibitem[{{Alca{\'{\i}}no} {et~al.}(1999){Alca{\'{\i}}no}, {Liller},
  {Alvarado}, {Mironov}, {Ipatov}, {Piskunov}, {Samus}, \&
  {Smirnov}}]{1999A&AS..136..461A}
{Alca{\'{\i}}no}, G., {Liller}, W., {Alvarado}, F., {et~al.} 1999, \aaps, 136,
  461

\bibitem[{{Alonso-Garc{\'{\i}}a} {et~al.}(2012){Alonso-Garc{\'{\i}}a}, {Mateo},
  {Sen}, {Banerjee}, {Catelan}, {Minniti}, \& {von
  Braun}}]{2012AJ....143...70A}
{Alonso-Garc{\'{\i}}a}, J., {Mateo}, M., {Sen}, B., {et~al.} 2012, \aj, 143, 70

\bibitem[{{Anderson} {et~al.}(2008){Anderson}, {Sarajedini}, {Bedin}, {King},
  {Piotto}, {Reid}, {Siegel}, {Majewski}, {Paust}, {Aparicio}, {Milone},
  {Chaboyer}, \& {Rosenberg}}]{2008AJ....135.2055A}
{Anderson}, J., {Sarajedini}, A., {Bedin}, L.~R., {et~al.} 2008, \aj, 135, 2055

\bibitem[{{Barbuy} {et~al.}(2018{\natexlab{a}}){Barbuy}, {Chiappini}, \&
  {Gerhard}}]{2018ARA&A..56..223B}
{Barbuy}, B., {Chiappini}, C., \& {Gerhard}, O. 2018{\natexlab{a}}, \araa, 56,
  223

\bibitem[{{Barbuy} {et~al.}(2018{\natexlab{b}}){Barbuy}, {Muniz}, {Ortolani},
  {Ernandes}, {Dias}, {Saviane}, {Kerber}, {Bica}, {P{\'e}rez-Villegas},
  {Rossi}, \& {Held}}]{2018A&A...619A.178B}
{Barbuy}, B., {Muniz}, L., {Ortolani}, S., {et~al.} 2018{\natexlab{b}}, \aap,
  619, A178

\bibitem[{{Bastian} \& {Lardo}(2018)}]{2018ARA&A..56...83B}
{Bastian}, N., \& {Lardo}, C. 2018, \araa, 56, 83

\bibitem[{{Baumgardt} \& {Hilker}(2018)}]{2018MNRAS.478.1520B}
{Baumgardt}, H., \& {Hilker}, M. 2018, \mnras, 478, 1520

\bibitem[{{Bedin} {et~al.}(2005){Bedin}, {Cassisi}, {Castelli}, {Piotto},
  {Anderson}, {Salaris}, {Momany}, \& {Pietrinferni}}]{2005MNRAS.357.1038B}
{Bedin}, L.~R., {Cassisi}, S., {Castelli}, F., {et~al.} 2005, \mnras, 357, 1038

\bibitem[{{Bedin} {et~al.}(2004){Bedin}, {Piotto}, {Anderson}, {Cassisi},
  {King}, {Momany}, \& {Carraro}}]{2004ApJ...605L.125B}
{Bedin}, L.~R., {Piotto}, G., {Anderson}, J., {et~al.} 2004, \apjl, 605, L125

\bibitem[{{Bellini} {et~al.}(2009){Bellini}, {Piotto}, {Bedin}, {Anderson},
  {Platais}, {Momany}, {Moretti}, {Milone}, \&
  {Ortolani}}]{2009A&A...493..959B}
{Bellini}, A., {Piotto}, G., {Bedin}, L.~R., {et~al.} 2009, \aap, 493, 959

\bibitem[{{Bellini} {et~al.}(2013){Bellini}, {Piotto}, {Milone}, {King},
  {Renzini}, {Cassisi}, {Anderson}, {Bedin}, {Nardiello}, {Pietrinferni}, \&
  {Sarajedini}}]{2013ApJ...765...32B}
{Bellini}, A., {Piotto}, G., {Milone}, A.~P., {et~al.} 2013, \apj, 765, 32

\bibitem[{{Bica} {et~al.}(2016){Bica}, {Ortolani}, \&
  {Barbuy}}]{2016PASA...33...28B}
{Bica}, E., {Ortolani}, S., \& {Barbuy}, B. 2016, \pasa, 33, e028

\bibitem[{{Biggs} {et~al.}(1994){Biggs}, {Bailes}, {Lyne}, {Goss}, \&
  {Fruchter}}]{1994MNRAS.267..125B}
{Biggs}, J.~D., {Bailes}, M., {Lyne}, A.~G., {Goss}, W.~M., \& {Fruchter},
  A.~S. 1994, \mnras, 267, 125

\bibitem[{{Bovy} {et~al.}(2019){Bovy}, {Leung}, {Hunt}, {Mackereth},
  {Garc{\'\i}a-Hern{\'a}ndez}, \& {Roman-Lopes}}]{2019MNRAS.490.4740B}
{Bovy}, J., {Leung}, H.~W., {Hunt}, J. A.~S., {et~al.} 2019, \mnras, 490, 4740

\bibitem[{{Bressan} {et~al.}(2012){Bressan}, {Marigo}, {Girardi}, {Salasnich},
  {Dal Cero}, {Rubele}, \& {Nanni}}]{2012MNRAS.427..127B}
{Bressan}, A., {Marigo}, P., {Girardi}, L., {et~al.} 2012, \mnras, 427, 127

\bibitem[{{Brocato} {et~al.}(1996){Brocato}, {Buonanno}, {Malakhova}, \&
  {Piersimoni}}]{1996A&A...311..778B}
{Brocato}, E., {Buonanno}, R., {Malakhova}, Y., \& {Piersimoni}, A.~M. 1996,
  \aap, 311, 778

\bibitem[{{Brown} {et~al.}(2016){Brown}, {Cassisi}, {D'Antona}, {Salaris},
  {Milone}, {Dalessandro}, {Piotto}, {Renzini}, {Sweigart}, {Bellini},
  {Ortolani}, {Sarajedini}, {Aparicio}, {Bedin}, {Anderson}, {Pietrinferni}, \&
  {Nardiello}}]{2016ApJ...822...44B}
{Brown}, T.~M., {Cassisi}, S., {D'Antona}, F., {et~al.} 2016, \apj, 822, 44
  (Paper VII)

\bibitem[{{Buck} {et~al.}(2018){Buck}, {Ness}, {Macci{\`o}}, {Obreja}, \&
  {Dutton}}]{2018ApJ...861...88B}
{Buck}, T., {Ness}, M.~K., {Macci{\`o}}, A.~V., {Obreja}, A., \& {Dutton},
  A.~A. 2018, \apj, 861, 88

\bibitem[{{Busso} {et~al.}(2007){Busso}, {Cassisi}, {Piotto}, {Castellani},
  {Romaniello}, {Catelan}, {Djorgovski}, {Recio Blanco}, {Renzini}, {Rich},
  {Sweigart}, \& {Zoccali}}]{2007A&A...474..105B}
{Busso}, G., {Cassisi}, S., {Piotto}, G., {et~al.} 2007, \aap, 474, 105

\bibitem[{{Campos} {et~al.}(2013){Campos}, {Kepler}, {Bonatto}, \&
  {Ducati}}]{2013MNRAS.433..243C}
{Campos}, F., {Kepler}, S.~O., {Bonatto}, C., \& {Ducati}, J.~R. 2013, \mnras,
  433, 243

\bibitem[{{Carretta} {et~al.}(2009){Carretta}, {Bragaglia}, {Gratton},
  {D'Orazi}, \& {Lucatello}}]{2009A-A...508..695C}
{Carretta}, E., {Bragaglia}, A., {Gratton}, R., {D'Orazi}, V., \& {Lucatello},
  S. 2009, \aap, 508, 695

\bibitem[{{Cassisi} {et~al.}(2013){Cassisi}, {Mucciarelli}, {Pietrinferni},
  {Salaris}, \& {Ferguson}}]{2013A&A...554A..19C}
{Cassisi}, S., {Mucciarelli}, A., {Pietrinferni}, A., {Salaris}, M., \&
  {Ferguson}, J. 2013, \aap, 554, A19

\bibitem[{{Cassisi} {et~al.}(2008){Cassisi}, {Salaris}, {Pietrinferni},
  {Piotto}, {Milone}, {Bedin}, \& {Anderson}}]{2008ApJ...672L.115C}
{Cassisi}, S., {Salaris}, M., {Pietrinferni}, A., {et~al.} 2008, \apjl, 672,
  L115

\bibitem[{{Chaboyer} {et~al.}(2000){Chaboyer}, {Sarajedini}, \&
  {Armandroff}}]{2000AJ....120.3102C}
{Chaboyer}, B., {Sarajedini}, A., \& {Armandroff}, T.~E. 2000, \aj, 120, 3102

\bibitem[{{Clement} {et~al.}(2001){Clement}, {Muzzin}, {Dufton}, {Ponnampalam},
  {Wang}, {Burford}, {Richardson}, {Rosebery}, {Rowe}, \&
  {Hogg}}]{2001AJ....122.2587C}
{Clement}, C.~M., {Muzzin}, A., {Dufton}, Q., {et~al.} 2001, \aj, 122, 2587

\bibitem[{{Clementini} {et~al.}(2003){Clementini}, {Gratton}, {Bragaglia},
  {Carretta}, {Di Fabrizio}, \& {Maio}}]{2003AJ....125.1309C}
{Clementini}, G., {Gratton}, R., {Bragaglia}, A., {et~al.} 2003, \aj, 125, 1309

\bibitem[{{Cohen} {et~al.}(2018){Cohen}, {Mauro}, {Alonso-Garc{\'{\i}}a},
  {Hempel}, {Sarajedini}, {Ordo{\~n}ez}, {Geisler}, \&
  {Kalirai}}]{2018AJ....156...41C}
{Cohen}, R.~E., {Mauro}, F., {Alonso-Garc{\'{\i}}a}, J., {et~al.} 2018, \aj,
  156, 41

\bibitem[{{Cohen} {et~al.}(2014){Cohen}, {Mauro}, {Geisler}, {Moni Bidin},
  {Dotter}, \& {Bonatto}}]{2014AJ....148...18C}
{Cohen}, R.~E., {Mauro}, F., {Geisler}, D., {et~al.} 2014, \aj, 148, 18

\bibitem[{{Conroy} {et~al.}(2018){Conroy}, {Villaume}, {van Dokkum}, \&
  {Lind}}]{2018ApJ...854..139C}
{Conroy}, C., {Villaume}, A., {van Dokkum}, P.~G., \& {Lind}, K. 2018, \apj,
  854, 139

\bibitem[{{Correnti} {et~al.}(2016){Correnti}, {Gennaro}, {Kalirai}, {Brown},
  \& {Calamida}}]{2016ApJ...823...18C}
{Correnti}, M., {Gennaro}, M., {Kalirai}, J.~S., {Brown}, T.~M., \& {Calamida},
  A. 2016, \apj, 823, 18

\bibitem[{{Correnti} {et~al.}(2018){Correnti}, {Gennaro}, {Kalirai}, {Cohen},
  \& {Brown}}]{2018ApJ...864..147C}
{Correnti}, M., {Gennaro}, M., {Kalirai}, J.~S., {Cohen}, R.~E., \& {Brown},
  T.~M. 2018, \apj, 864, 147

\bibitem[{{Crestani} {et~al.}(2019){Crestani}, {Alves-Brito}, {Bono}, {Puls},
  \& {Alonso-Garc{\'\i}a}}]{2019MNRAS.tmp.1639C}
{Crestani}, J., {Alves-Brito}, A., {Bono}, G., {Puls}, A.~A., \&
  {Alonso-Garc{\'\i}a}, J. 2019, \mnras, 1639

\bibitem[{{Dambis} {et~al.}(2013){Dambis}, {Berdnikov}, {Kniazev}, {Kravtsov},
  {Rastorguev}, {Sefako}, \& {Vozyakova}}]{2013MNRAS.435.3206D}
{Dambis}, A.~K., {Berdnikov}, L.~N., {Kniazev}, A.~Y., {et~al.} 2013, \mnras,
  435, 3206

\bibitem[{{D'Antona} {et~al.}(2018){D'Antona}, {Caloi}, \&
  {Tailo}}]{2018NatAs...2..270D}
{D'Antona}, F., {Caloi}, V., \& {Tailo}, M. 2018, Nature Astronomy, 2, 270

\bibitem[{{D'Antona} {et~al.}(2017){D'Antona}, {Milone}, {Tailo}, {Ventura},
  {Vesperini}, \& {di Criscienzo}}]{2017NatAs...1E.186D}
{D'Antona}, F., {Milone}, A.~P., {Tailo}, M., {et~al.} 2017, Nature Astronomy,
  1, 0186

\bibitem[{{D'Antona} {et~al.}(2016){D'Antona}, {Vesperini}, {D'Ercole},
  {Ventura}, {Milone}, {Marino}, \& {Tailo}}]{2016MNRAS.458.2122D}
{D'Antona}, F., {Vesperini}, E., {D'Ercole}, A., {et~al.} 2016, \mnras, 458,
  2122

\bibitem[{{De Angeli} {et~al.}(2005){De Angeli}, {Piotto}, {Cassisi}, {Busso},
  {Recio-Blanco}, {Salaris}, {Aparicio}, \& {Rosenberg}}]{2005AJ....130..116D}
{De Angeli}, F., {Piotto}, G., {Cassisi}, S., {et~al.} 2005, \aj, 130, 116

\bibitem[{{DeCesar} {et~al.}(2015){DeCesar}, {Ransom}, {Kaplan}, {Ray}, \&
  {Geller}}]{2015ApJ...807L..23D}
{DeCesar}, M.~E., {Ransom}, S.~M., {Kaplan}, D.~L., {Ray}, P.~S., \& {Geller},
  A.~M. 2015, \apjl, 807, L23

\bibitem[{{Decressin} {et~al.}(2007){Decressin}, {Meynet}, {Charbonnel},
  {Prantzos}, \& {Ekstr{\"o}m}}]{2007A-A...464.1029D}
{Decressin}, T., {Meynet}, G., {Charbonnel}, C., {Prantzos}, N., \&
  {Ekstr{\"o}m}, S. 2007, \aap, 464, 1029

\bibitem[{{D'Ercole} {et~al.}(2008){D'Ercole}, {Vesperini}, {D'Antona},
  {McMillan}, \& {Recchi}}]{2008MNRAS.391..825D}
{D'Ercole}, A., {Vesperini}, E., {D'Antona}, F., {McMillan}, S. L.~W., \&
  {Recchi}, S. 2008, \mnras, 391, 825

\bibitem[{{Dias} {et~al.}(2016){Dias}, {Barbuy}, {Saviane}, {Held}, {Da Costa},
  {Ortolani}, {Gullieuszik}, \& {V{\'a}squez}}]{2016A&A...590A...9D}
{Dias}, B., {Barbuy}, B., {Saviane}, I., {et~al.} 2016, \aap, 590, A9

\bibitem[{{Dias} {et~al.}(2015){Dias}, {Barbuy}, {Saviane}, {Held}, {Da Costa},
  {Ortolani}, {Vasquez}, {Gullieuszik}, \& {Katz}}]{2015A&A...573A..13D}
---. 2015, \aap, 573, A13

\bibitem[{{Dotter} {et~al.}(2008){Dotter}, {Chaboyer}, {Jevremovi{\'c}},
  {Kostov}, {Baron}, \& {Ferguson}}]{2008ApJS..178...89D}
{Dotter}, A., {Chaboyer}, B., {Jevremovi{\'c}}, D., {et~al.} 2008, \apjs, 178,
  89

\bibitem[{{Dotter} {et~al.}(2010){Dotter}, {Sarajedini}, {Anderson},
  {Aparicio}, {Bedin}, {Chaboyer}, {Majewski}, {Mar{\'{\i}}n-Franch}, {Milone},
  {Paust}, {Piotto}, {Reid}, {Rosenberg}, \& {Siegel}}]{2010ApJ...708..698D}
{Dotter}, A., {Sarajedini}, A., {Anderson}, J., {et~al.} 2010, \apj, 708, 698

\bibitem[{{Faria} \& {Feltzing}(2002)}]{2002ASPC..274..373F}
{Faria}, D., \& {Feltzing}, S. 2002, in Astronomical Society of the Pacific
  Conference Series, Vol. 274, Observed HR Diagrams and Stellar Evolution, ed.
  T.~{Lejeune} \& J.~{Fernandes}, 373

\bibitem[{{Feltzing} {et~al.}(2009){Feltzing}, {Primas}, \&
  {Johnson}}]{2009A-A...493..913F}
{Feltzing}, S., {Primas}, F., \& {Johnson}, R.~A. 2009, \aap, 493, 913

\bibitem[{{Ferraro} {et~al.}(2016){Ferraro}, {Massari}, {Dalessandro},
  {Lanzoni}, {Origlia}, {Rich}, \& {Mucciarelli}}]{2016ApJ...828...75F}
{Ferraro}, F.~R., {Massari}, D., {Dalessandro}, E., {et~al.} 2016, \apj, 828,
  75

\bibitem[{{Foreman-Mackey} {et~al.}(2013){Foreman-Mackey}, {Hogg}, {Lang}, \&
  {Goodman}}]{2013PASP..125..306F}
{Foreman-Mackey}, D., {Hogg}, D.~W., {Lang}, D., \& {Goodman}, J. 2013, \pasp,
  125, 306

\bibitem[{{Fullton} {et~al.}(1995){Fullton}, {Carney}, {Olszewski}, {Zinn},
  {Demarque}, {Janes}, {Da Costa}, \& {Seitzer}}]{1995AJ....110..652F}
{Fullton}, L.~K., {Carney}, B.~W., {Olszewski}, E.~W., {et~al.} 1995, \aj, 110,
  652

\bibitem[{{Gaia Collaboration} {et~al.}(2016){Gaia Collaboration}, {Prusti},
  {de Bruijne}, {Brown}, {Vallenari}, {Babusiaux}, {Bailer-Jones}, {Bastian},
  {Biermann}, {Evans}, \& et~al.}]{2016A&A...595A...1G}
{Gaia Collaboration}, {Prusti}, T., {de Bruijne}, J.~H.~J., {et~al.} 2016,
  \aap, 595, A1

\bibitem[{{Gaia Collaboration} {et~al.}(2017){Gaia Collaboration},
  {Clementini}, {Eyer}, {Ripepi}, {Marconi}, {Muraveva}, {Garofalo}, {Sarro},
  {Palmer}, {Luri}, \& et~al.}]{2017A&A...605A..79G}
{Gaia Collaboration}, {Clementini}, G., {Eyer}, L., {et~al.} 2017, \aap, 605,
  A79

\bibitem[{{Gaia Collaboration} {et~al.}(2018{\natexlab{a}}){Gaia
  Collaboration}, {Brown}, {Vallenari}, {Prusti}, {de Bruijne}, {Babusiaux},
  {Bailer-Jones}, {Biermann}, {Evans}, {Eyer}, \& et~al.}]{2018A&A...616A...1G}
{Gaia Collaboration}, {Brown}, A.~G.~A., {Vallenari}, A., {et~al.}
  2018{\natexlab{a}}, \aap, 616, A1

\bibitem[{{Gaia Collaboration} {et~al.}(2018{\natexlab{b}}){Gaia
  Collaboration}, {Helmi}, {van Leeuwen}, {McMillan}, {Massari}, {Antoja},
  {Robin}, {Lindegren}, {Bastian}, \& {Arenou}}]{2018A&A...616A..12G}
{Gaia Collaboration}, {Helmi}, A., {van Leeuwen}, F., {et~al.}
  2018{\natexlab{b}}, \aap, 616, A12

\bibitem[{{Gieles} {et~al.}(2018){Gieles}, {Charbonnel}, {Krause},
  {H{\'e}nault-Brunet}, {Agertz}, {Lamers}, {Bastian}, {Gualand ris}, {Zocchi},
  \& {Petts}}]{2018MNRAS.478.2461G}
{Gieles}, M., {Charbonnel}, C., {Krause}, M. G.~H., {et~al.} 2018, \mnras, 478,
  2461

\bibitem[{{Goranskii}(1979)}]{1979SvA....23..284G}
{Goranskii}, V.~P. 1979, \sovast, 23, 510

\bibitem[{{Gratton}(1987)}]{1987A-A...177..177G}
{Gratton}, R.~C. 1987, \aap, 177, 177

\bibitem[{{Gratton} {et~al.}(2012){Gratton}, {Carretta}, \&
  {Bragaglia}}]{2012A&ARv..20...50G}
{Gratton}, R.~G., {Carretta}, E., \& {Bragaglia}, A. 2012, \aapr, 20, 50

\bibitem[{{Gratton} {et~al.}(2015){Gratton}, {Lucatello}, {Sollima},
  {Carretta}, {Bragaglia}, {Momany}, {D'Orazi}, {Salaris}, {Cassisi}, \&
  {Stetson}}]{Gratton15}
{Gratton}, R.~G., {Lucatello}, S., {Sollima}, A., {et~al.} 2015, \aap, 573, A92

\bibitem[{{Guillot} {et~al.}(2009){Guillot}, {Rutledge}, {Brown}, {Pavlov}, \&
  {Zavlin}}]{2009ApJ...699.1418G}
{Guillot}, S., {Rutledge}, R.~E., {Brown}, E.~F., {Pavlov}, G.~G., \& {Zavlin},
  V.~E. 2009, \apj, 699, 1418

\bibitem[{{Guillot} {et~al.}(2013){Guillot}, {Servillat}, {Webb}, \&
  {Rutledge}}]{2013ApJ...772....7G}
{Guillot}, S., {Servillat}, M., {Webb}, N.~A., \& {Rutledge}, R.~E. 2013, \apj,
  772, 7

\bibitem[{{Harris}(1996)}]{1996AJ....112.1487H}
{Harris}, W.~E. 1996, \aj, 112, 1487

\bibitem[{{Heasley} {et~al.}(2000){Heasley}, {Janes}, {Zinn}, {Demarque}, {Da
  Costa}, \& {Christian}}]{2000AJ....120..879H}
{Heasley}, J.~N., {Janes}, K.~A., {Zinn}, R., {et~al.} 2000, \aj, 120, 879

\bibitem[{{Hesser} {et~al.}(1977){Hesser}, {Hartwick}, \&
  {McClure}}]{1977ApJS...33..471H}
{Hesser}, J.~E., {Hartwick}, F.~D.~A., \& {McClure}, R.~D. 1977, \apjs, 33, 471

\bibitem[{{Hidalgo} {et~al.}(2018){Hidalgo}, {Pietrinferni}, {Cassisi},
  {Salaris}, {Mucciarelli}, {Savino}, {Aparicio}, {Silva Aguirre}, \&
  {Verma}}]{2018ApJ...856..125H}
{Hidalgo}, S.~L., {Pietrinferni}, A., {Cassisi}, S., {et~al.} 2018, \apj, 856,
  125

\bibitem[{{J{\"o}nsson} {et~al.}(2017){J{\"o}nsson}, {Ryde}, {Schultheis}, \&
  {Zoccali}}]{2017A&A...598A.101J}
{J{\"o}nsson}, H., {Ryde}, N., {Schultheis}, M., \& {Zoccali}, M. 2017, \aap,
  598, A101

\bibitem[{{Joo} \& {Lee}(2013)}]{2013ApJ...762...36J}
{Joo}, S.~J., \& {Lee}, Y.~W. 2013, \apj, 762, 36

\bibitem[{{Kerber} {et~al.}(2018){Kerber}, {Nardiello}, {Ortolani}, {Barbuy},
  {Bica}, {Cassisi}, {Libralato}, \& {Vieira}}]{2018ApJ...853...15K}
{Kerber}, L.~O., {Nardiello}, D., {Ortolani}, S., {et~al.} 2018, \apj, 853, 15

\bibitem[{{Kerber} {et~al.}(2007){Kerber}, {Santiago}, \&
  {Brocato}}]{2007A&A...462..139K}
{Kerber}, L.~O., {Santiago}, B.~X., \& {Brocato}, E. 2007, \aap, 462, 139

\bibitem[{{Kerber} {et~al.}(2019){Kerber}, {Libralato}, {Souza}, {Oliveira},
  {Ortolani}, {P{\'e}rez-Villegas}, {Barbuy}, {Dias}, {Bica}, \&
  {Nardiello}}]{2019MNRAS.484.5530K}
{Kerber}, L.~O., {Libralato}, M., {Souza}, S.~O., {et~al.} 2019, \mnras, 484,
  5530

\bibitem[{{Kraft}(1994)}]{1994PASP..106..553K}
{Kraft}, R.~P. 1994, \pasp, 106, 553

\bibitem[{{Lagioia} {et~al.}(2014){Lagioia}, {Milone}, {Stetson}, {Bono},
  {Prada Moroni}, {Dall'Ora}, {Aparicio}, {Buonanno}, {Calamida}, {Ferraro},
  {Gilmozzi}, {Iannicola}, {Matsunaga}, {Monelli}, \&
  {Walker}}]{2014ApJ...782...50L}
{Lagioia}, E.~P., {Milone}, A.~P., {Stetson}, P.~B., {et~al.} 2014, \apj, 782,
  50

\bibitem[{{Lagioia} {et~al.}(2018){Lagioia}, {Milone}, {Marino}, {Cassisi},
  {Aparicio}, {Piotto}, {Anderson}, {Barbuy}, {Bedin}, {Bellini}, {Brown},
  {D'Antona}, {Nardiello}, {Ortolani}, {Pietrinferni}, {Renzini}, {Salaris},
  {Sarajedini}, {van der Marel}, \& {Vesperini}}]{2018MNRAS.475.4088L}
{Lagioia}, E.~P., {Milone}, A.~P., {Marino}, A.~F., {et~al.} 2018, \mnras, 475,
  4088 (Paper XII)

\bibitem[{{Lee}(2007)}]{2007RMxAC..28..120L}
{Lee}, J.-W. 2007, in Revista Mexicana de Astronomia y Astrofisica Conference
  Series, Vol.~28, Revista Mexicana de Astronomia y Astrofisica Conference
  Series, ed. S.~{Kurtz}, 120--124

\bibitem[{{Lee} {et~al.}(1999){Lee}, {Joo}, {Sohn}, {Rey}, {Lee}, \&
  {Walker}}]{1999Natur.402...55L}
{Lee}, Y.~W., {Joo}, J.~M., {Sohn}, Y.~J., {et~al.} 1999, \nat, 402, 55

\bibitem[{{Lee} {et~al.}(2013){Lee}, {Han}, {Joo}, {Jang}, {Na}, {Okamoto},
  {Arimoto}, {Lim}, {Kim}, \& {Yoon}}]{2013ApJ...778L..13L}
{Lee}, Y.-W., {Han}, S.-I., {Joo}, S.-J., {et~al.} 2013, \apjl, 778, L13

\bibitem[{{Libralato} {et~al.}(2019){Libralato}, {Bellini}, {Piotto},
  {Nardiello}, {van der Marel}, {Anderson}, {Bedin}, \&
  {Vesperini}}]{2019ApJ...873..109L}
{Libralato}, M., {Bellini}, A., {Piotto}, G., {et~al.} 2019, \apj, 873, 109
  (Paper XVIII)

\bibitem[{{Lynch} {et~al.}(2012){Lynch}, {Freire}, {Ransom}, \&
  {Jacoby}}]{2012ApJ...745..109L}
{Lynch}, R.~S., {Freire}, P.~C.~C., {Ransom}, S.~M., \& {Jacoby}, B.~A. 2012,
  \apj, 745, 109

\bibitem[{{Mar{\'{\i}}n-Franch} {et~al.}(2009){Mar{\'{\i}}n-Franch},
  {Aparicio}, {Piotto}, {Rosenberg}, {Chaboyer}, {Sarajedini}, {Siegel},
  {Anderson}, {Bedin}, {Dotter}, {Hempel}, {King}, {Majewski}, {Milone},
  {Paust}, \& {Reid}}]{2009ApJ...694.1498M}
{Mar{\'{\i}}n-Franch}, A., {Aparicio}, A., {Piotto}, G., {et~al.} 2009, \apj,
  694, 1498

\bibitem[{{Marino} {et~al.}(2012{\natexlab{a}}){Marino}, {Milone}, {Piotto},
  {Cassisi}, {D'Antona}, {Anderson}, {Aparicio}, {Bedin}, {Renzini}, \&
  {Villanova}}]{2012ApJ...746...14M}
{Marino}, A.~F., {Milone}, A.~P., {Piotto}, G., {et~al.} 2012{\natexlab{a}},
  \apj, 746, 14

\bibitem[{{Marino} {et~al.}(2012{\natexlab{b}}){Marino}, {Milone}, {Sneden},
  {Bergemann}, {Kraft}, {Wallerstein}, {Cassisi}, {Aparicio}, {Asplund},
  {Bedin}, {Hilker}, {Lind}, {Momany}, {Piotto}, {Roederer}, {Stetson}, \&
  {Zoccali}}]{2012A&A...541A..15M}
{Marino}, A.~F., {Milone}, A.~P., {Sneden}, C., {et~al.} 2012{\natexlab{b}},
  \aap, 541, A15

\bibitem[{{Massari} {et~al.}(2017){Massari}, {Mucciarelli}, {Dalessandro},
  {Bellazzini}, {Cassisi}, {Fiorentino}, {Ibata}, {Lardo}, \&
  {Salaris}}]{2017MNRAS.468.1249M}
{Massari}, D., {Mucciarelli}, A., {Dalessandro}, E., {et~al.} 2017, \mnras,
  468, 1249

\bibitem[{{Meissner} \& {Weiss}(2006)}]{2006A&A...456.1085M}
{Meissner}, F., \& {Weiss}, A. 2006, \aap, 456, 1085

\bibitem[{{Milone} {et~al.}(2008){Milone}, {Bedin}, {Piotto}, {Anderson},
  {King}, {Sarajedini}, {Dotter}, {Chaboyer}, {Mar{\'{\i}}n-Franch},
  {Majewski}, {Aparicio}, {Hempel}, {Paust}, {Reid}, {Rosenberg}, \&
  {Siegel}}]{2008ApJ...673..241M}
{Milone}, A.~P., {Bedin}, L.~R., {Piotto}, G., {et~al.} 2008, \apj, 673, 241

\bibitem[{{Milone} {et~al.}(2012{\natexlab{a}}){Milone}, {Piotto}, {Bedin},
  {Aparicio}, {Anderson}, {Sarajedini}, {Marino}, {Moretti}, {Davies},
  {Chaboyer}, {Dotter}, {Hempel}, {Mar{\'{\i}}n-Franch}, {Majewski}, {Paust},
  {Reid}, {Rosenberg}, \& {Siegel}}]{2012A&A...540A..16M}
{Milone}, A.~P., {Piotto}, G., {Bedin}, L.~R., {et~al.} 2012{\natexlab{a}},
  \aap, 540, A16

\bibitem[{{Milone} {et~al.}(2012{\natexlab{b}}){Milone}, {Piotto}, {Bedin},
  {King}, {Anderson}, {Marino}, {Bellini}, {Gratton}, {Renzini}, {Stetson},
  {Cassisi}, {Aparicio}, {Bragaglia}, {Carretta}, {D'Antona}, {Di Criscienzo},
  {Lucatello}, {Monelli}, \& {Pietrinferni}}]{2012ApJ...744...58M}
---. 2012{\natexlab{b}}, \apj, 744, 58

\bibitem[{{Milone} {et~al.}(2014){Milone}, {Marino}, {Dotter}, {Norris},
  {Jerjen}, {Piotto}, {Cassisi}, {Bedin}, {Recio Blanco}, {Sarajedini},
  {Asplund}, {Monelli}, \& {Aparicio}}]{2014ApJ...785...21M}
{Milone}, A.~P., {Marino}, A.~F., {Dotter}, A., {et~al.} 2014, \apj, 785, 21

\bibitem[{{Milone} {et~al.}(2015){Milone}, {Marino}, {Piotto}, {Renzini},
  {Bedin}, {Anderson}, {Cassisi}, {D'Antona}, {Bellini}, {Jerjen},
  {Pietrinferni}, \& {Ventura}}]{2015ApJ...808...51M}
{Milone}, A.~P., {Marino}, A.~F., {Piotto}, G., {et~al.} 2015, \apj, 808, 51
  (Paper III)

\bibitem[{{Milone} {et~al.}(2017){Milone}, {Piotto}, {Renzini}, {Marino},
  {Bedin}, {Vesperini}, {D'Antona}, {Nardiello}, {Anderson}, {King}, {Yong},
  {Bellini}, {Aparicio}, {Barbuy}, {Brown}, {Cassisi}, {Ortolani}, {Salaris},
  {Sarajedini}, \& {van der Marel}}]{2017MNRAS.464.3636M}
{Milone}, A.~P., {Piotto}, G., {Renzini}, A., {et~al.} 2017, \mnras, 464, 3636
  (Paper IX)

\bibitem[{{Milone} {et~al.}(2018{\natexlab{a}}){Milone}, {Marino}, {Di
  Criscienzo}, {D'Antona}, {Bedin}, {Da Costa}, {Piotto}, {Tailo}, {Dotter},
  {Angeloni}, {Anderson}, {Jerjen}, {Li}, {Dupree}, {Granata}, {Lagioia},
  {Mackey}, {Nardiello}, \& {Vesperini}}]{2018MNRAS.477.2640M}
{Milone}, A.~P., {Marino}, A.~F., {Di Criscienzo}, M., {et~al.}
  2018{\natexlab{a}}, \mnras, 477, 2640

\bibitem[{{Milone} {et~al.}(2018{\natexlab{b}}){Milone}, {Marino}, {Renzini},
  {D'Antona}, {Anderson}, {Barbuy}, {Bedin}, {Bellini}, {Brown}, {Cassisi},
  {Cordoni}, {Lagioia}, {Nardiello}, {Ortolani}, {Piotto}, {Sarajedini},
  {Tailo}, {van der Marel}, \& {Vesperini}}]{2018MNRAS.481.5098M}
{Milone}, A.~P., {Marino}, A.~F., {Renzini}, A., {et~al.} 2018{\natexlab{b}},
  \mnras, 481, 5098 (Paper XVI)

\bibitem[{{Morris} \& {Mitchel}(2015)}]{2015AAS...22534523M}
{Morris}, D.~C., \& {Mitchel}, R. 2015, in American Astronomical Society
  Meeting Abstracts, Vol. 225, American Astronomical Society Meeting Abstracts
  \#225, 345.23

\bibitem[{{Mucciarelli} {et~al.}(2016){Mucciarelli}, {Dalessandro}, {Massari},
  {Bellazzini}, {Ferraro}, {Lanzoni}, {Lardo}, {Salaris}, \&
  {Cassisi}}]{2016ApJ...824...73M}
{Mucciarelli}, A., {Dalessandro}, E., {Massari}, D., {et~al.} 2016, \apj, 824,
  73

\bibitem[{{Muraveva} {et~al.}(2018){Muraveva}, {Delgado}, {Clementini},
  {Sarro}, \& {Garofalo}}]{2018MNRAS.481.1195M}
{Muraveva}, T., {Delgado}, H.~E., {Clementini}, G., {Sarro}, L.~M., \&
  {Garofalo}, A. 2018, \mnras, 481, 1195

\bibitem[{{Nardiello} {et~al.}(2015){Nardiello}, {Piotto}, {Milone}, {Marino},
  {Bedin}, {Anderson}, {Aparicio}, {Bellini}, {Cassisi}, {D'Antona}, {Hidalgo},
  {Ortolani}, {Pietrinferni}, {Renzini}, {Salaris}, {Marel}, \&
  {Vesperini}}]{2015MNRAS.451..312N}
{Nardiello}, D., {Piotto}, G., {Milone}, A.~P., {et~al.} 2015, \mnras, 451, 312
  (Paper IV)

\bibitem[{{Nardiello} {et~al.}(2018){Nardiello}, {Libralato}, {Piotto},
  {Anderson}, {Bellini}, {Aparicio}, {Bedin}, {Cassisi}, {Granata}, {King},
  {Lucertini}, {Marino}, {Milone}, {Ortolani}, {Platais}, \& {van der
  Marel}}]{2018MNRAS.481.3382N}
{Nardiello}, D., {Libralato}, M., {Piotto}, G., {et~al.} 2018, \mnras, 481,
  3382 (Paper XVII)

\bibitem[{{O'Malley} {et~al.}(2017){O'Malley}, {Gilligan}, \&
  {Chaboyer}}]{2017ApJ...838..162O}
{O'Malley}, E.~M., {Gilligan}, C., \& {Chaboyer}, B. 2017, \apj, 838, 162

\bibitem[{{Ortolani} {et~al.}(1999){Ortolani}, {Barbuy}, \&
  {Bica}}]{1999A&AS..136..237O}
{Ortolani}, S., {Barbuy}, B., \& {Bica}, E. 1999, \aaps, 136, 237

\bibitem[{{Ortolani} {et~al.}(1994){Ortolani}, {Bica}, \&
  {Barbuy}}]{1994A&A...286..444O}
{Ortolani}, S., {Bica}, E., \& {Barbuy}, B. 1994, \aap, 286, 444

\bibitem[{{Ortolani} {et~al.}(2000){Ortolani}, {Momany}, {Bica}, \&
  {Barbuy}}]{2000A&A...357..495O}
{Ortolani}, S., {Momany}, Y., {Bica}, E., \& {Barbuy}, B. 2000, \aap, 357, 495

\bibitem[{{Ortolani} {et~al.}(2019){Ortolani}, {Held}, {Nardiello}, {Souza},
  {Barbuy}, {P{\'e}rez-Villegas}, {Cassisi}, {Bica}, {Momany}, \&
  {Saviane}}]{2019A&A...627A.145O}
{Ortolani}, S., {Held}, E.~V., {Nardiello}, D., {et~al.} 2019, \aap, 627, A145

\bibitem[{{Osborn}(1971)}]{1971Obs....91..223O}
{Osborn}, W. 1971, The Observatory, 91, 223

\bibitem[{{Pancino} {et~al.}(2010){Pancino}, {Rejkuba}, {Zoccali}, \&
  {Carrera}}]{2010A&A...524A..44P}
{Pancino}, E., {Rejkuba}, M., {Zoccali}, M., \& {Carrera}, R. 2010, \aap, 524,
  A44

\bibitem[{{Paust} {et~al.}(2010){Paust}, {Reid}, {Piotto}, {Aparicio},
  {Anderson}, {Sarajedini}, {Bedin}, {Chaboyer}, {Dotter}, {Hempel},
  {Majewski}, {Mar{\'{\i}}n-Franch}, {Milone}, {Rosenberg}, \&
  {Siegel}}]{2010AJ....139..476P}
{Paust}, N.~E.~Q., {Reid}, I.~N., {Piotto}, G., {et~al.} 2010, \aj, 139, 476

\bibitem[{{P{\'e}rez-Villegas} {et~al.}(2020){P{\'e}rez-Villegas}, {Barbuy},
  {Kerber}, {Ortolani}, {Souza}, \& {Bica}}]{2019MNRAS.tmp.2773P}
{P{\'e}rez-Villegas}, A., {Barbuy}, B., {Kerber}, L., {et~al.} 2020, \mnras,
  491, 3251

\bibitem[{{Peuten} {et~al.}(2014){Peuten}, {Brockamp}, {K{\"u}pper}, \&
  {Kroupa}}]{2014ApJ...795..116P}
{Peuten}, M., {Brockamp}, M., {K{\"u}pper}, A.~H.~W., \& {Kroupa}, P. 2014,
  \apj, 795, 116

\bibitem[{{Pietrinferni} {et~al.}(2006){Pietrinferni}, {Cassisi}, {Salaris}, \&
  {Castelli}}]{2006ApJ...642..797P}
{Pietrinferni}, A., {Cassisi}, S., {Salaris}, M., \& {Castelli}, F. 2006, \apj,
  642, 797

\bibitem[{{Pilachowski} {et~al.}(1982){Pilachowski}, {Leep}, {Wallerstein}, \&
  {Peterson}}]{1982ApJ...263..187P}
{Pilachowski}, C., {Leep}, E.~M., {Wallerstein}, G., \& {Peterson}, R.~C. 1982,
  \apj, 263, 187

\bibitem[{{Piotto}(2009)}]{2009IAUS..258..233P}
{Piotto}, G. 2009, in IAU Symposium, Vol. 258, The Ages of Stars, ed. E.~E.
  {Mamajek}, D.~R. {Soderblom}, \& R.~F.~G. {Wyse}, 233--244

\bibitem[{{Piotto} {et~al.}(2005){Piotto}, {Villanova}, {Bedin}, {Gratton},
  {Cassisi}, {Momany}, {Recio-Blanco}, {Lucatello}, {Anderson}, {King},
  {Pietrinferni}, \& {Carraro}}]{2005ApJ...621..777P}
{Piotto}, G., {Villanova}, S., {Bedin}, L.~R., {et~al.} 2005, \apj, 621, 777

\bibitem[{{Piotto} {et~al.}(2015){Piotto}, {Milone}, {Bedin}, {Anderson},
  {King}, {Marino}, {Nardiello}, {Aparicio}, {Barbuy}, {Bellini}, {Brown},
  {Cassisi}, {Cool}, {Cunial}, {Dalessandro}, {D'Antona}, {Ferraro}, {Hidalgo},
  {Lanzoni}, {Monelli}, {Ortolani}, {Renzini}, {Salaris}, {Sarajedini}, {van
  der Marel}, {Vesperini}, \& {Zoccali}}]{2015AJ....149...91P}
{Piotto}, G., {Milone}, A.~P., {Bedin}, L.~R., {et~al.} 2015, \aj, 149, 91
  (Paper I)

\bibitem[{{Pulone} {et~al.}(2003){Pulone}, {De Marchi}, {Covino}, \&
  {Paresce}}]{2003A&A...399..121P}
{Pulone}, L., {De Marchi}, G., {Covino}, S., \& {Paresce}, F. 2003, \aap, 399,
  121

\bibitem[{{Renzini} {et~al.}(2015){Renzini}, {D'Antona}, {Cassisi}, {King},
  {Milone}, {Ventura}, {Anderson}, {Bedin}, {Bellini}, {Brown}, {Piotto}, {van
  der Marel}, {Barbuy}, {Dalessandro}, {Hidalgo}, {Marino}, {Ortolani},
  {Salaris}, \& {Sarajedini}}]{2015MNRAS.454.4197R}
{Renzini}, A., {D'Antona}, F., {Cassisi}, S., {et~al.} 2015, \mnras, 454, 4197
  (Paper V)

\bibitem[{{Renzini} {et~al.}(2018){Renzini}, {Gennaro}, {Zoccali}, {Brown},
  {Anderson}, {Minniti}, {Sahu}, {Valenti}, \&
  {VandenBerg}}]{2018ApJ...863...16R}
{Renzini}, A., {Gennaro}, M., {Zoccali}, M., {et~al.} 2018, \apj, 863, 16

\bibitem[{{Rich} {et~al.}(1997){Rich}, {Sosin}, {Djorgovski}, {Piotto}, {King},
  {Renzini}, {Phinney}, {Dorman}, {Liebert}, \& {Meylan}}]{1997ApJ...484L..25R}
{Rich}, R.~M., {Sosin}, C., {Djorgovski}, S.~G., {et~al.} 1997, \apjl, 484, L25

\bibitem[{{Roediger} {et~al.}(2014){Roediger}, {Courteau}, {Graves}, \&
  {Schiavon}}]{2014ApJS..210...10R}
{Roediger}, J.~C., {Courteau}, S., {Graves}, G., \& {Schiavon}, R.~P. 2014,
  \apjs, 210, 10

\bibitem[{{Roh} {et~al.}(2011){Roh}, {Lee}, {Joo}, {Han}, {Sohn}, \&
  {Lee}}]{2011ApJ...733L..45R}
{Roh}, D.-G., {Lee}, Y.-W., {Joo}, S.-J., {et~al.} 2011, \apjl, 733, L45

\bibitem[{{Rojas-Arriagada} {et~al.}(2016){Rojas-Arriagada}, {Zoccali},
  {V{\'a}squez}, {Ripepi}, {Musella}, {Marconi}, {Grado}, \&
  {Limatola}}]{RojasArriagada16}
{Rojas-Arriagada}, A., {Zoccali}, M., {V{\'a}squez}, S., {et~al.} 2016, \aap,
  587, A95

\bibitem[{{Rosenberg} {et~al.}(2000){Rosenberg}, {Piotto}, {Saviane}, \&
  {Aparicio}}]{2000A&AS..144....5R}
{Rosenberg}, A., {Piotto}, G., {Saviane}, I., \& {Aparicio}, A. 2000, \aaps,
  144, 5

\bibitem[{{Salinas} {et~al.}(2019){Salinas}, {Vivas}, \& {Contreras
  Ramos}}]{2019AJ....157...47S}
{Salinas}, R., {Vivas}, A.~K., \& {Contreras Ramos}, R. 2019, \aj, 157, 47

\bibitem[{{Sandage}(1993)}]{1993AJ....106..703S}
{Sandage}, A. 1993, \aj, 106, 703

\bibitem[{{Saracino} {et~al.}(2016){Saracino}, {Dalessandro}, {Ferraro},
  {Geisler}, {Mauro}, {Lanzoni}, {Origlia}, {Miocchi}, {Cohen}, {Villanova}, \&
  {Moni Bidin}}]{2016ApJ...832...48S}
{Saracino}, S., {Dalessandro}, E., {Ferraro}, F.~R., {et~al.} 2016, \apj, 832,
  48

\bibitem[{{Saracino} {et~al.}(2019){Saracino}, {Dalessandro}, {Ferraro},
  {Lanzoni}, {Geisler}, {Cohen}, {Bellini}, {Vesperini}, {Salaris}, {Cassisi},
  {Pietrinferni}, {Origlia}, {Mauro}, {Villanova}, \& {Moni
  Bidin}}]{2019ApJ...874...86S}
---. 2019, \apj, 874, 86

\bibitem[{{Sarajedini} \& {Norris}(1994)}]{1994ApJS...93..161S}
{Sarajedini}, A., \& {Norris}, J.~E. 1994, \apjs, 93, 161

\bibitem[{{Sarajedini} {et~al.}(2007){Sarajedini}, {Bedin}, {Chaboyer},
  {Dotter}, {Siegel}, {Anderson}, {Aparicio}, {King}, {Majewski},
  {Mar{\'{\i}}n-Franch}, {Piotto}, {Reid}, \&
  {Rosenberg}}]{2007AJ....133.1658S}
{Sarajedini}, A., {Bedin}, L.~R., {Chaboyer}, B., {et~al.} 2007, \aj, 133, 1658

\bibitem[{{Sbordone} {et~al.}(2011){Sbordone}, {Salaris}, {Weiss}, \&
  {Cassisi}}]{2011A&A...534A...9S}
{Sbordone}, L., {Salaris}, M., {Weiss}, A., \& {Cassisi}, S. 2011, \aap, 534,
  A9

\bibitem[{{Sirianni} {et~al.}(2005){Sirianni}, {Jee}, {Ben{\'{\i}}tez},
  {Blakeslee}, {Martel}, {Meurer}, {Clampin}, {De Marchi}, {Ford}, {Gilliland},
  {Hartig}, {Illingworth}, {Mack}, \& {McCann}}]{2005PASP..117.1049S}
{Sirianni}, M., {Jee}, M.~J., {Ben{\'{\i}}tez}, N., {et~al.} 2005, \pasp, 117,
  1049

\bibitem[{{Soszy{\'n}ski} {et~al.}(2014){Soszy{\'n}ski}, {Udalski},
  {Szyma{\'n}ski}, {Pietrukowicz}, {Mr{\'o}z}, {Skowron}, {Koz{\l}owski},
  {Poleski}, {Skowron}, {Pietrzy{\'n}ski}, {Wyrzykowski}, {Ulaczyk}, \&
  {Kubiak}}]{2014AcA....64..177S}
{Soszy{\'n}ski}, I., {Udalski}, A., {Szyma{\'n}ski}, M.~K., {et~al.} 2014,
  \actaa, 64, 177

\bibitem[{{Souza} {et~al.}(2020){Souza}, {Kerber}, {Barbuy},
  {P{\'e}rez-Villegas}, {Oliveira}, \& {Nardiello}}]{2020arXiv200102697S}
{Souza}, S.~O., {Kerber}, L.~O., {Barbuy}, B., {et~al.} 2020, arXiv e-prints,
  arXiv:2001.02697

\bibitem[{{Stacey} {et~al.}(2012){Stacey}, {Heinke}, {Cohn}, {Lugger}, \&
  {Bahramian}}]{2012ApJ...751...62S}
{Stacey}, W.~S., {Heinke}, C.~O., {Cohn}, H.~N., {Lugger}, P.~M., \&
  {Bahramian}, A. 2012, \apj, 751, 62

\bibitem[{{Valenti} {et~al.}(2005){Valenti}, {Origlia}, \&
  {Ferraro}}]{2005MNRAS.361..272V}
{Valenti}, E., {Origlia}, L., \& {Ferraro}, F.~R. 2005, \mnras, 361, 272

\bibitem[{{Valenti} {et~al.}(2011){Valenti}, {Origlia}, \&
  {Rich}}]{2011MNRAS.414.2690V}
{Valenti}, E., {Origlia}, L., \& {Rich}, R.~M. 2011, \mnras, 414, 2690

\bibitem[{{VandenBerg} {et~al.}(2013){VandenBerg}, {Brogaard}, {Leaman}, \&
  {Casagrande}}]{2013ApJ...775..134V}
{VandenBerg}, D.~A., {Brogaard}, K., {Leaman}, R., \& {Casagrande}, L. 2013,
  \apj, 775, 134

\bibitem[{{V{\'a}squez} {et~al.}(2018){V{\'a}squez}, {Saviane}, {Held}, {Da
  Costa}, {Dias}, {Gullieuszik}, {Barbuy}, {Ortolani}, \&
  {Zoccali}}]{2018A-A...619A..13V}
{V{\'a}squez}, S., {Saviane}, I., {Held}, E.~V., {et~al.} 2018, \aap, 619, A13

\bibitem[{{Villanova} {et~al.}(2014){Villanova}, {Geisler}, {Gratton}, \&
  {Cassisi}}]{2014ApJ...791..107V}
{Villanova}, S., {Geisler}, D., {Gratton}, R.~G., \& {Cassisi}, S. 2014, \apj,
  791, 107

\bibitem[{{Wagner-Kaiser} {et~al.}(2017){Wagner-Kaiser}, {Sarajedini}, {von
  Hippel}, {Stenning}, {van Dyk}, {Jeffery}, {Robinson}, {Stein}, {Anderson},
  \& {Jefferys}}]{2017MNRAS.468.1038W}
{Wagner-Kaiser}, R., {Sarajedini}, A., {von Hippel}, T., {et~al.} 2017, \mnras,
  468, 1038

\bibitem[{{Weiland} {et~al.}(1994){Weiland}, {Arendt}, {Berriman}, {Dwek},
  {Freudenreich}, {Hauser}, {Kelsall}, {Lisse}, {Mitra}, {Moseley}, {Odegard},
  {Silverberg}, {Sodroski}, {Spiesman}, \& {Stemwedel}}]{1994ApJ...425L..81W}
{Weiland}, J.~L., {Arendt}, R.~G., {Berriman}, G.~B., {et~al.} 1994, \apjl,
  425, L81

\end{thebibliography}

\end{document}